Technology & Engineering Division

# Engineering Study of Sector Magnet for the Daeδalus Experiment

# Preliminary Design and Analysis

September 21, 2012

MIT – Plasma Science and Fusion Center
Technology and Engineering Division

Contributing Authors:

MIT – PSFC

Technology and Engineering Division

Joseph Minervini

Mike Cheadle

Val Fishman

Craig Miller

Alexi Radovinsky

Brad Smith

## Table of Contents

## List of Figures

# List of Tables

## 1. Introduction

The Daedalus experiment seeks to evaluate neutrino scattering effects that go beyond the standard model. Modular accelerators are employed to produce 800 Mev proton beams at the megawatt power level directed toward a target, producing neutrinos. The Superconducting Ring Cyclotron (SRC) consists of identical sectors (currently 6) of superconducting dipole magnets with iron return frames. The Daedalus Collaboration has produced a conceptual design for the magnet, which, after several iterations, is the current best design that achieves the physics requirements of the experiment. This design is still evolving. Scientists at MIT are members of the Daedalus Collaboration.

The Technology and Engineering Division (T&ED) of the MIT Plasma Science and Fusion Center was awarded with a contract by the Daedalus team to further develop the magnet conceptual design and to provide very preliminary cost estimates to fabricate a single magnet sector and to produce the complete 6-sector ring.

T&ED work on this project started with a preliminary stage, during which alternative magnetic designs were evaluated from various perspectives, including manufacturability, and the baseline magnetic design was selected. This stage took two months between the kickoff meeting on March 8, 2012 and May 9, 2012 when the decision on the final choice of the baseline magnetic design was made. This is the date when actual engineering work by T&ED began.

The main purpose of the analytical effort, results of which are presented in this report, is to develop a viable engineering design satisfying requirements to the superconductor, as well as structural and cryogenic requirements. The work reported here includes proposed conceptual approaches, solid modeling and analyses for the conductor and winding pack design, high temperature superconductor (HTS) and copper current leads for the magnet, structural design of the magnet cold mass, cryostat and warm-to-cold supports, cryogenic design of the magnet cooling system, and magnet power supply sizing.

Cost estimates are based on the point conceptual design that has been developed to date.

# 2. Engineering Considerations for Fabrication, and Challenges for the Future

## 2.1. Magnetic Design

The initial proposed magnetic design was the 8-segment Erice design shown in Figs 2-1 and 2-2. This design raised concerns primarily due to its complexity and the consequent high price of winding a coil with variable shape and cross section.

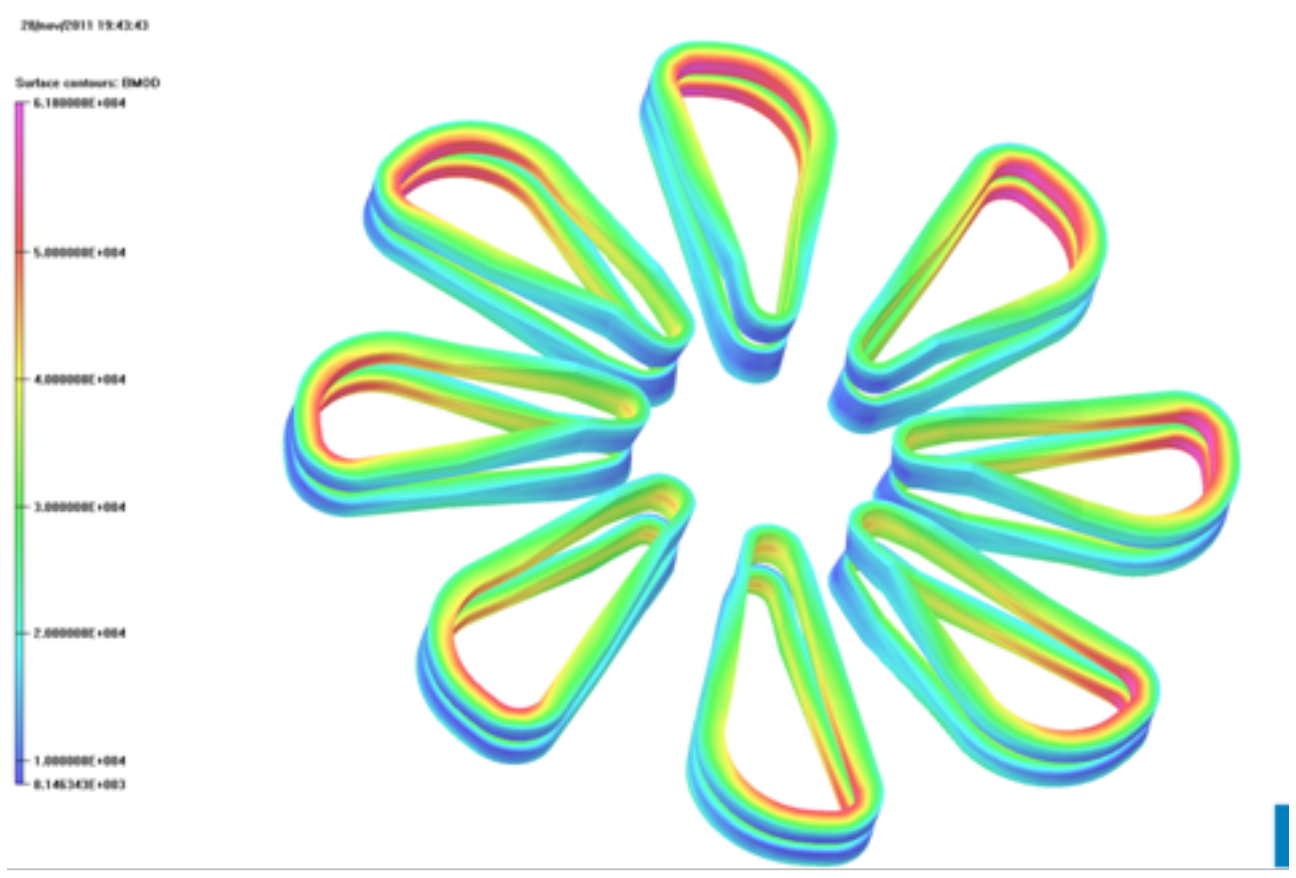

Fig. 2-1. 8-segment Erice coil arrangement

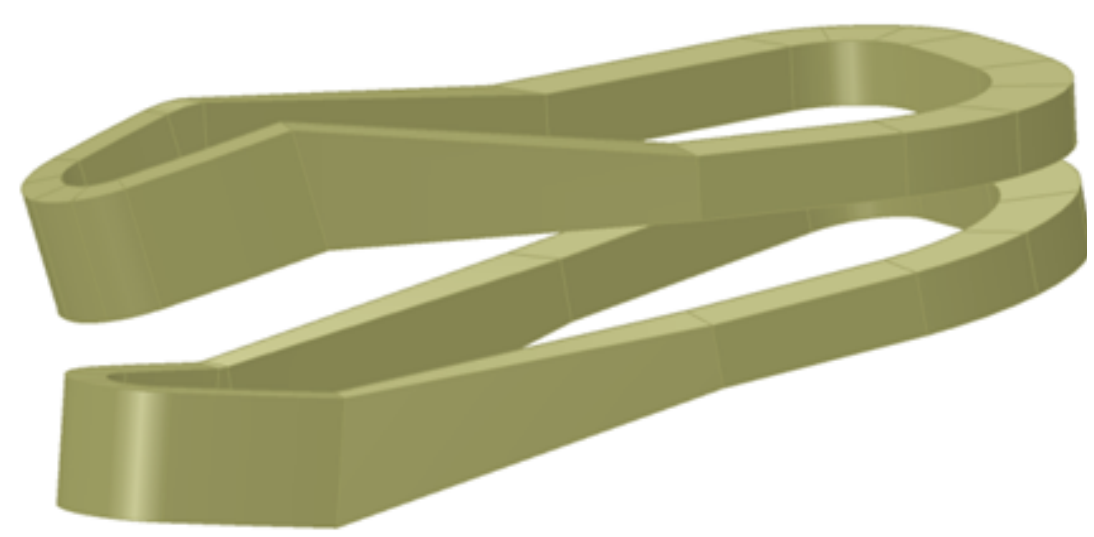

Fig. 2-2. Erice split coil segment

Alternatively two flat coil designs were proposed, one with tilted coil pairs arranged in 8 segments and the other – with parallel coils in 6 segments. They were labeled as Solutions A and B respectively. They are shown in Fig. 2-3 and Fig. 2-4.

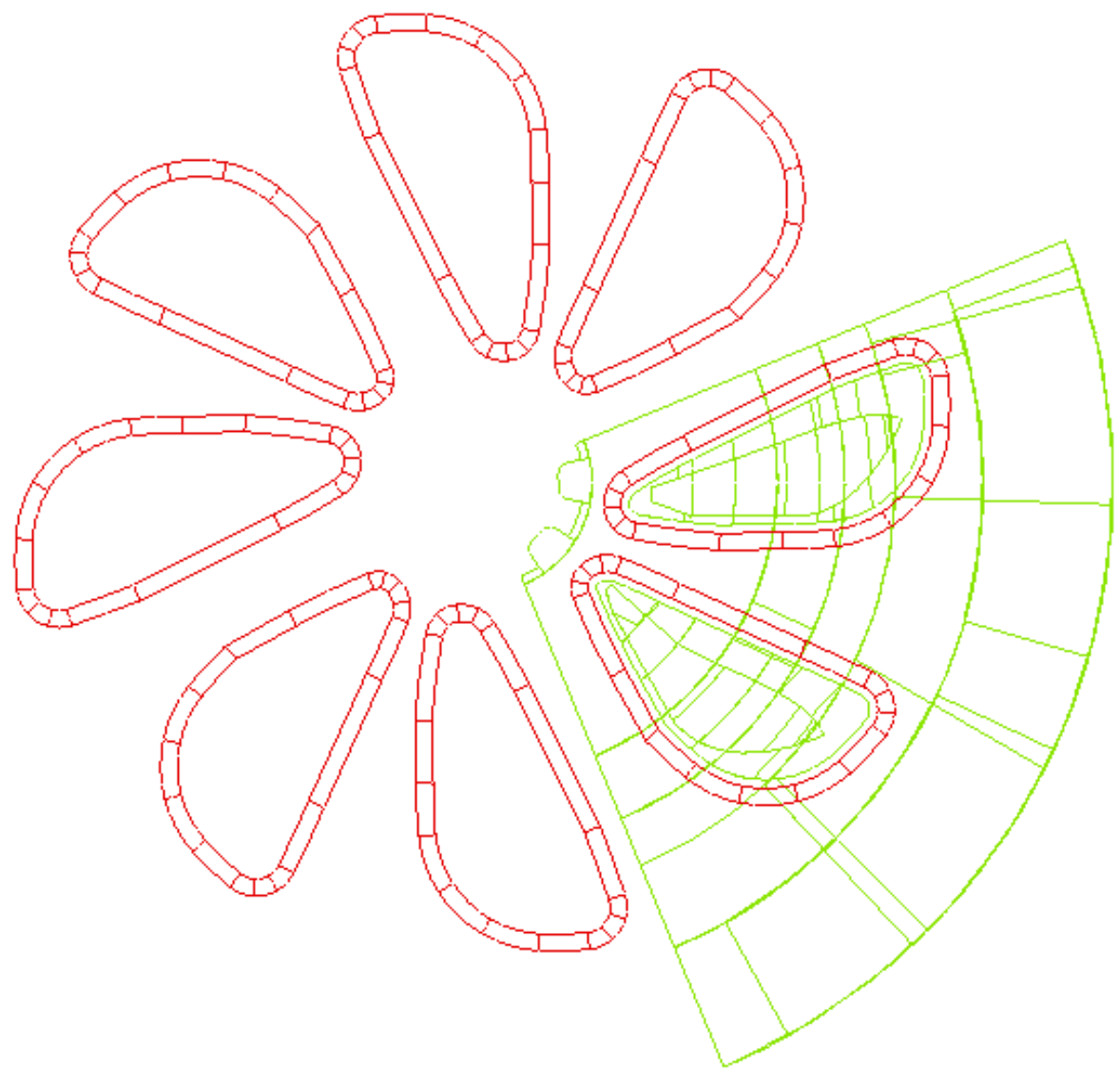

Fig. 2-3. 8-segment arrangement (Solution A)

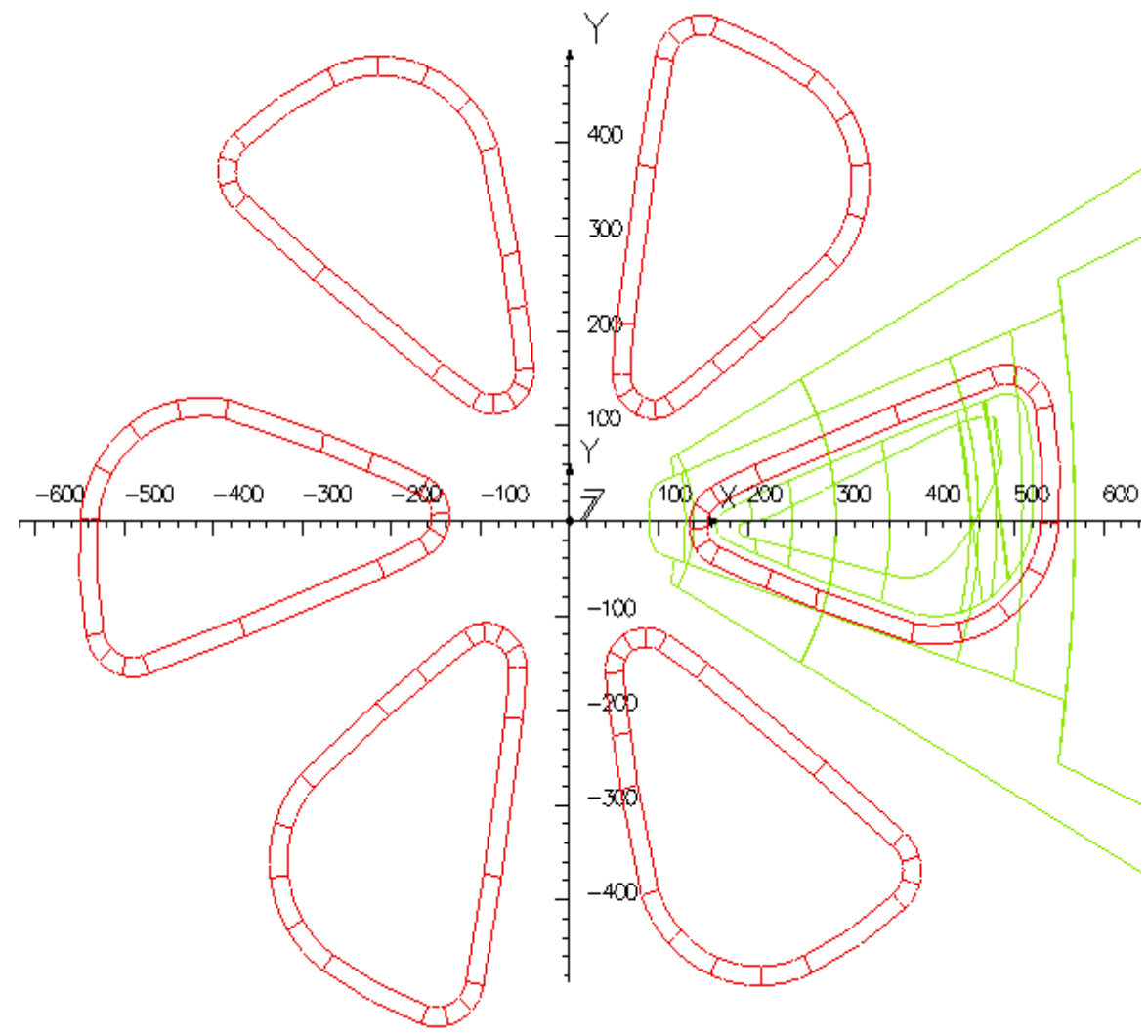

Fig. 2-4. 6-segment arrangement (Solution B)

Between Solutions A and B the preference was given to Solution B with 6 segments, which is shown in Fig. 2-5. The arguments in favor of this decision were:

- lower price of building a smaller number of magnets (~6/8=75%)
- more space around the winding to accommodate the structural coil case and the cryostat. (8-segment design required a small in-plane distance between the coils shown in Fig. 2-6. This calls for placing 2 coil pairs into one cryostat. This may be possible but definitely is more complex than one spit pair per cryostat.)
- arrangements with tilted coils are acceptable but add complexity, which is better avoided if possible.

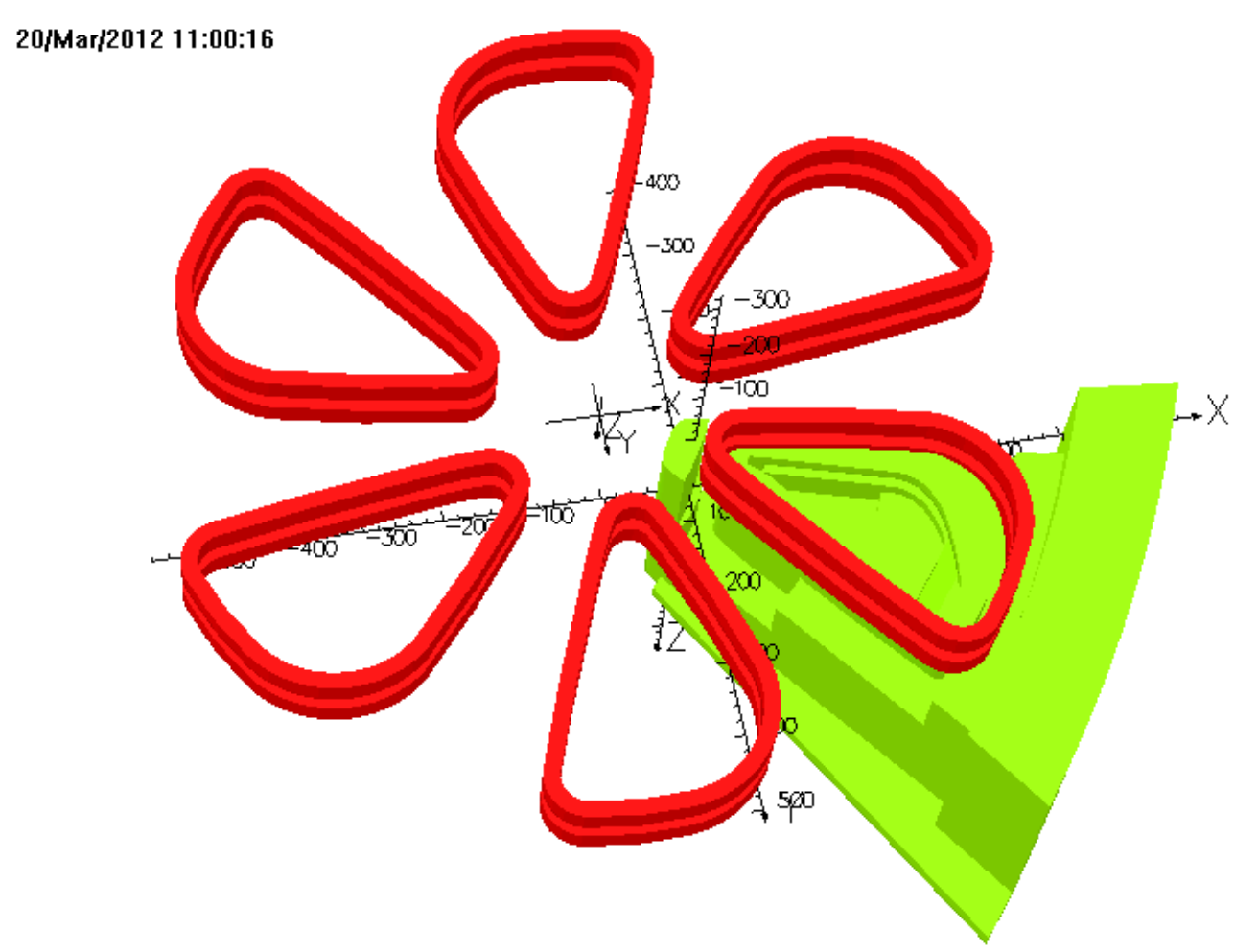

Fig. 2-5. Baseline 6-segment arrangement.

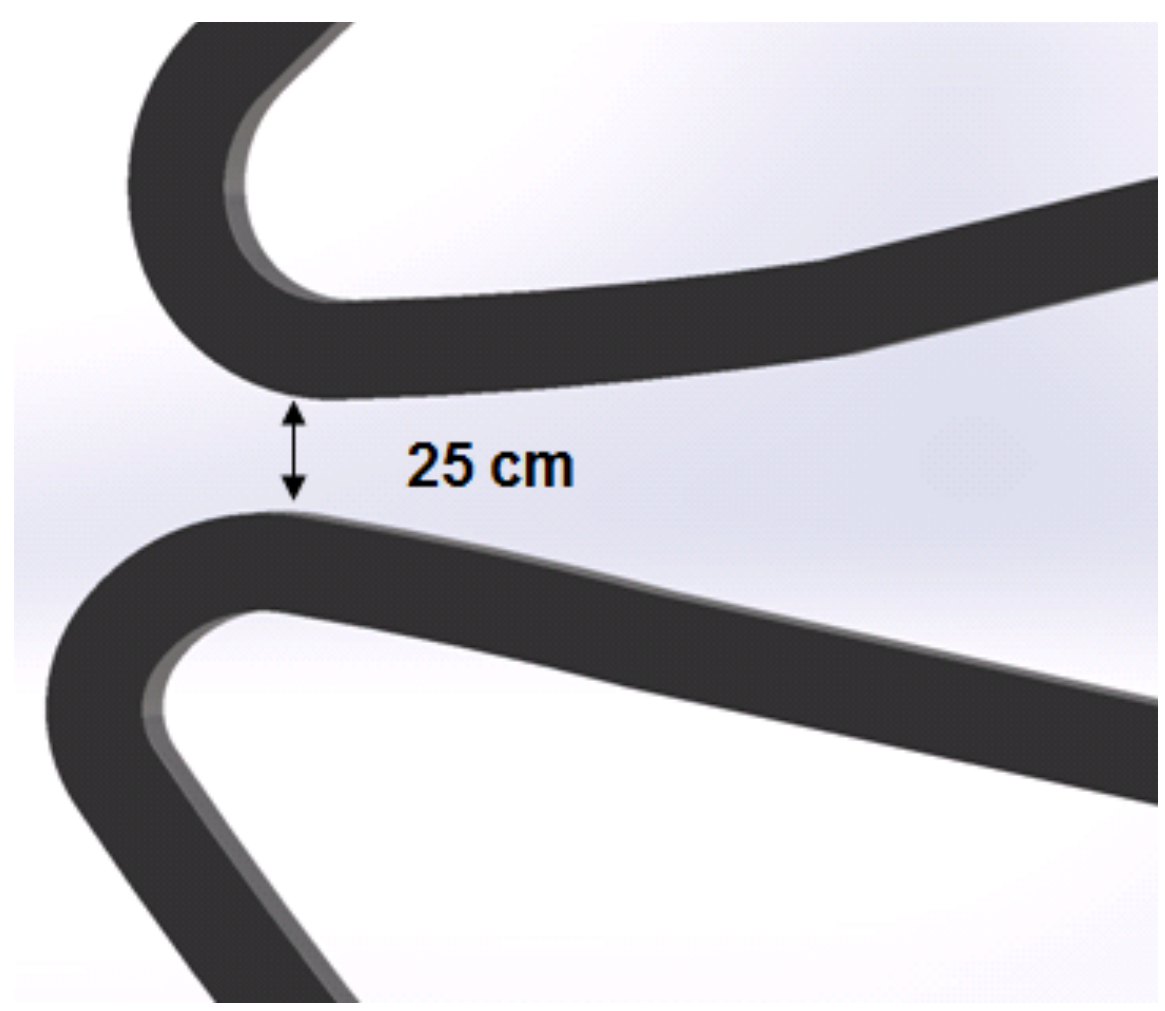

Fig. 2-6. In-plane distance between the coils in 8-segment design

Initially the 6-segment magnetic design was provided as VF Opera model a60_38_2. Examination of this model revealed multiple incompatibilities between the shapes of the coil and the iron yoke vs. the limitations imposed by practical engineering requirements. Consequently the following design rules for flat and arched coils were formulated. (Definitions are customized for the terminology used in VF Opera.)

### 2.1.1. Coil design rules

- On the top view (shown in Fig. 2-7) the top and bottom coil surfaces coincide. Side walls are formed by cylindrical surfaces (as shown in Fig. 2-8).
- Any cross section perpendicular to the edge lines is a XX cm x YY cm rectangle.
- Edge lines have a smooth tangent, no break points.

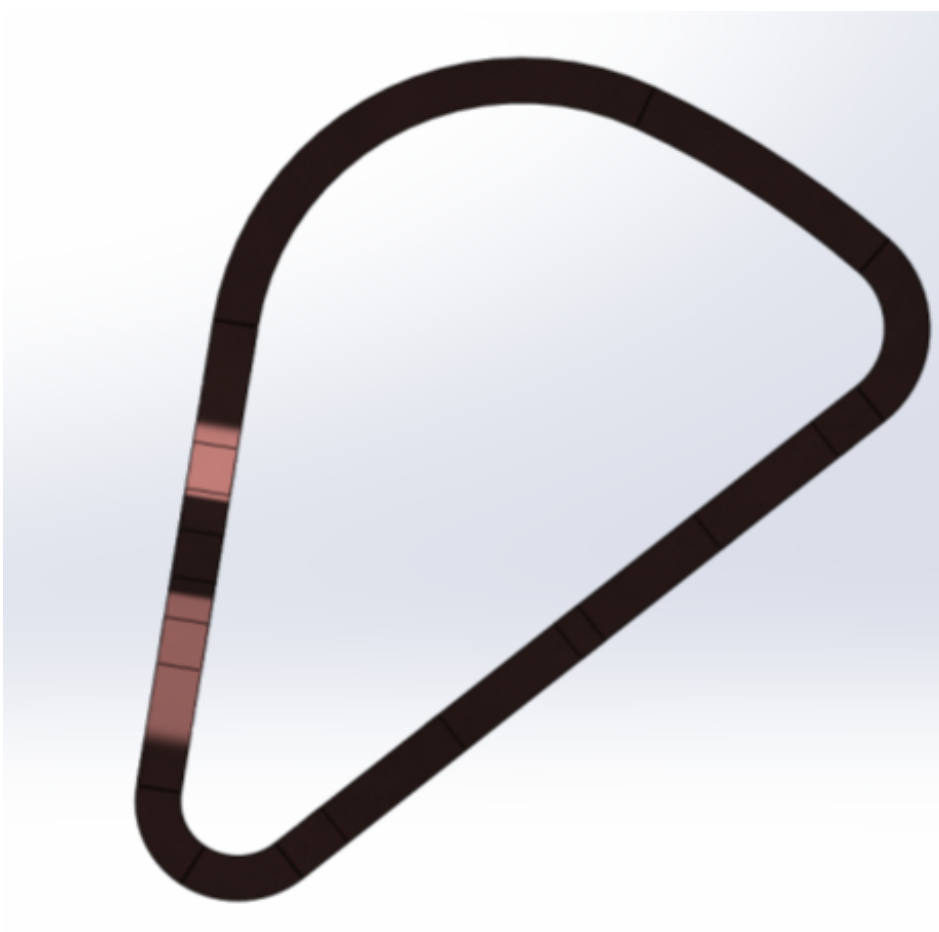

Fig. 2-7. Top view

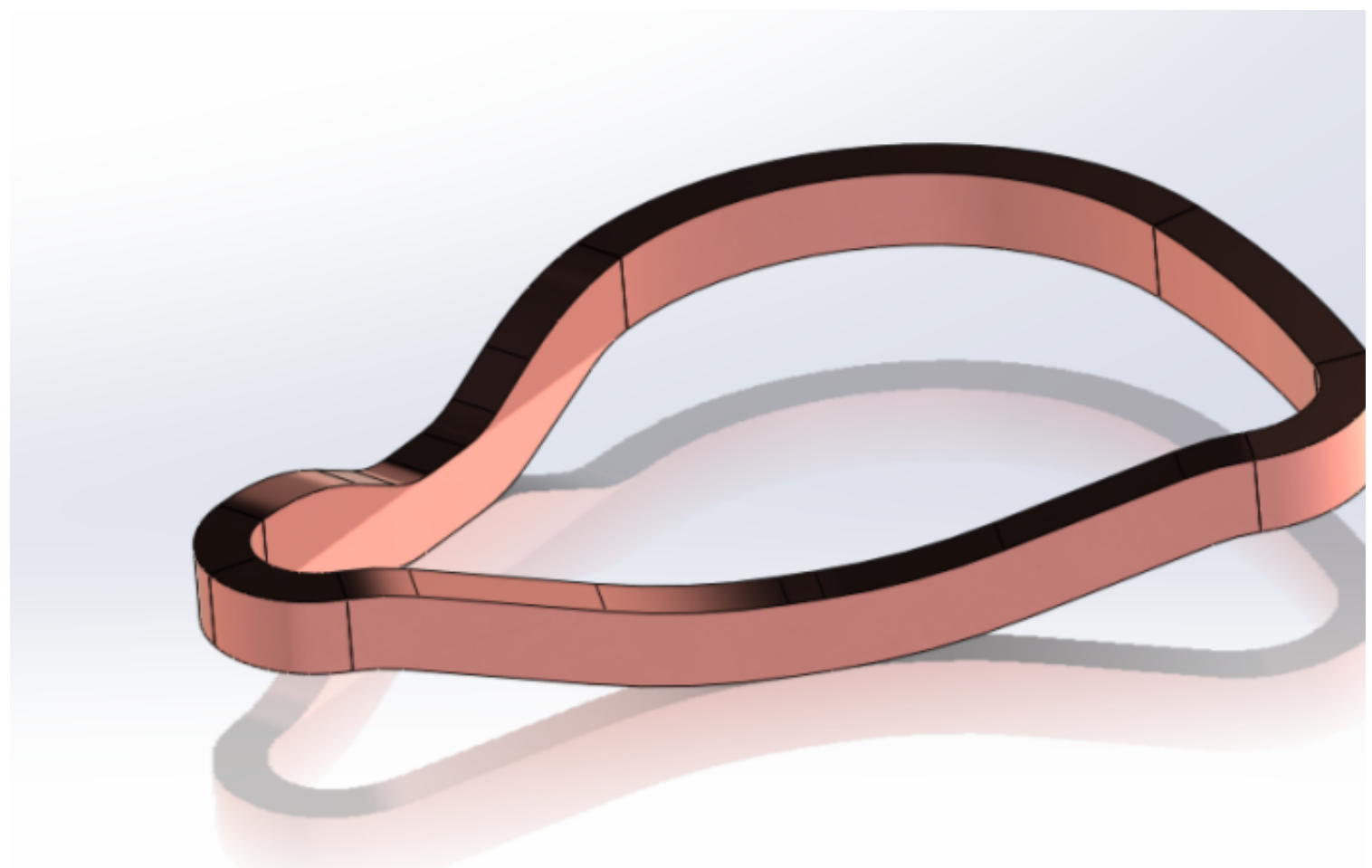

Fig. 2-8. Isometric view

- The coil is comprised of 2 End Turns and 2 Side Legs (shown in Fig. 2-9)
- End Turns and Side Legs are comprised of ARCs and rectangular 8-node BRICKS, Opera conductors
- Top/bottom surfaces of the End Turns are parallel to Local XY plane
- For system assembly the Local XY plane can be tilted with respect to the Global Mid-plane

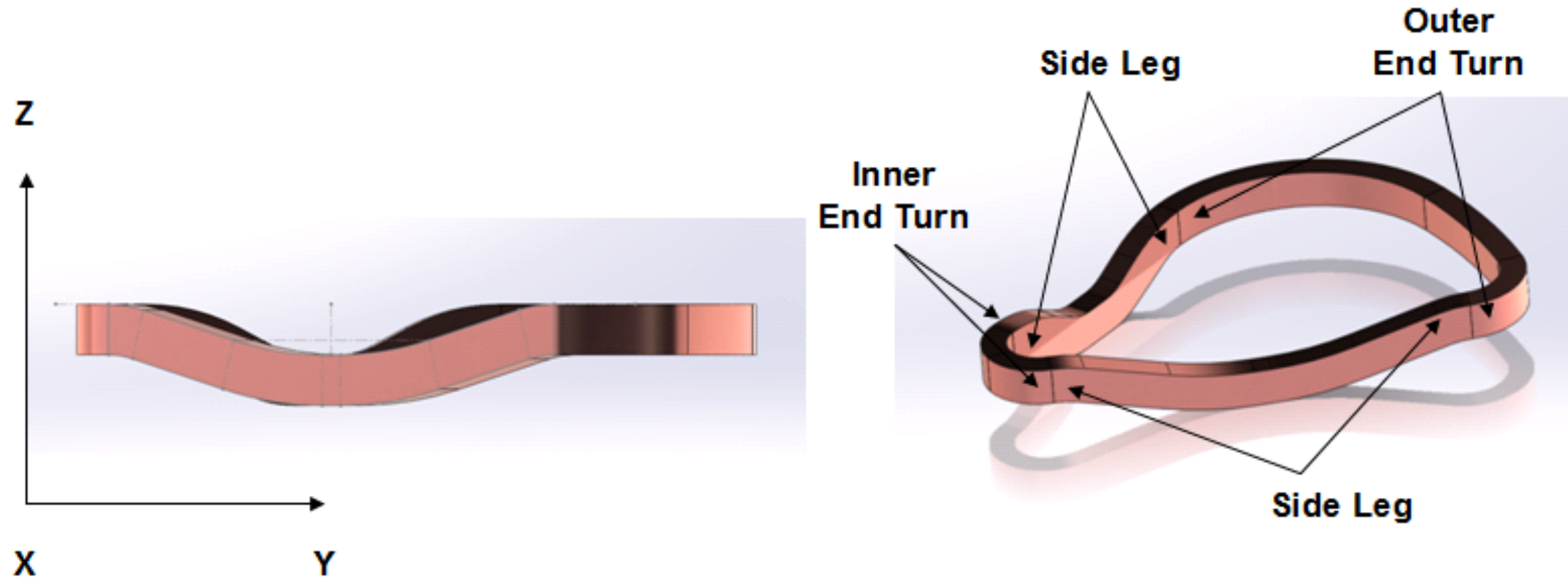

Fig. 2-9. Coil composition

Definitions of BRICKs and ARCs comprising arched coil (see Fig. 2-10):

- End Turns are made of BRICKs and ARCs with axes parallel to the Local Z-axis (ARC-Z). Upper/lower surfaces of the End Turns are parallel to the Local XY-plane. Side walls of the End Turns are parallel to the Local Z-axis.
- Side Legs are made of BRICKs and ARCs with axes parallel to the Local XY-plane. Side walls of the Side Legs are parallel to the Local Z-axis.

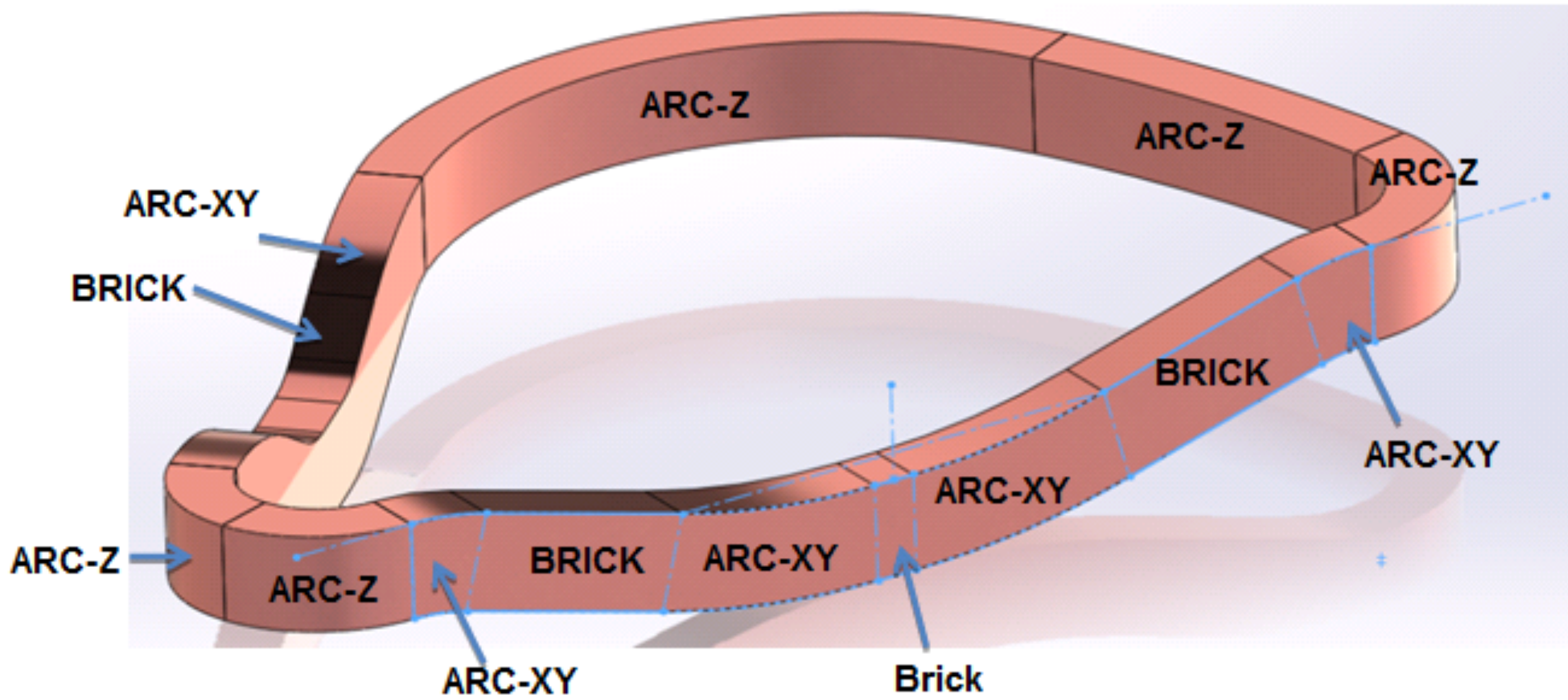

Fig. 2-10. Coil composition. Continued

Requirements for the gaps between the coil winding and the iron yoke were also formulated:

- At least 100 mm between the Side Walls of the Winding and the Iron
- 120 mm axially between the Mid-plane and the Top/Bottom (nearest surface) of the Winding
- 200 mm Iron-to-Iron axial gap for Tie Plate. 70 mm to the Iron on Top/Bottom sides of a 60-mm thick Tie Plate.

Eventually magnet design q68 was presented by the Daedalus team. It failed to comply with some of the above requirements but after small fixes described in the structural part of the report it was converted into a working baseline magnetic design.

The summary of the major characteristics of this design is shown in Table 2-1.

**Table 2-1. Model q68 Magnetic Design Parameters**

| Model | Units | q68 | |
|---|---|---|---|
| Number of Sectors | | 6 | |
| Coils per Sector | | 2 | |
| Coil Current Density | A/mm^2 | 34.68 | |
| Coil X-section | cm x cm | 16 x 31 | |
| Coil X-section Area | m^2 | 0.0496 | |
| Coil Average Length | m | 10.85 | |
| Coil Volume | m^3 | 0.5383 | |
| Data Source | | Opera | ANSYS |
| Coil Peak Field | T | 4.37 | 4.45 |
| Total Energy | MJ | 319.72 | 381.12 |
| Energy/Sector | MJ | 53.29 | 63.52 |

It shall be noted that all magnetic VF Opera models were developed and provided to us by Alessandra Calanna. These models were converted to solid models in SAT format and imported into SolidWorks and used as the baseline for the engineering and analytical work.

## 2.2. Engineering Design

The engineering magnet design proposed in this report has similarities with the RIKEN design [1]. This is due to the similarities in the topology of the magnet arrangements in the RIKEN and the current implementation of the SRS projects. However, many design details differ from RIKEN significantly. In particular, the present design uses:

- NbTi Sc cable in Cu channel;
- Double-pancake winding;
- Conduction cooled sandwich coil scheme with pairs of double pancakes (quadro pancakes) interlaced by Stainless Steel (SS) cooling plates cooled by He flow in parallel channels;
- VPI of the stack comprised of the quadro pancakes and the cooling plates and wrapped into the ground insulation to form a solid coil winding block;
- A special procedure of clamping and welding the walls of the SS coil case around the coil to eliminate the gaps and possible slipping at the coil to coil case interfaces.
- SS cooling and structural elements of the cold mass and copper conductor stabilizer to minimize the differential thermal contraction of the materials comprising the cold mass;
- Structural SS Coil Case reinforced by stiffening boxes to reduce sagging over long unsupported coil spans bridging over the beam chamber space;
- Cold mass support using only SS support struts.

Details of these design components and rationale behind this choice is presented in the corresponding chapters of this report.

The following sequence of figures illustrates the design of the major parts of the magnet and shows the sequence of their assembly.

Note that only parts related to the cold mass, cryostat and the support structure are considered in this sequence. Since the design at this stage is conceptual, many details, such as electrical and hydraulic connections are shown schematically. Everything related to the design of the radiation shield and multi-layer heat insulation (MLI) is omitted. These details are usually developed during later stages of the design effort. The intercepts at the cold ends of support struts are shown schematically. Their design has to be addressed at later stages of the project.

---

[1] A. Goto, et. Al "Sector Magnets for the RIKEN Superconducting Ring Cyclotron," IEEE TRANSACTIONS ON APPLIED SUPERCONDUCTIVITY, VOL. 14, NO. 2, JUNE 2004

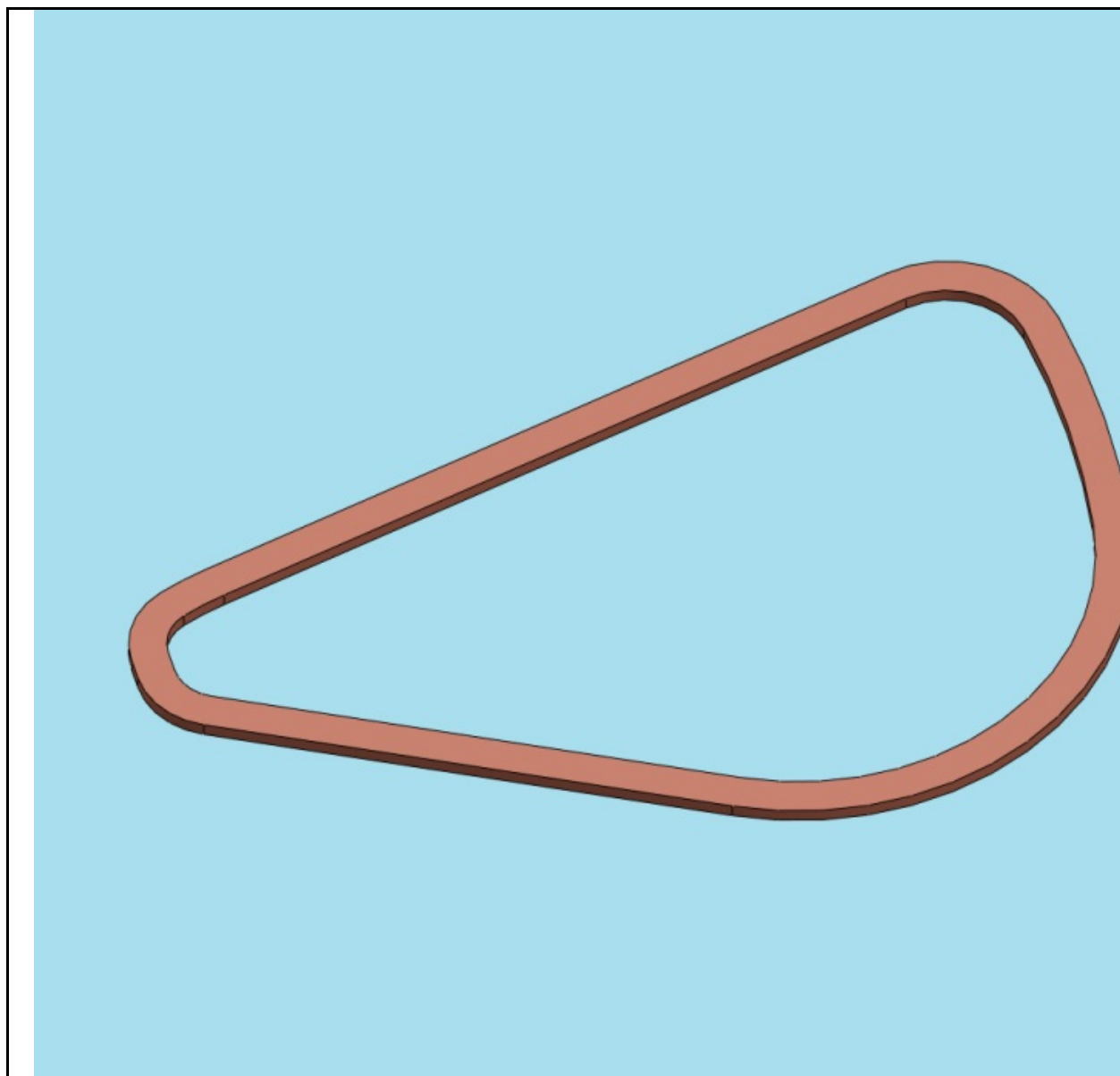

1. Wind and vacuum-pressure impregnate (VPI) quadro pancakes. 4 per coil.

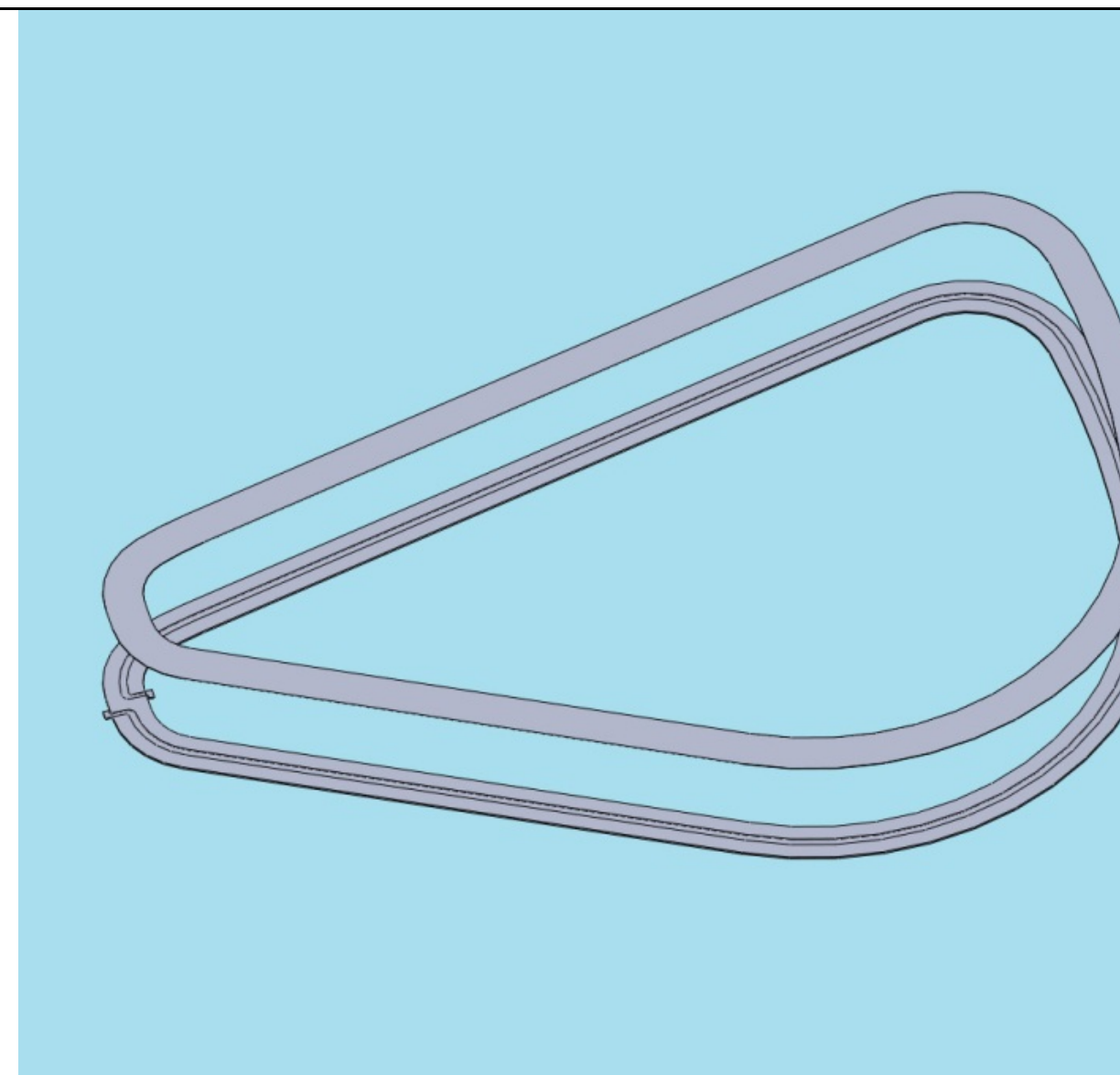

2. Make winding cooling plates, each comprised of a thick base plate with grooved He channels and a top plate. 5 pairs per coil.

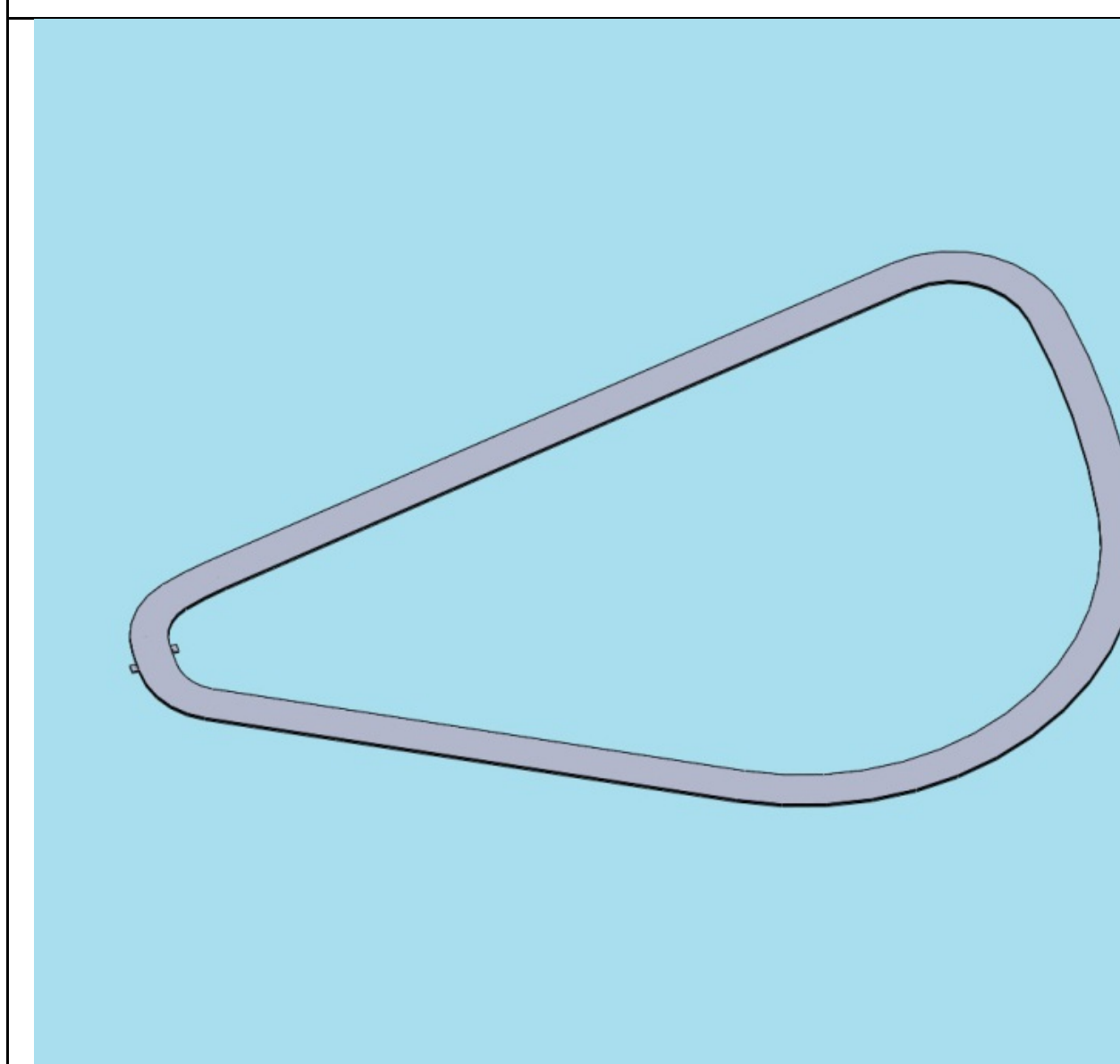

3. Thick base plate with grooved He channels and a top plate are welded along ID and OD. 5 plates per coil.

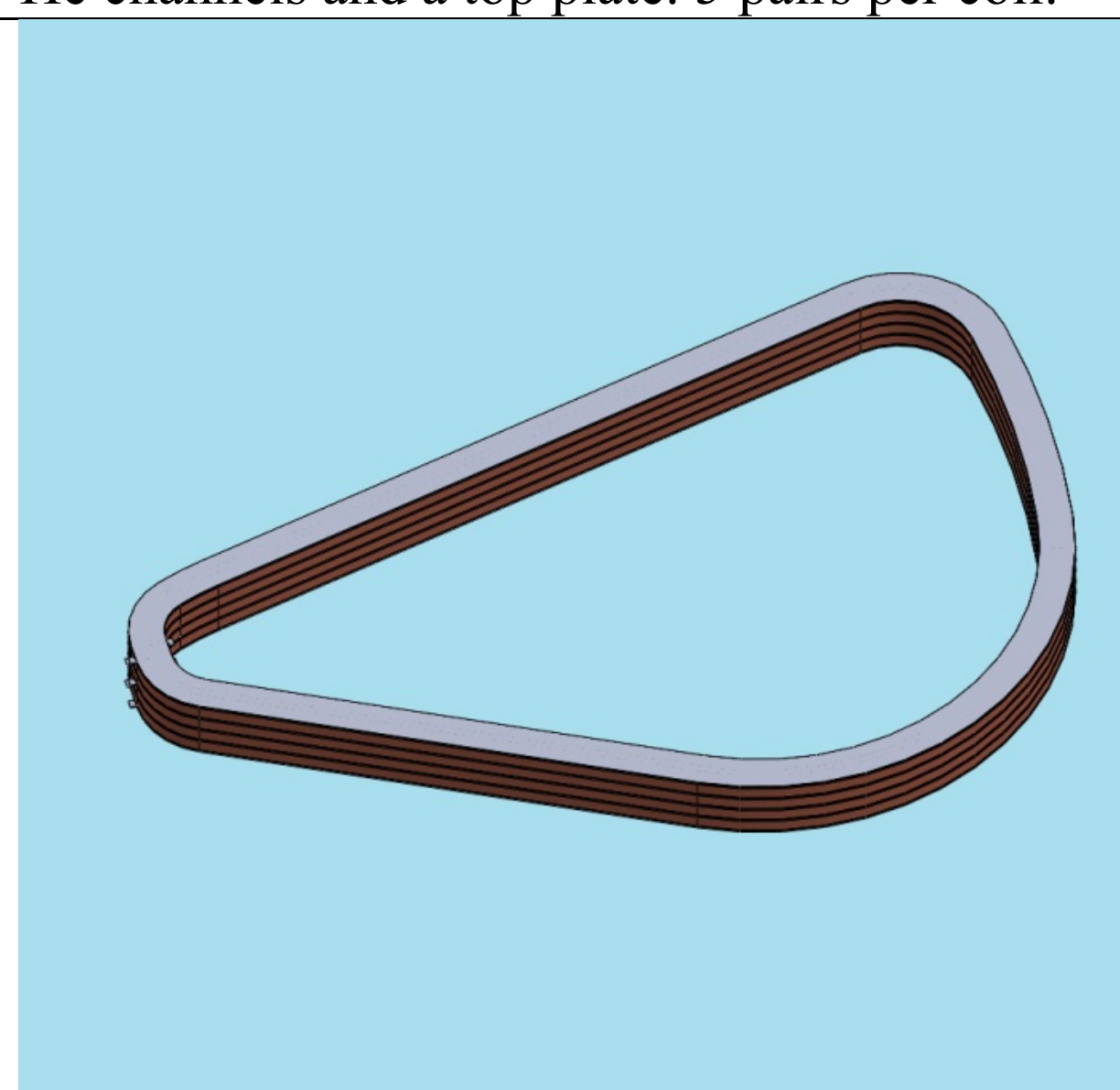

4. Stack 4 quadro pancakes and 5 cooling plates into a coil sandwich assembly. VPI with ground wrap.

| | |
|---|---|
| 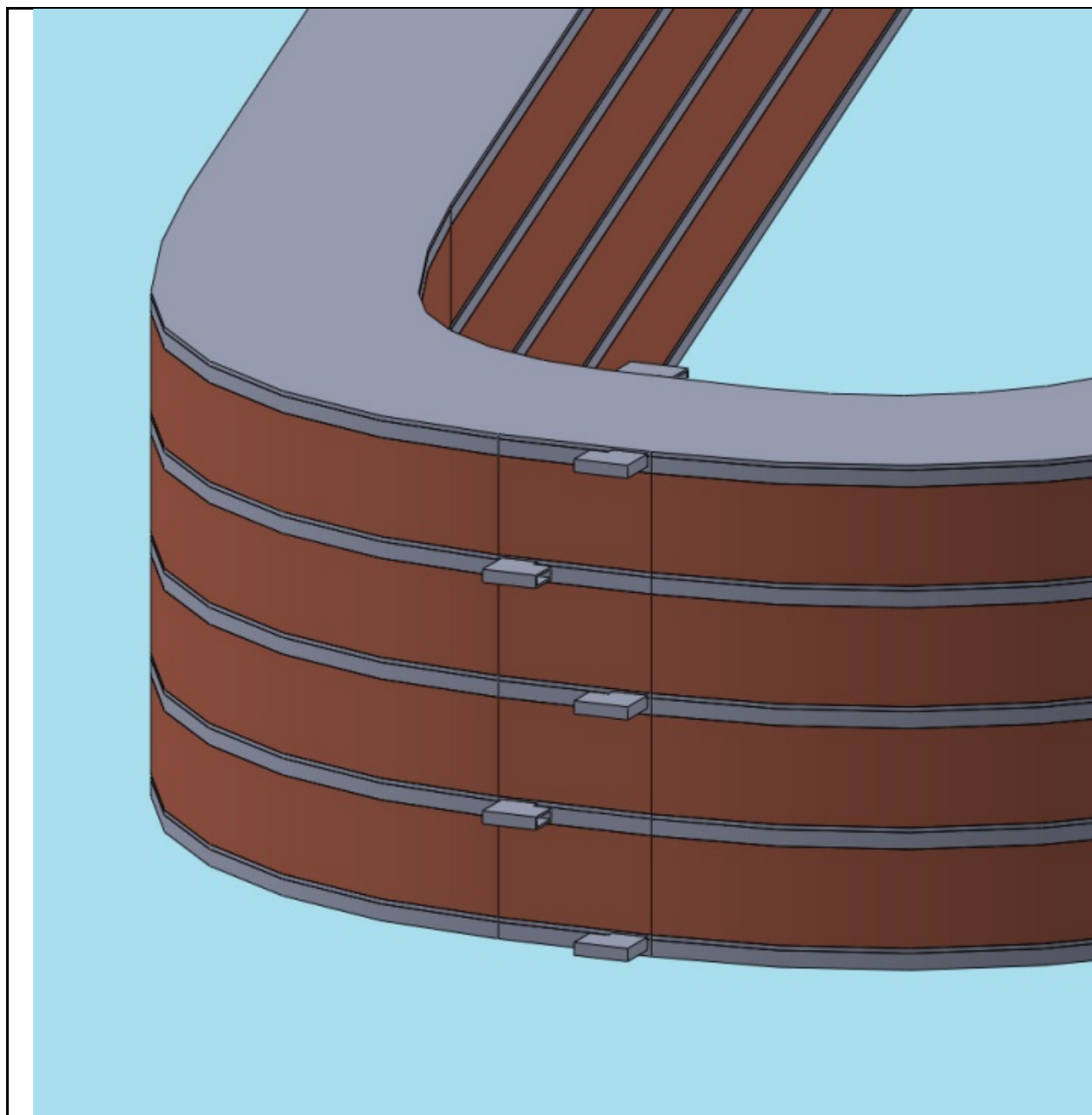<br>5. Expanded View. Coil sandwich assembly showing 5 cooling plates with He inlet and outlet tabs (schematic) and 4 quadro pancakes. | 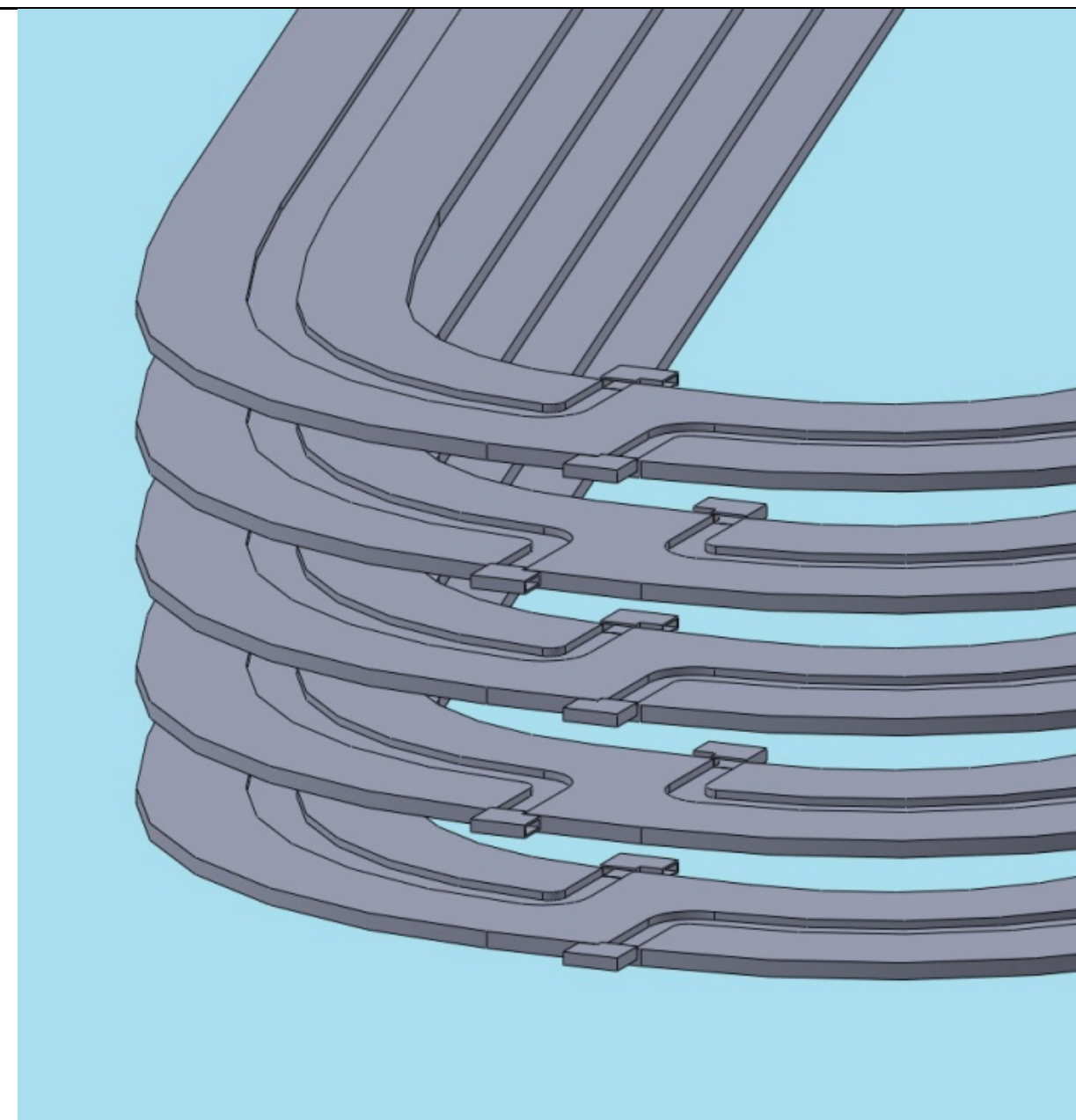<br>6. Detail: Base cooling plates in a stack. Note alternating directions of He channels from inlets at ID to outlets at OD. |
| 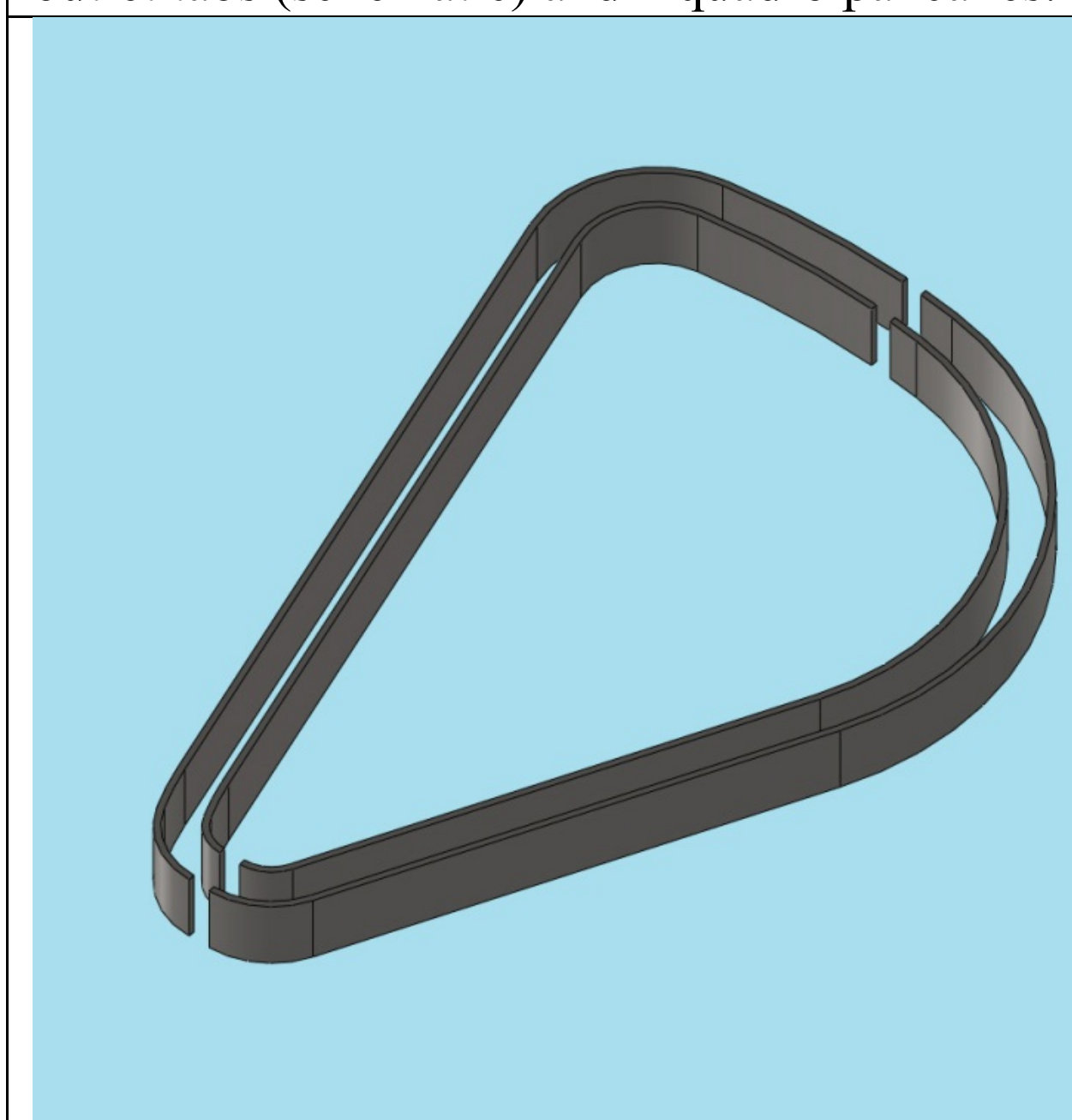<br>7. Make vertical walls of the coil case, 2 at the ID and 2 at the OD. | 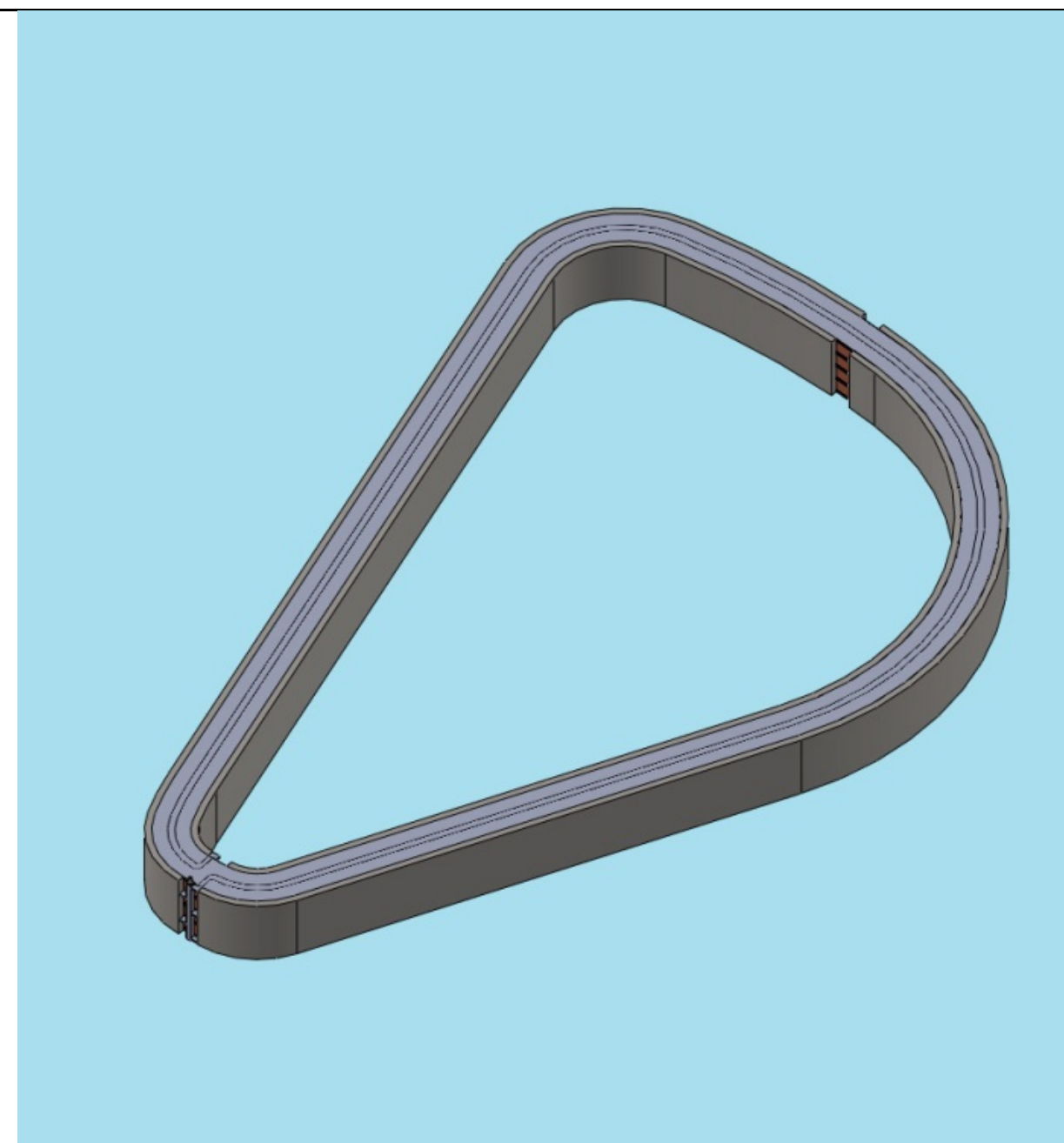<br>8. Clamp vertical walls of the coil case around the coil assembly. |

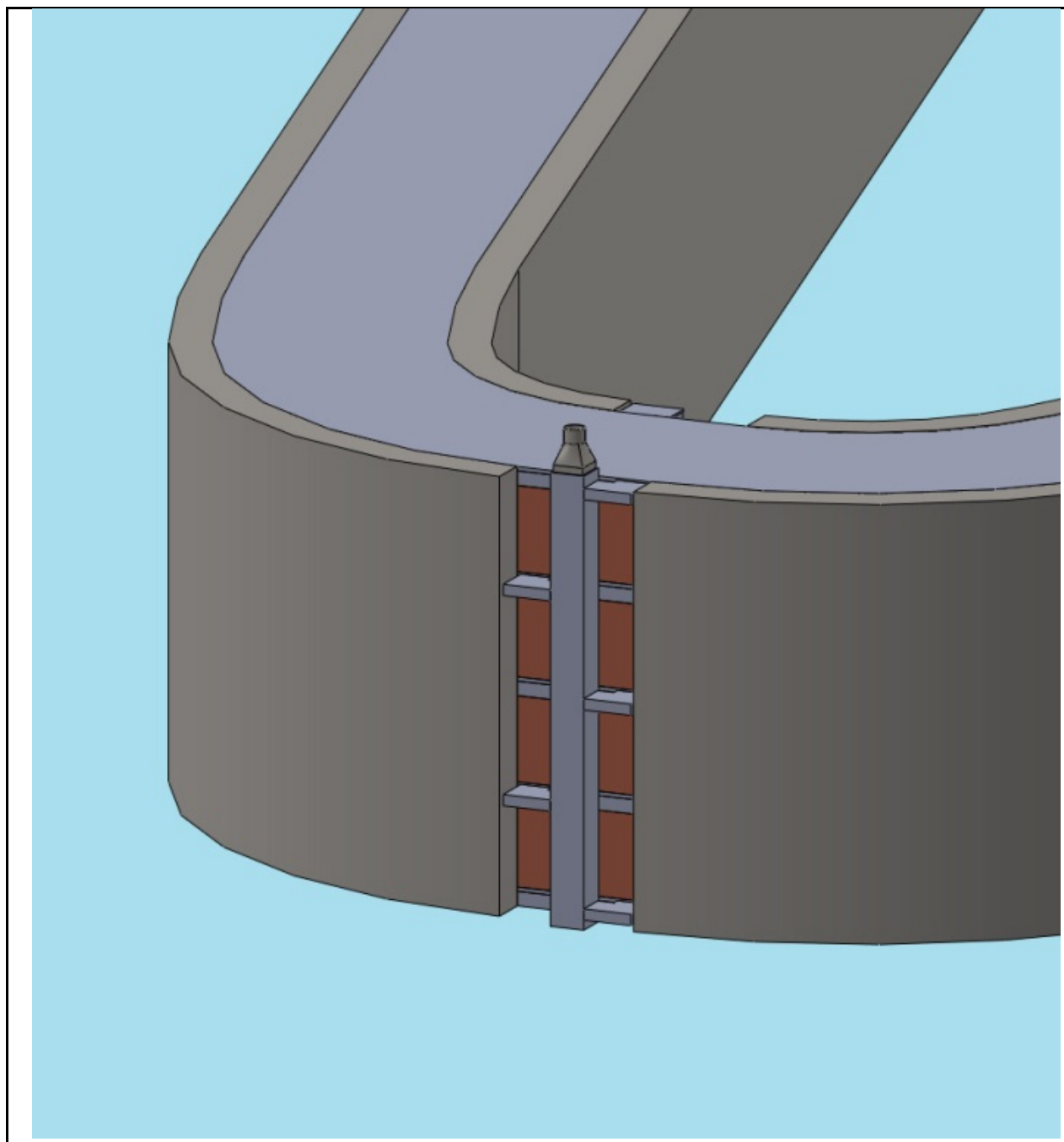

9. Detail: He inlet and outlet manifolds in the gap between vertical walls of the coil case.

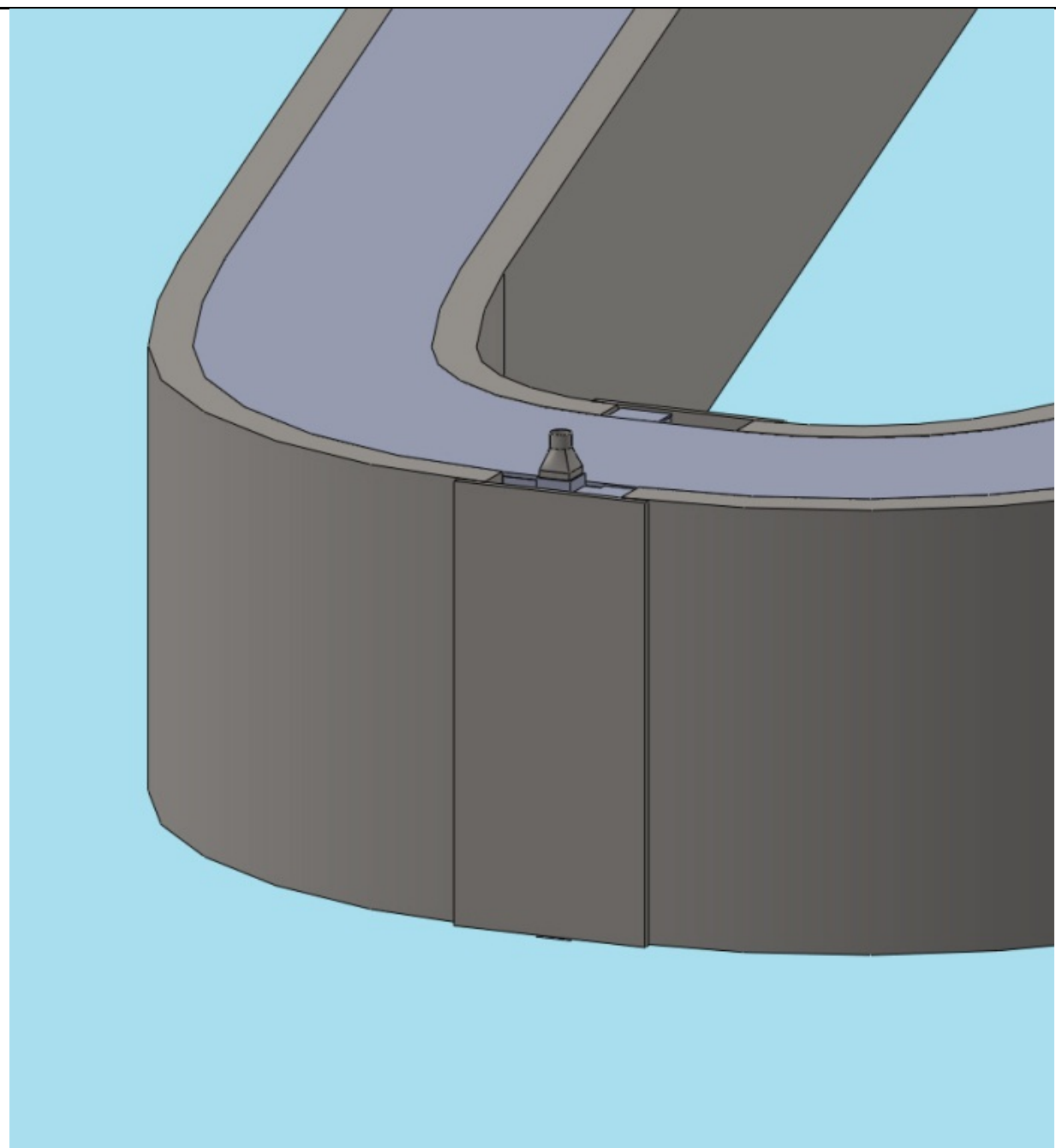

10. Tie vertical walls of the coil case by welding in 4 bridge plates..

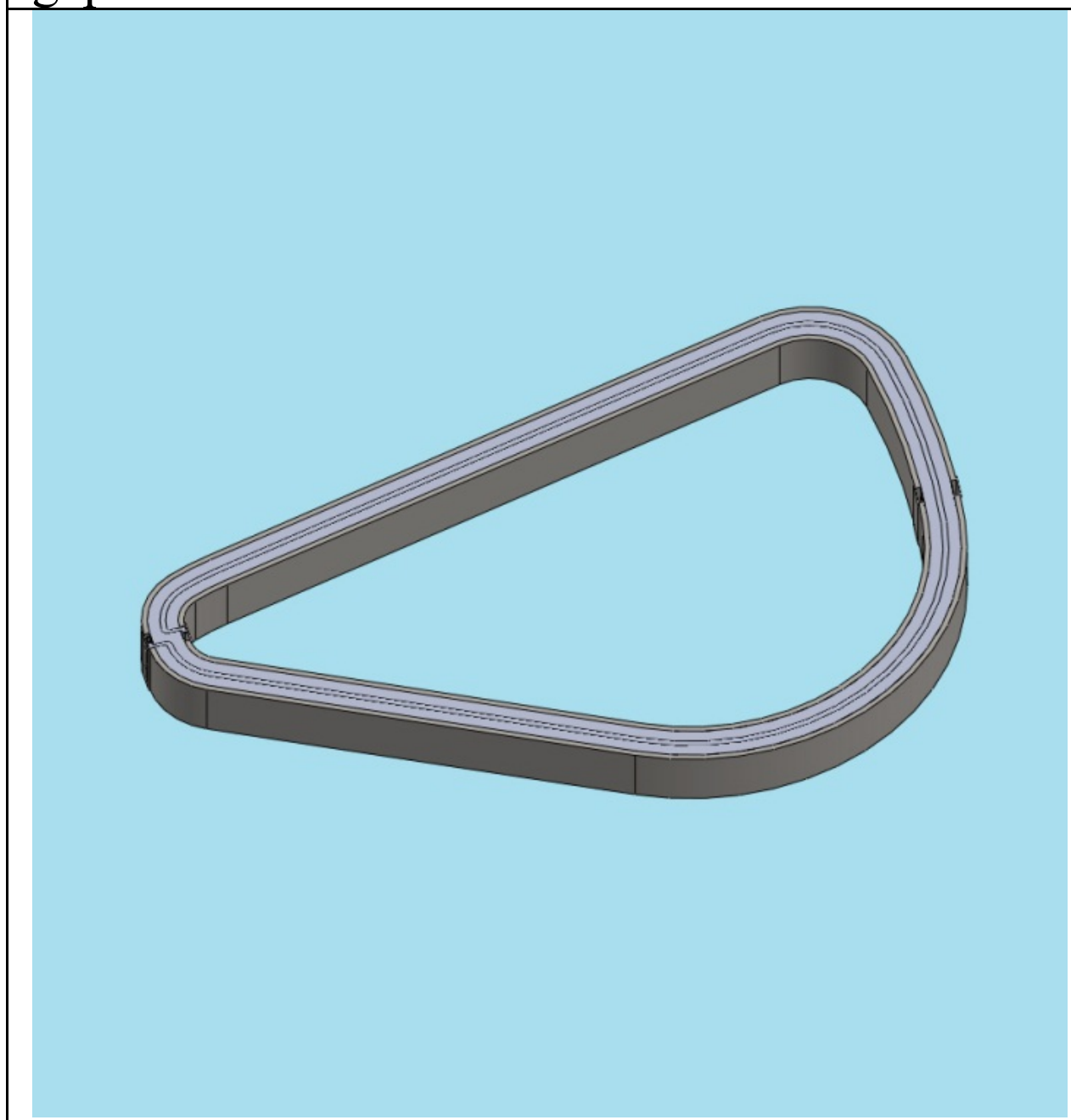

11. View: Winding assembly encased in vertical walls of the coil case.

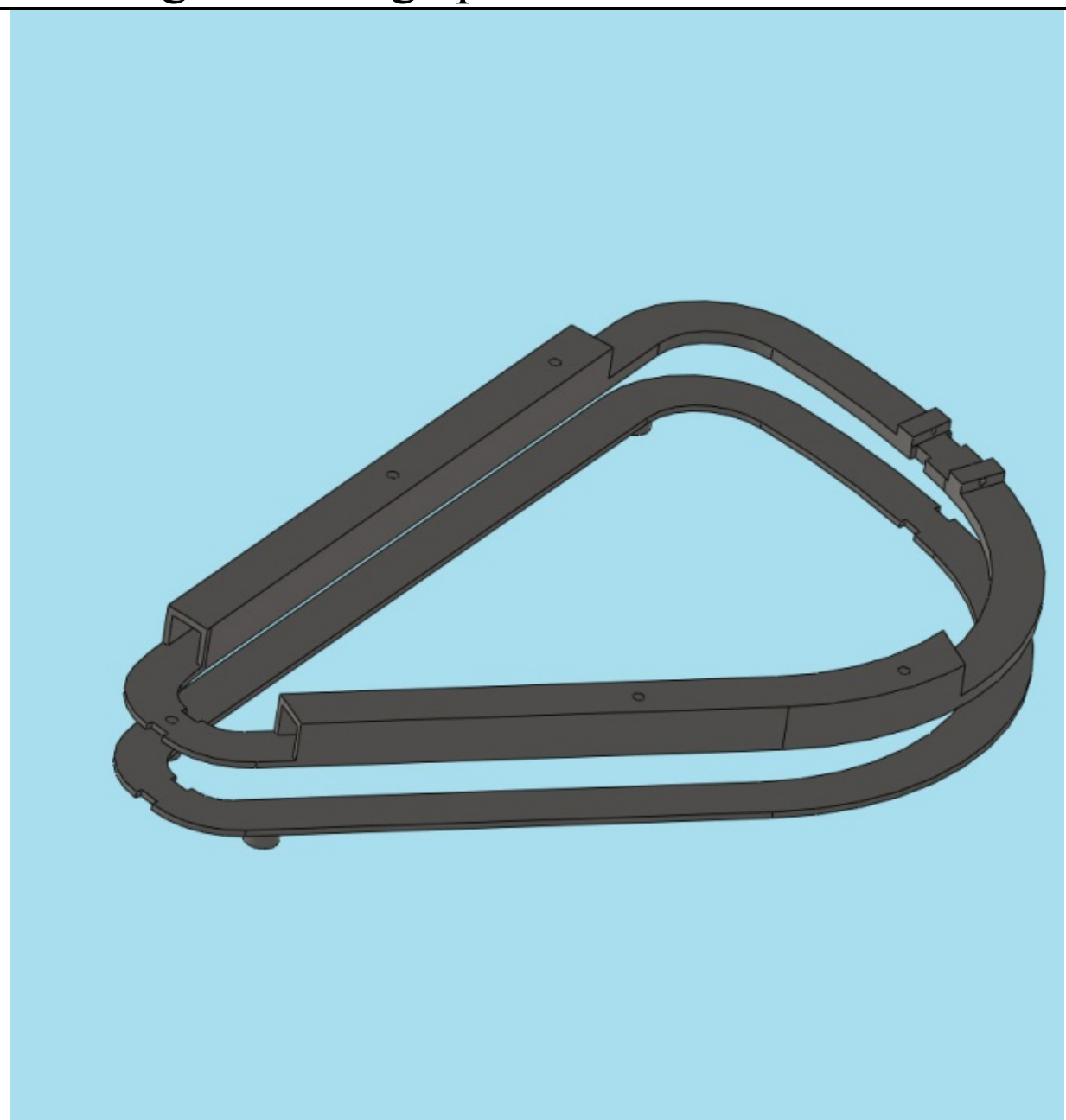

12. Coil case top and bottom plate preassemblies.

| | |
|---|---|
| 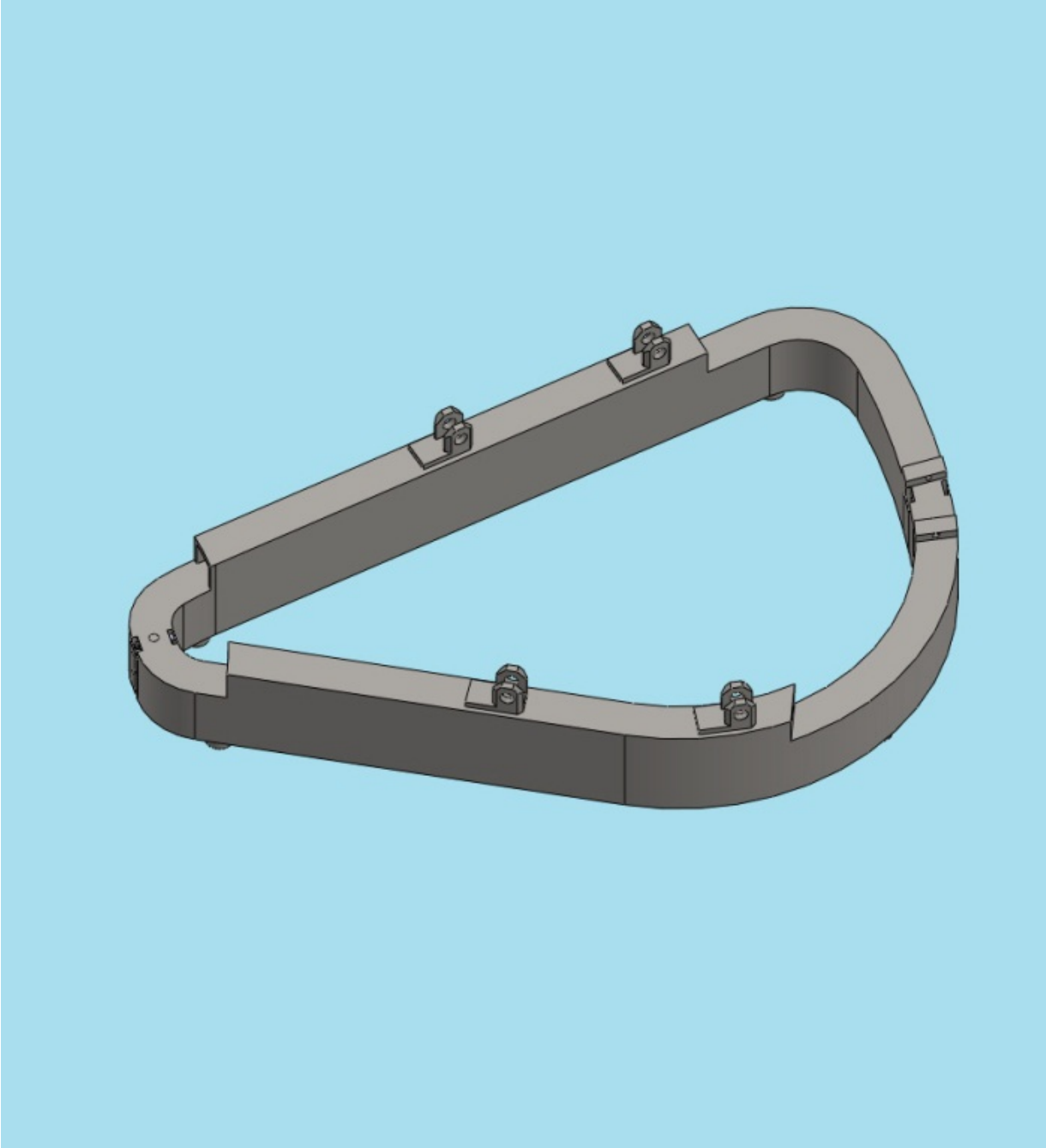 13. Top and bottom plates of the coil case welded to the vertical walls of the coil case . | 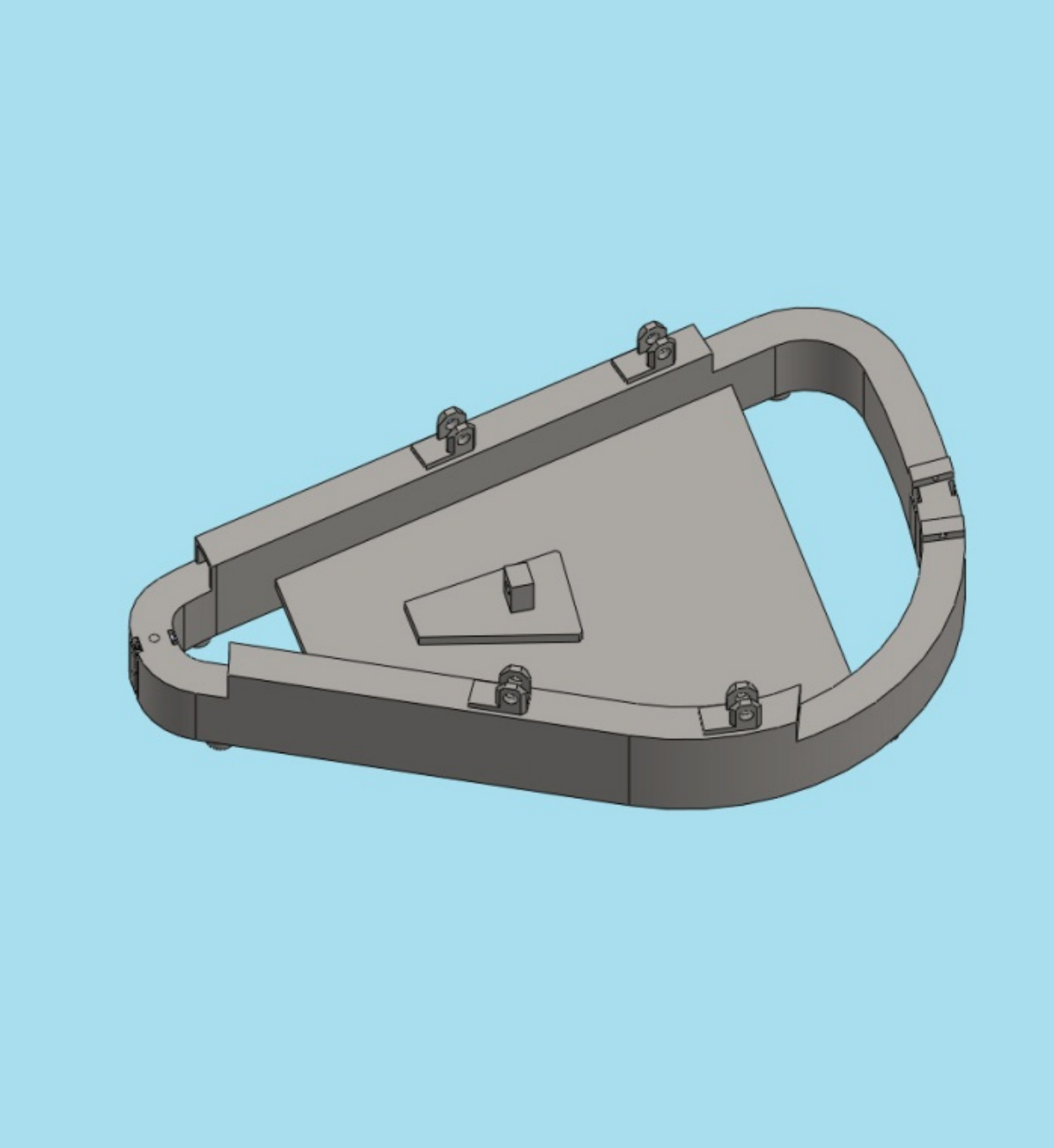 14. Tie plate assembly welded into the coil case. |
| 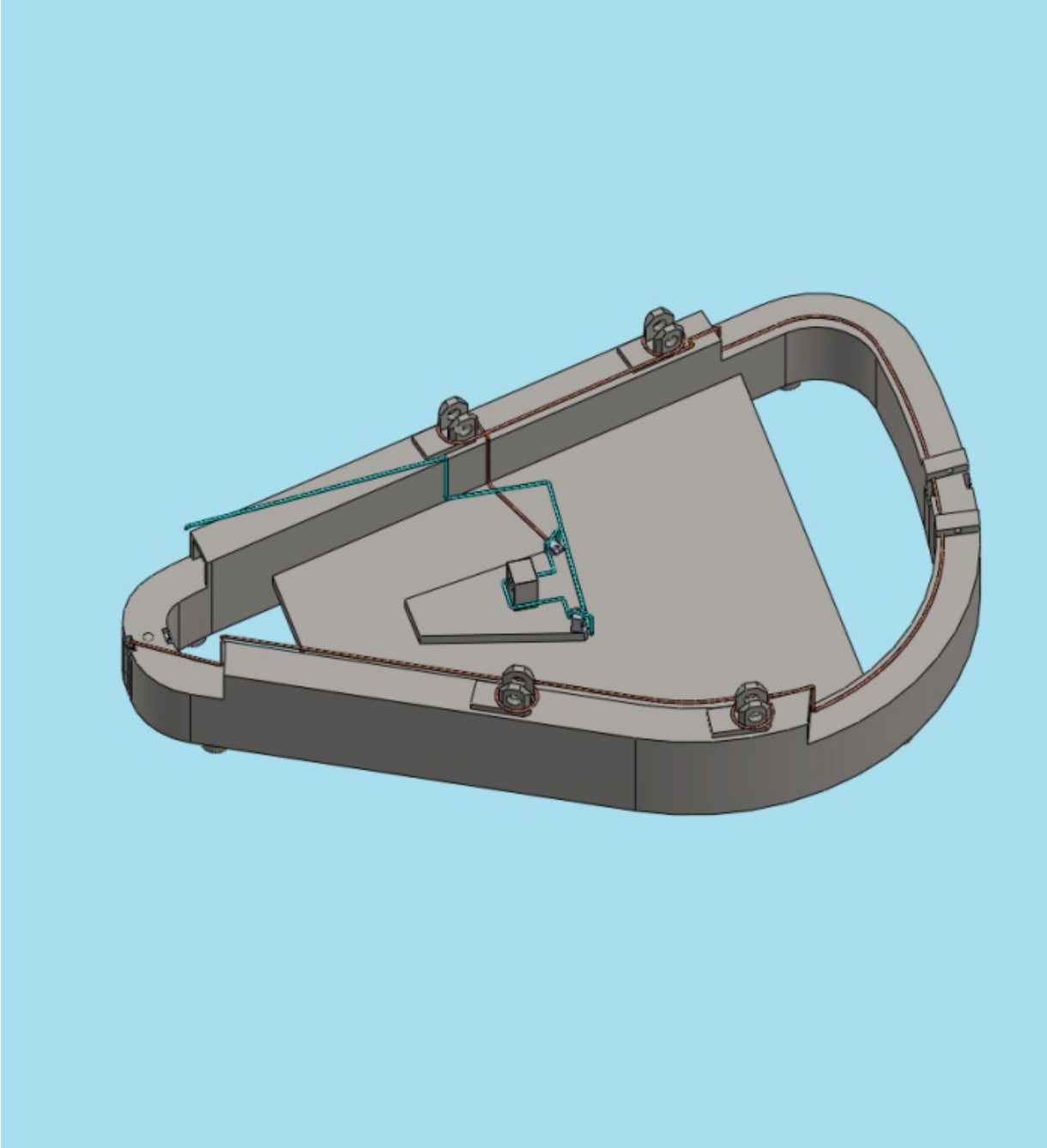 15. Cold mass wiring and plumbing arrangement schematic. | 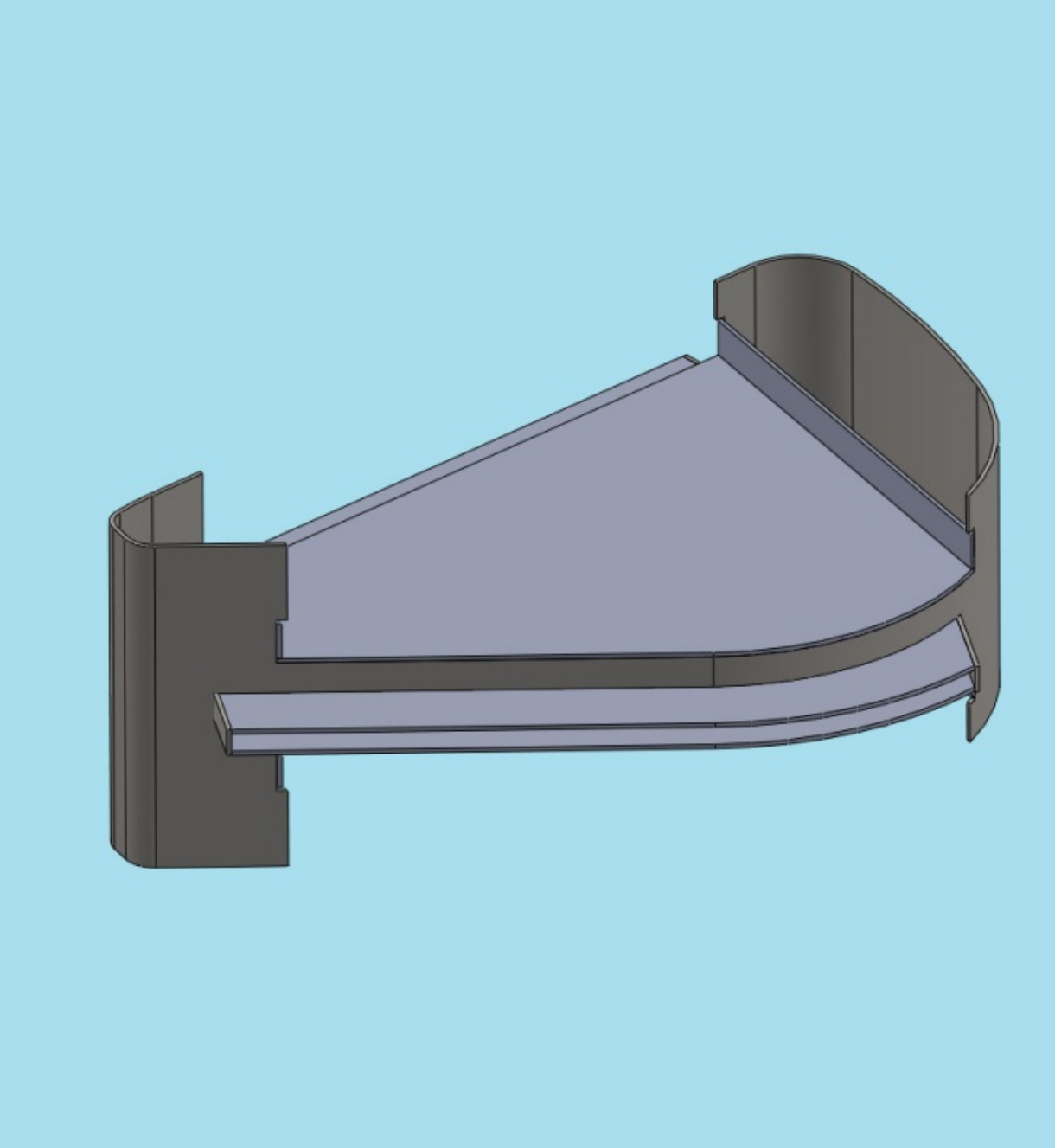 16. Cryostat preassembly. |
| 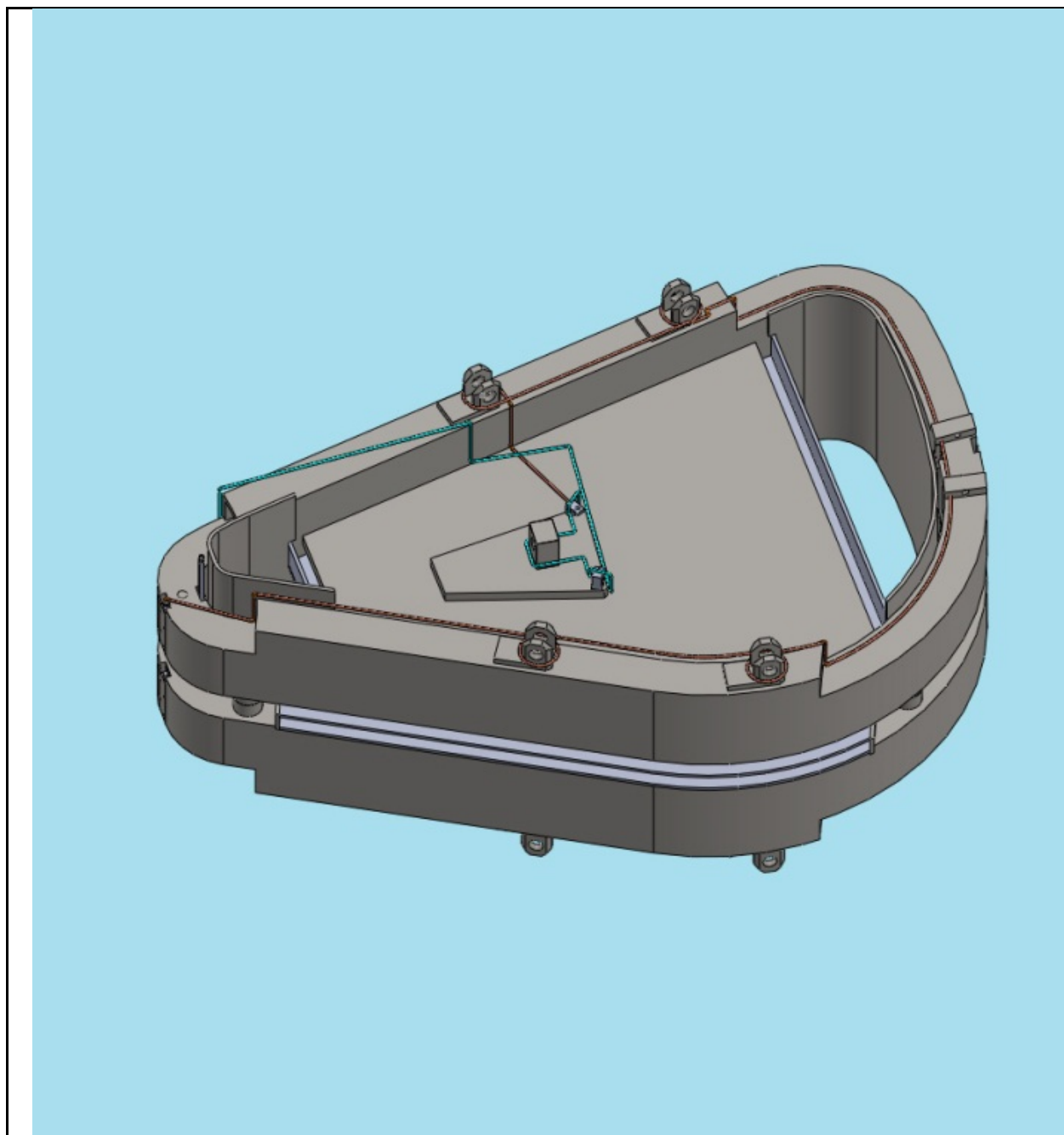 17. Top and bottom cold mass assemblies installed in the cryostat preassembly. | 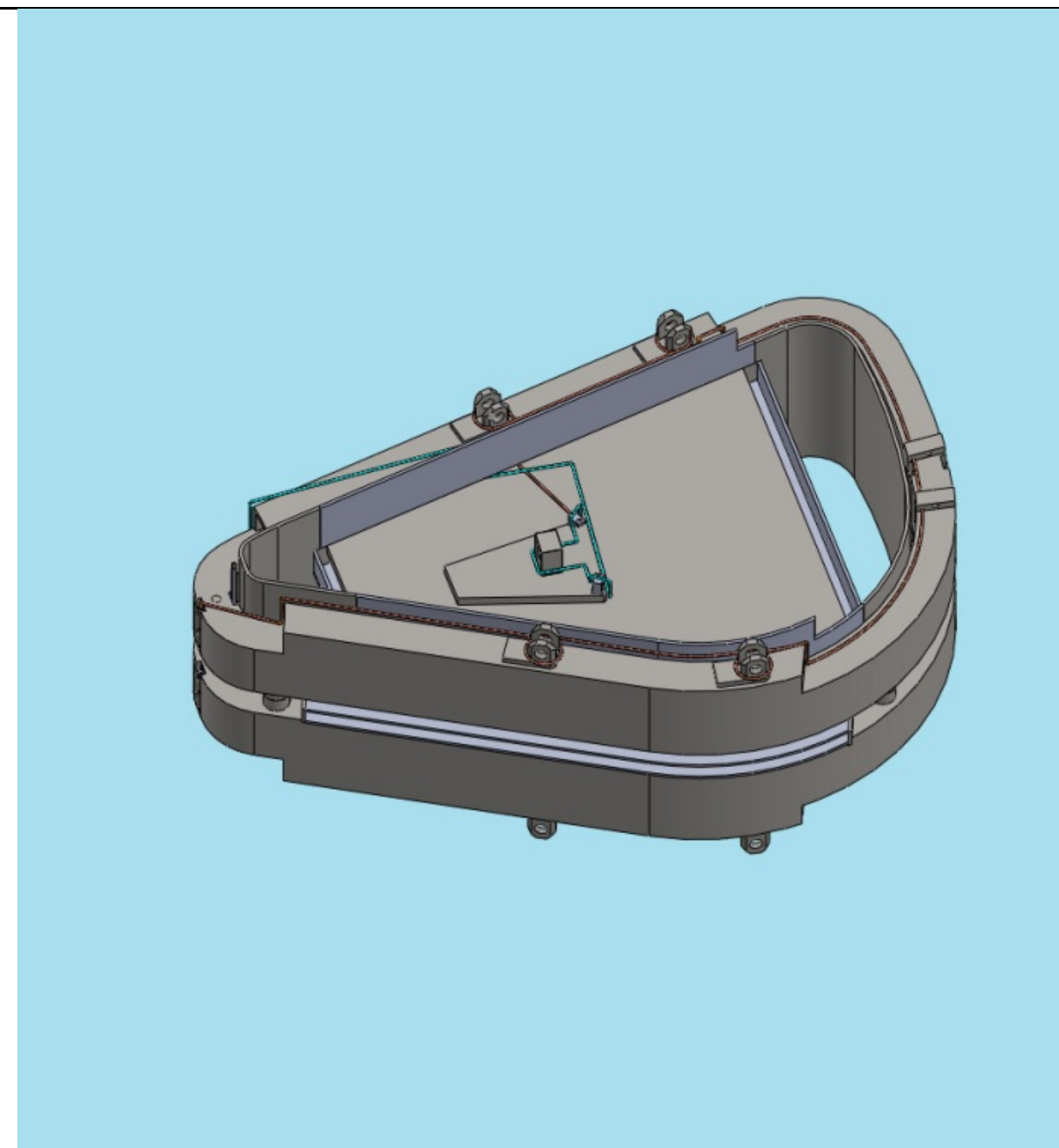 18. Inner cryostat wall cutout plates welded in. |
| 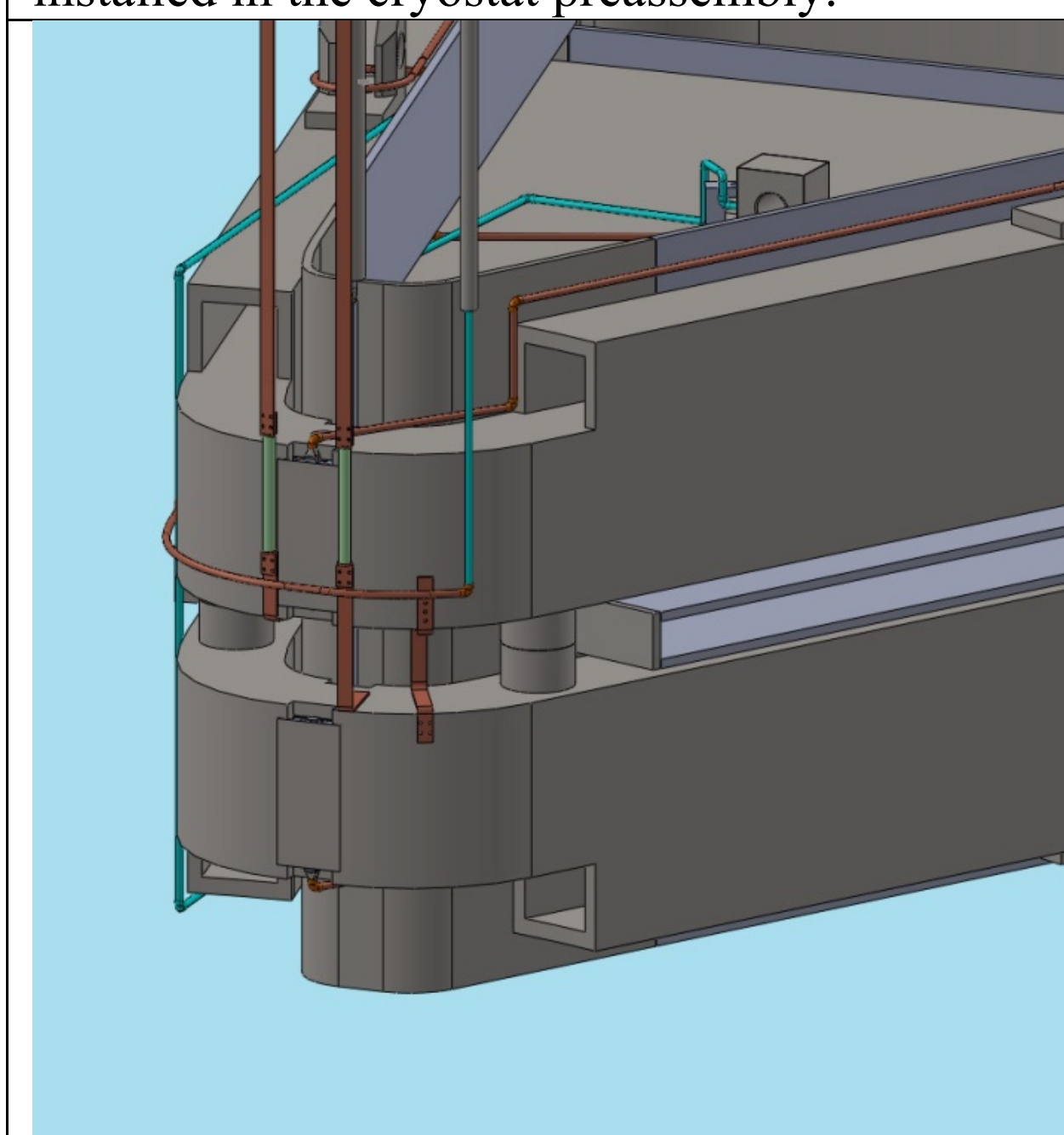 19. Top and bottom coils He plumbing and cabling connected. | 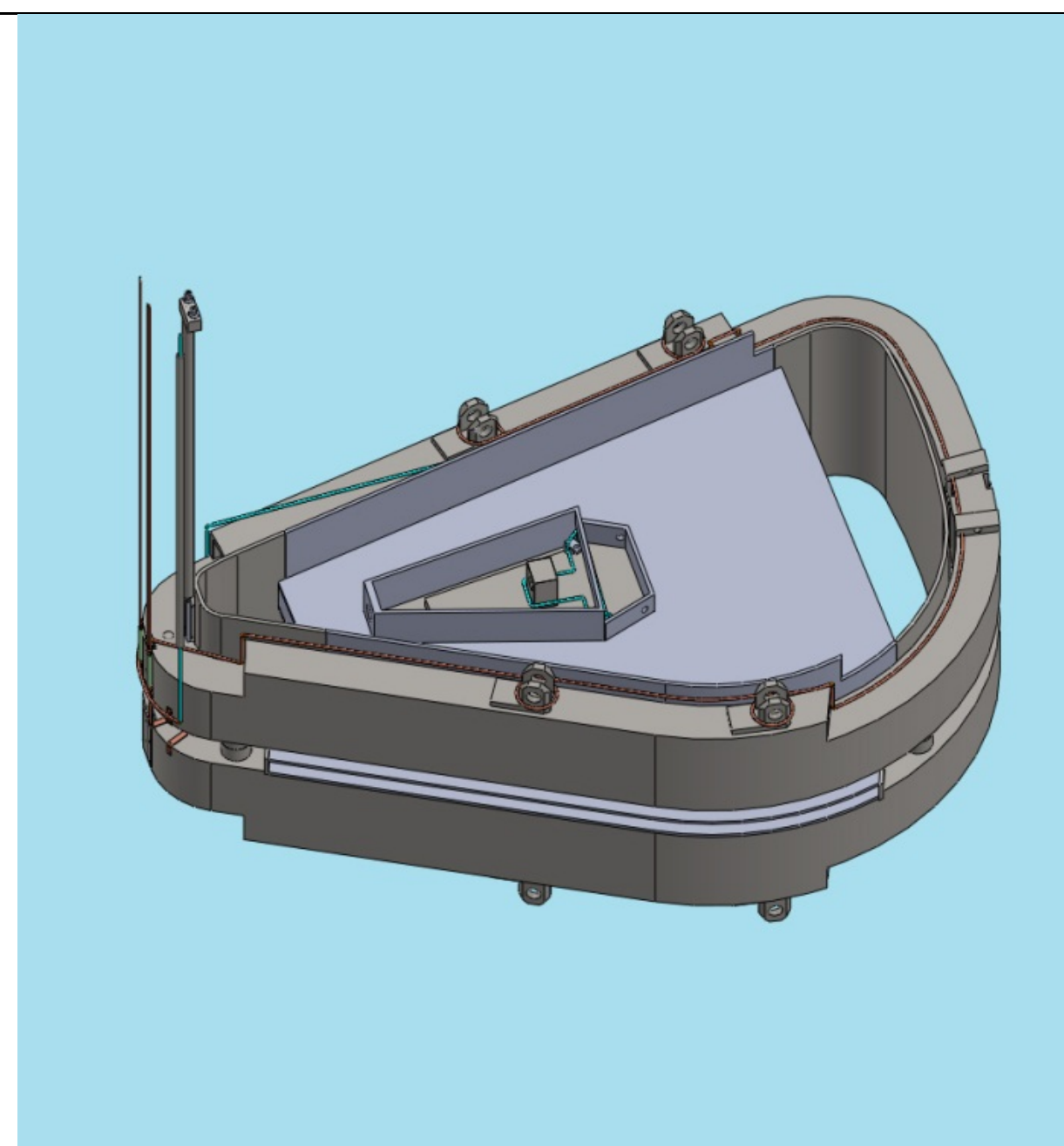 20 Cryostat top plate covering cold mass tie plate welded in. |

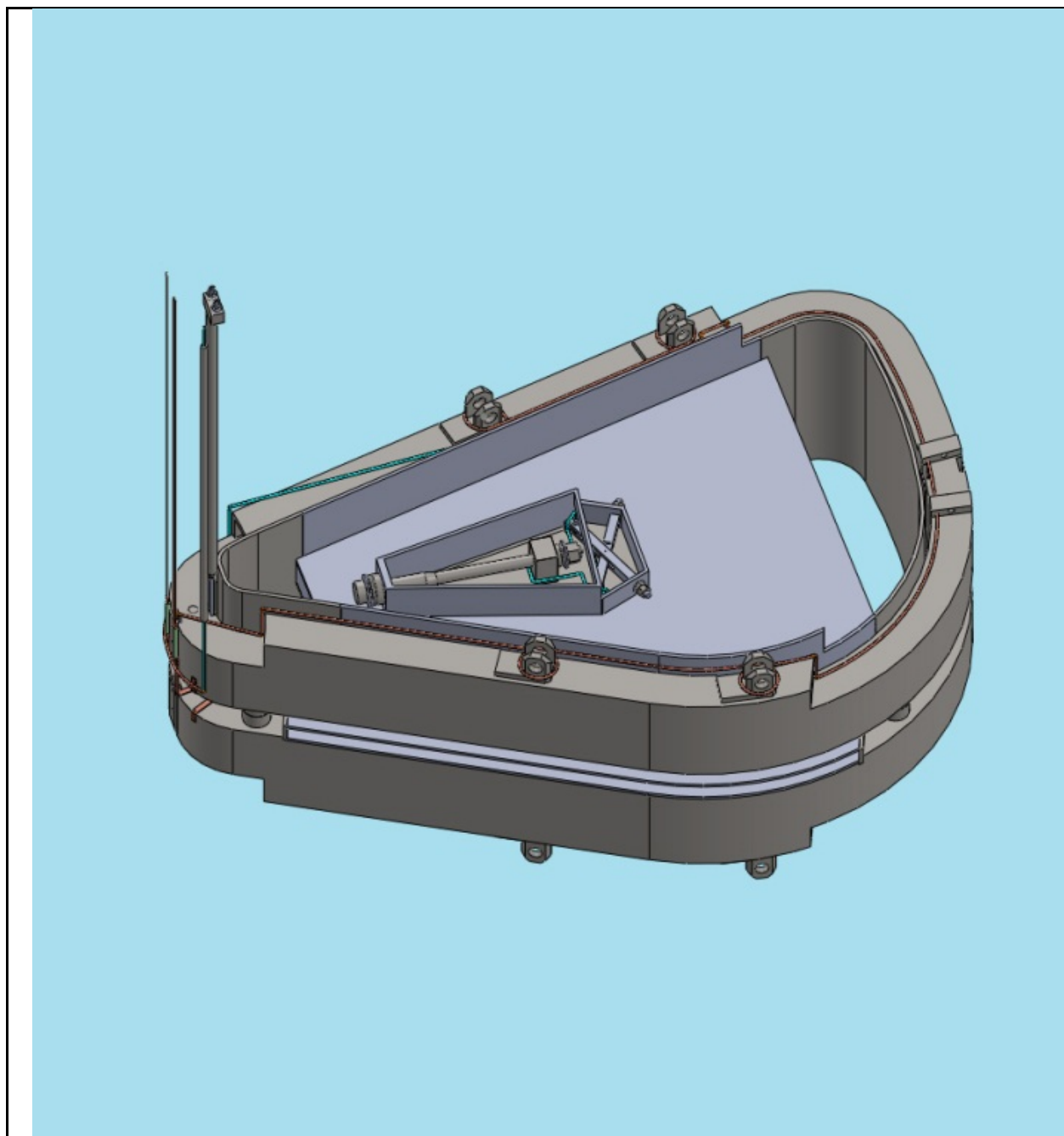

21. Radial and lateral tie rods of the central cluster installed.

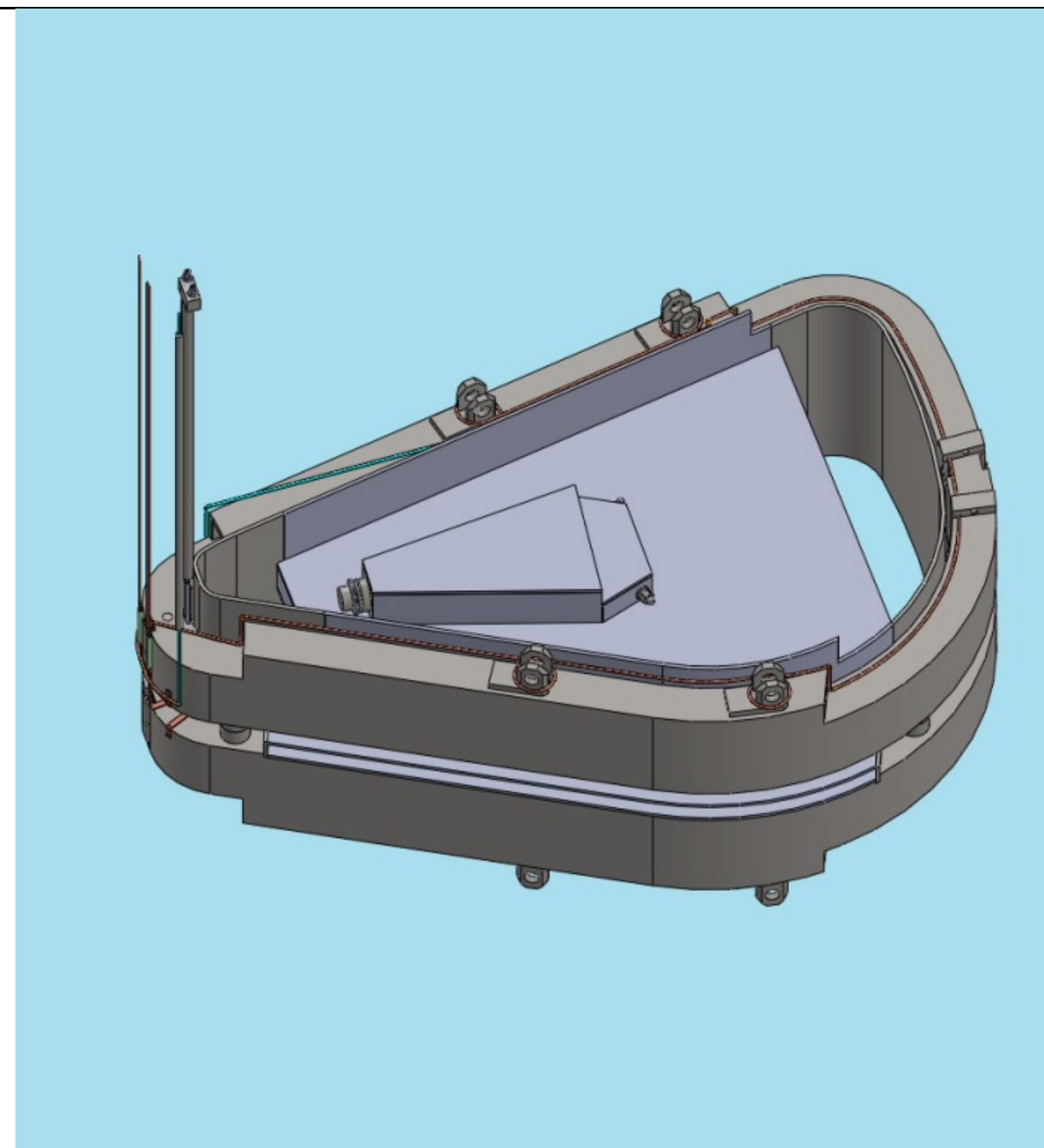

22. Cryostat box around the central support cluster complete.

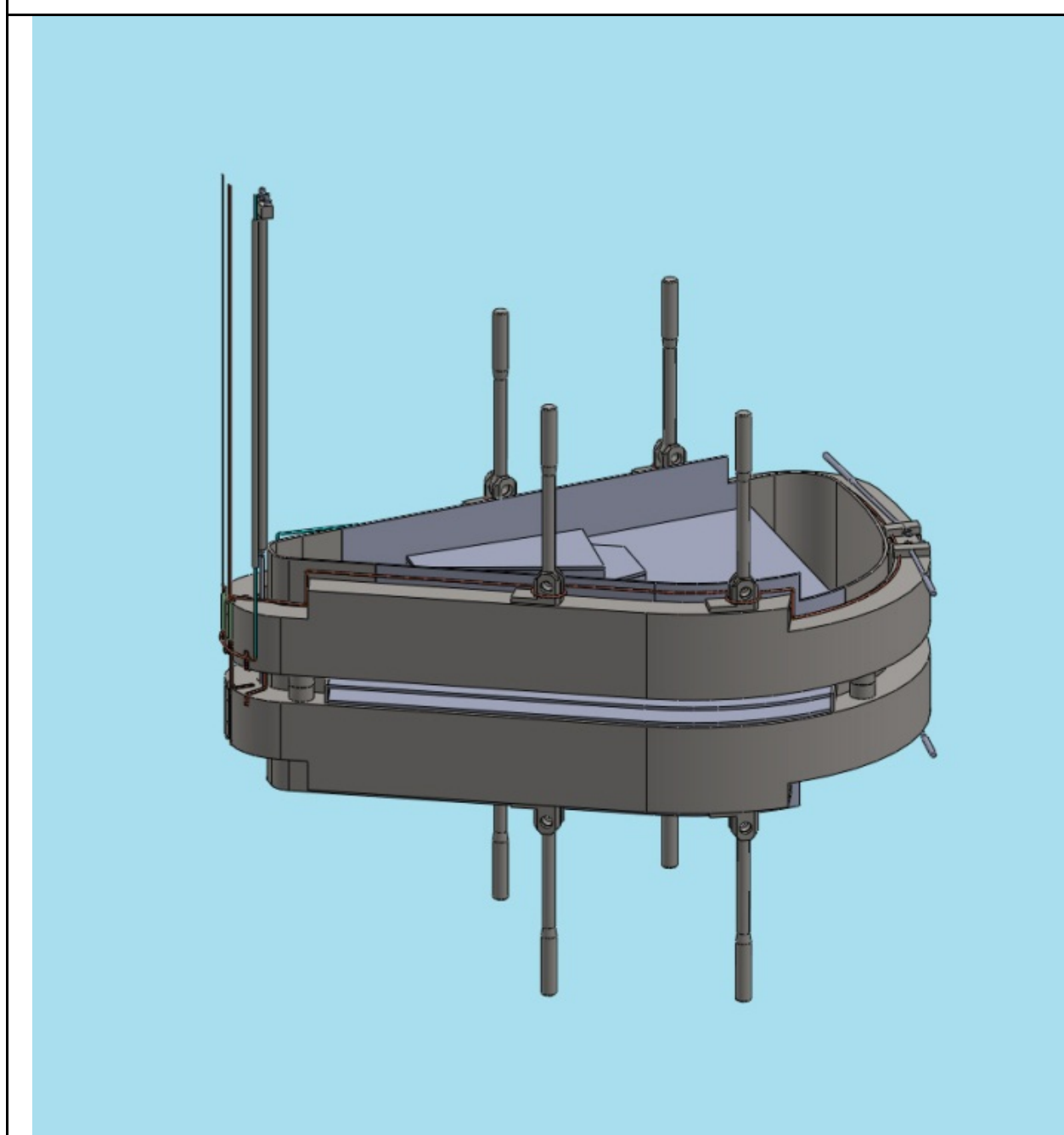

23. Axial and lateral OD support rods attached.

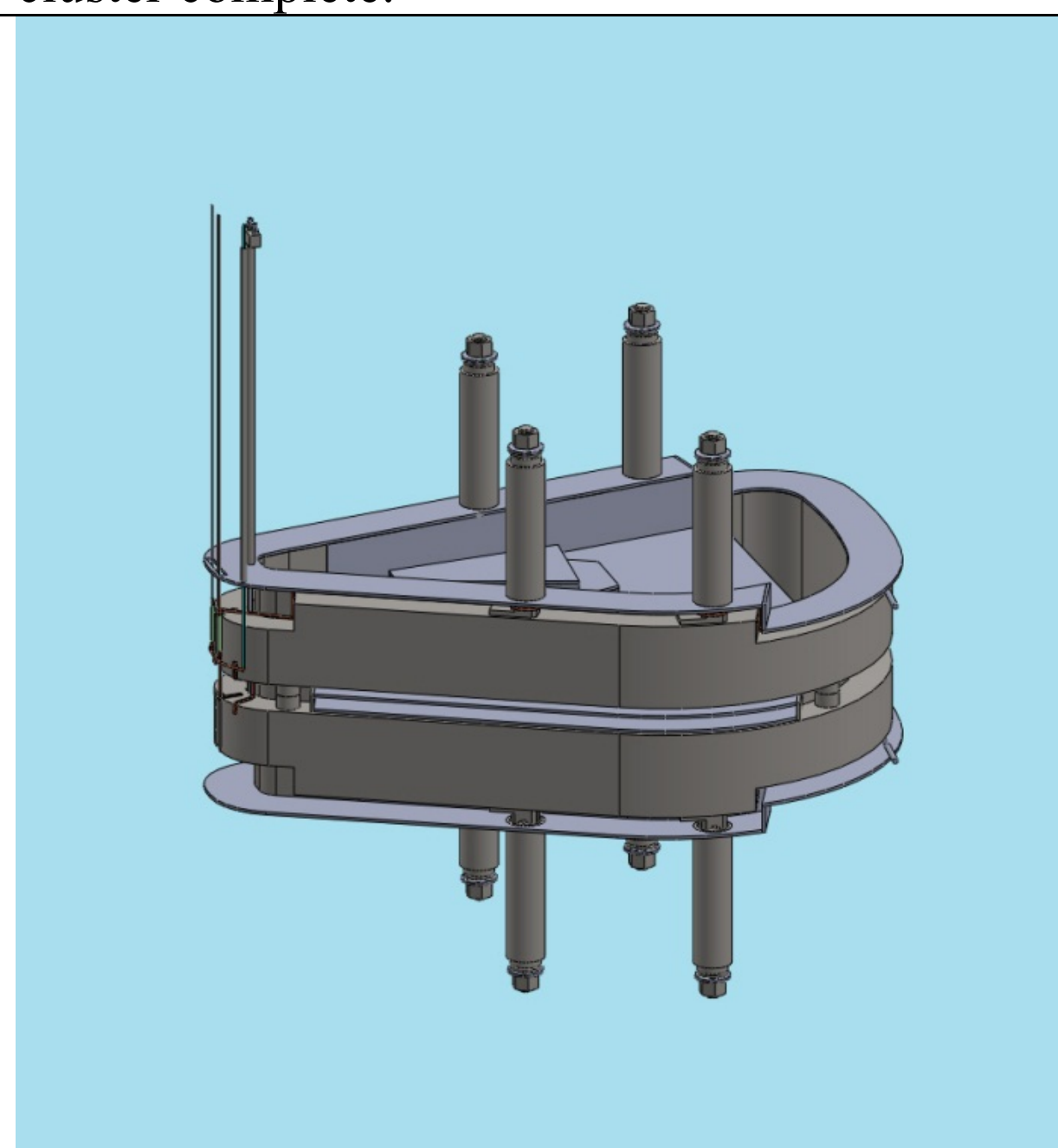

24. Top and bottom cryostat plates with axial support rod encasements welded to the cryostat ID wall. Axial supports complete.

| | |
|---|---|
| 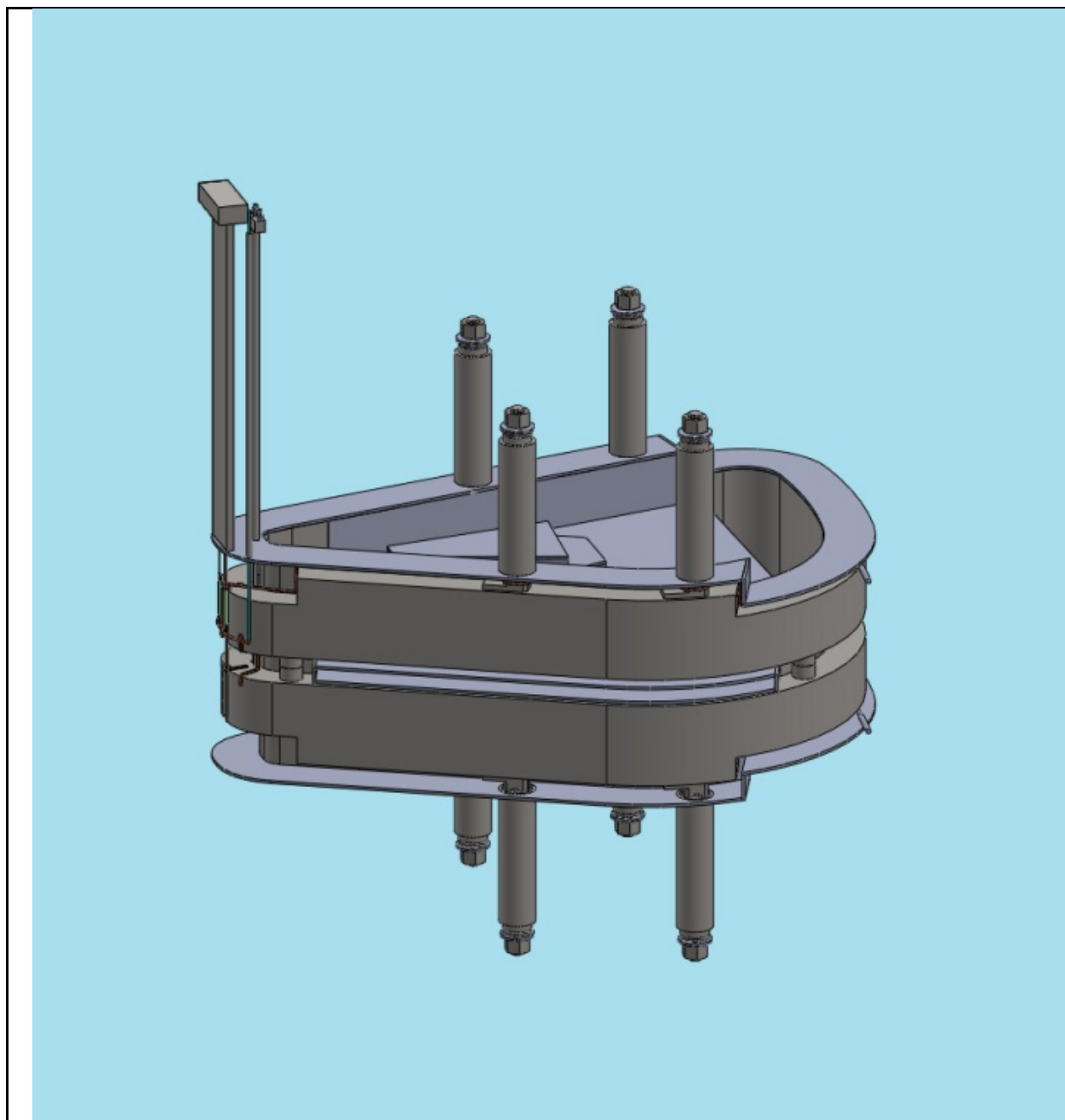 25. Cryostat chimney case with He plumbing and electrical connections arranged. | 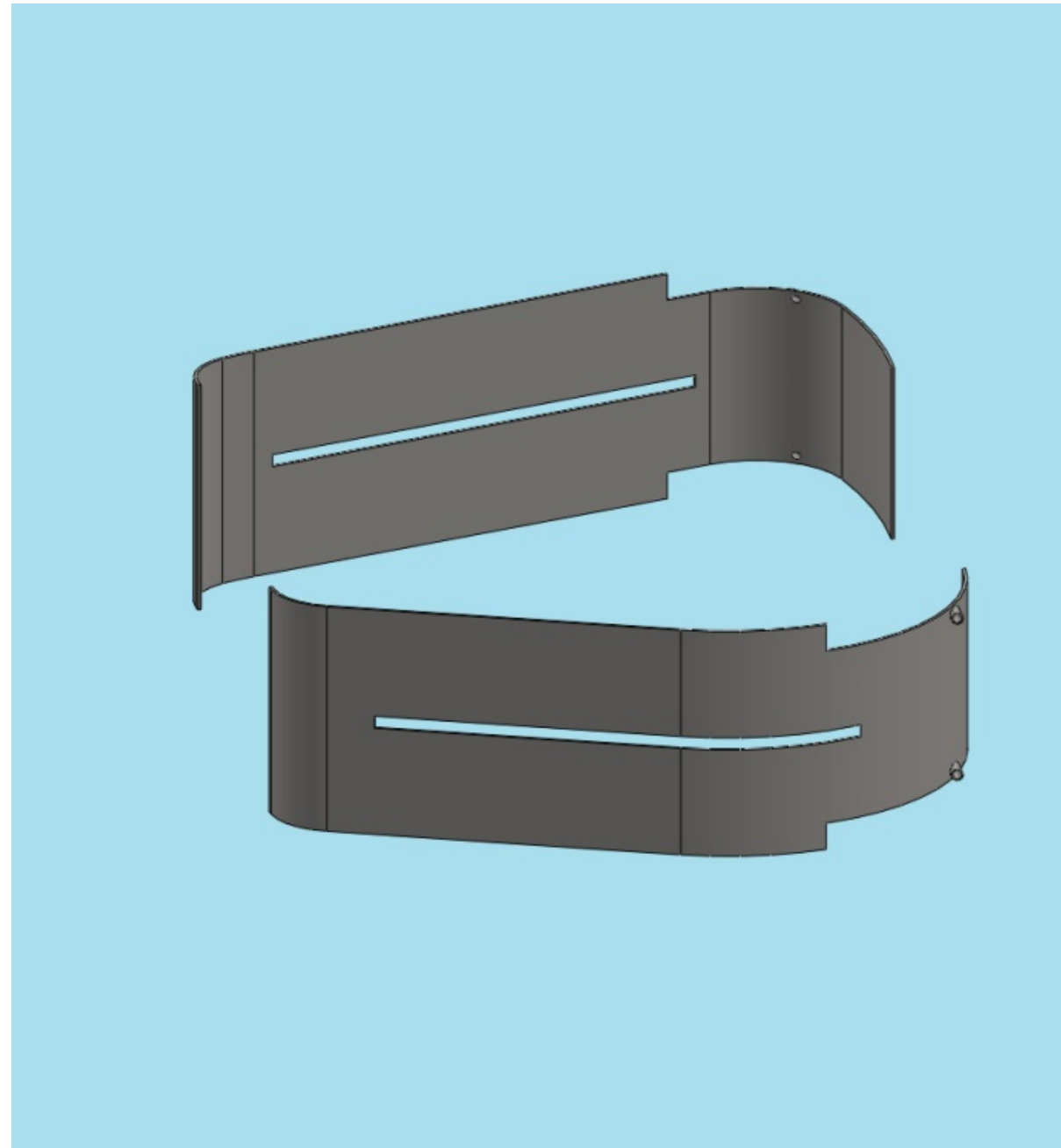 26. Cryostat OD wall comprises of 2 shells. |
| 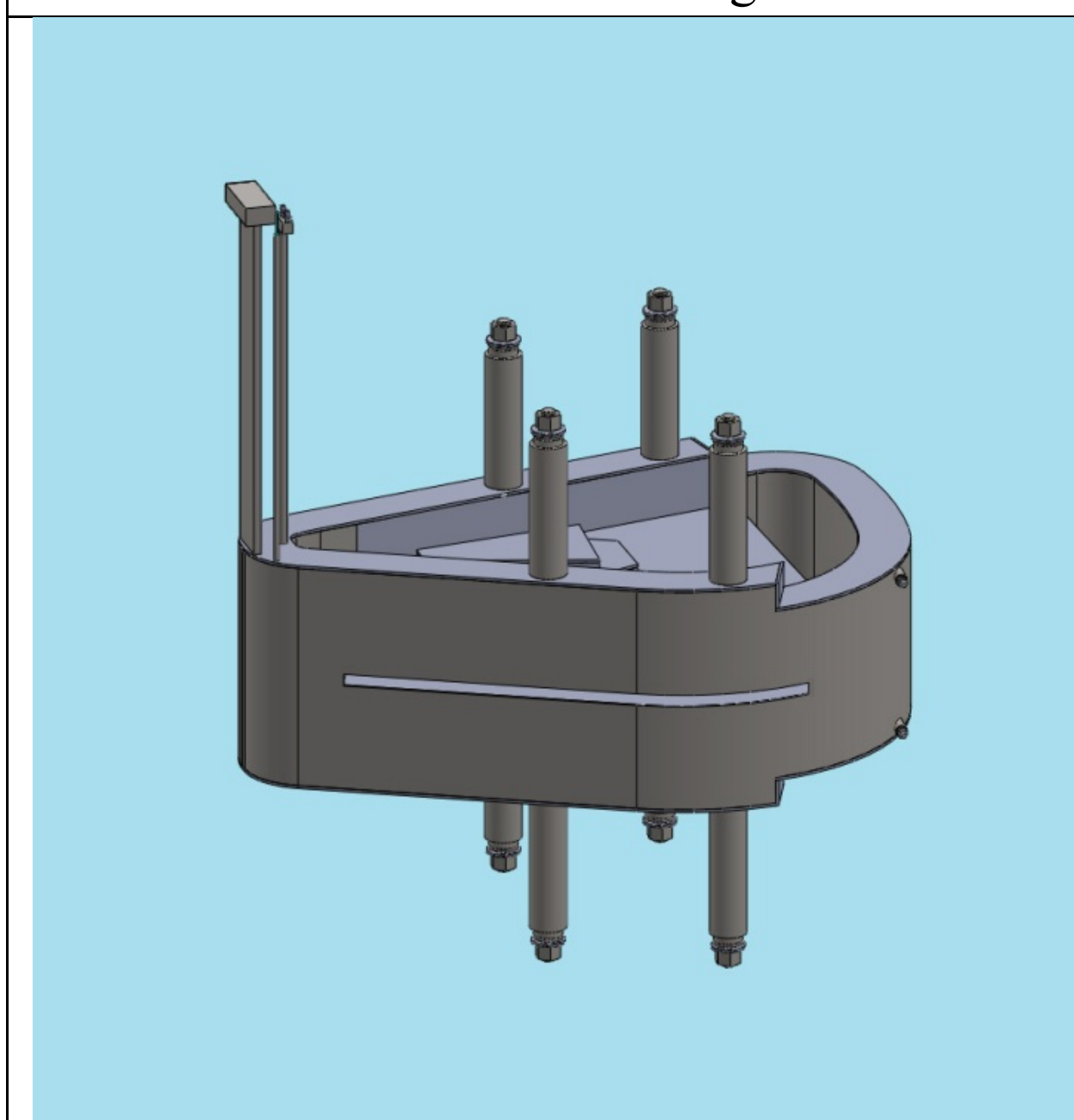 27. Cryostat OD wall shells welded to the cryostat assembly. Lateral OD supports complete. | 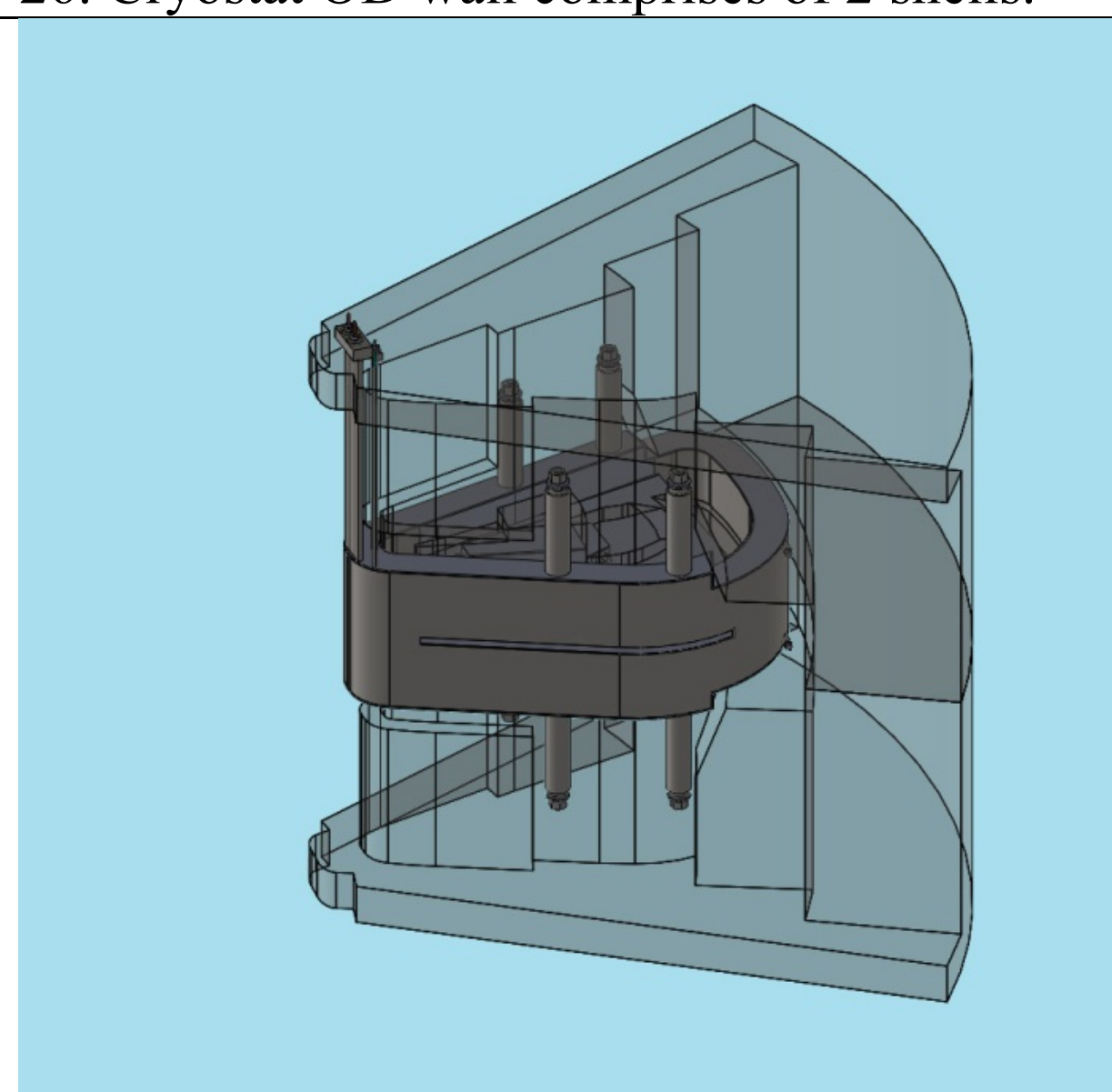 28. View: Magnet in the iron (transparent). |

## 2.3.Issues to be addressed in the future

As was said before some design details were beyond the scope of this work and will have to be investigated at later stages of the project. These include but are not limited to:

- He and LN2 thermal intercepts of support struts present a challenge because of the difficulty of cooling through the whole cross section of large, up to 10-cm, diameter support struts. The design of the intercepts has to be developed and experimentally tested to ensure their efficiency.

- Assembling the iron yoke around the magnet was not considered. Usually it is not very challenging. EM attraction between the coil and the iron is transferred through support struts to the cryostat, which is restrained against the iron by spacers. In this particular design there is a clear challenge associated with the necessity of placing two 16-cm iron plates in the space between the cryostat around the coil case tie plates and the later installed beam chamber as shown in Fig. 2-11. The minimum gap between the cryostat walls in the space accommodated for the beam chamber is only 8 cm. This means that either each plate has to be made of two 8-cm thick halves and then assembled in place or it has to be placed in the central part of the cryostat preassembly prior to integrating the coils into the cryostat.

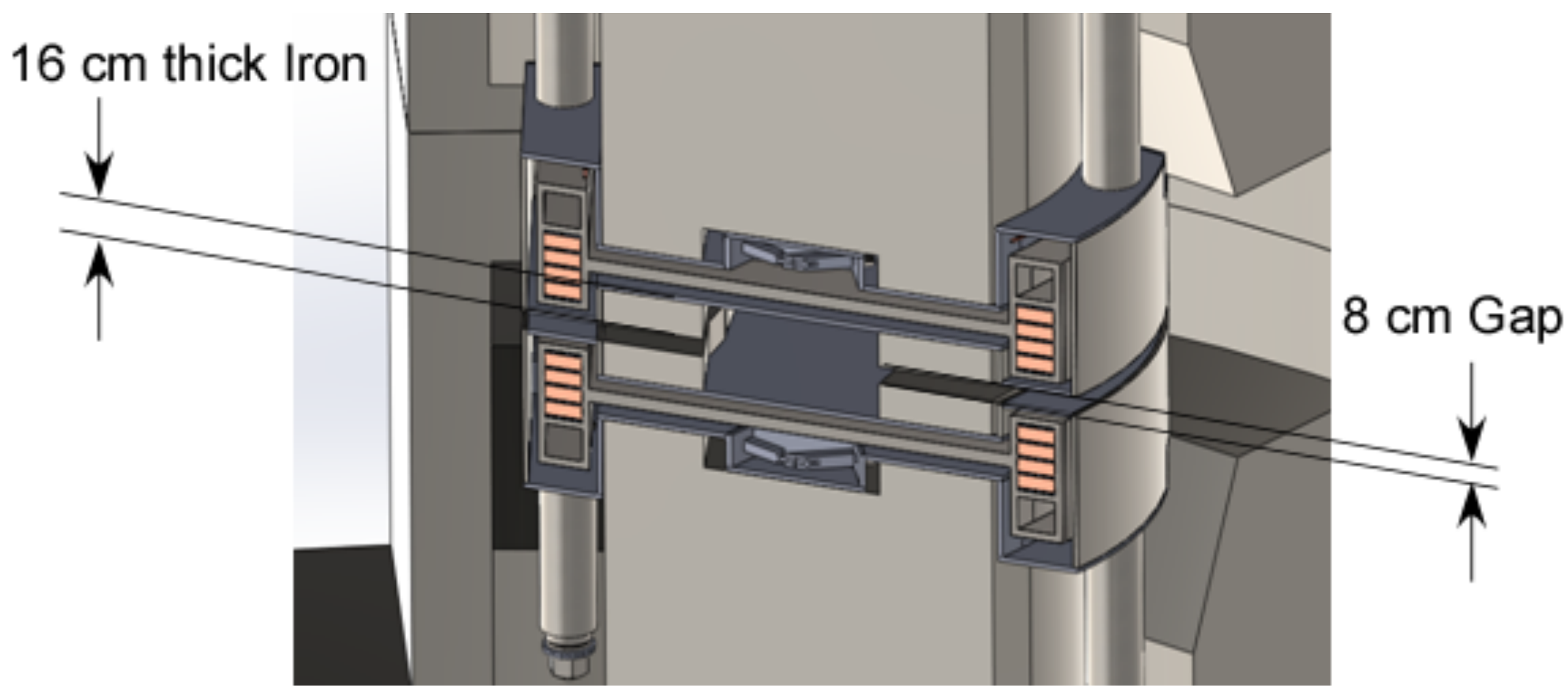

Fig. 2-11. Iron yoke around the beam chamber

- A large portion of heat losses come through the vertical support struts, which are sized primarily to withstand a fault involving a single coil short during the fast current dump usually resulting from a coil quench. The conservative rationale behind this design is given in the structural chapter. The probability of this fault, its consequences and the possible mitigation of these consequences should be assessed by the Collaboration. If this fault is deemed non-credible, it could lead to a significantly less conservative and consequently less costly design.

- The design of the support struts was shown schematically. A more detailed design facilitating a convenient way of adjusting the position of the cold mass within the cryostat after assembling the system and field mapping at the operating current requires consideration of (a) special , possibly accommodating bellows, design of the RT ends of the support struts and (b) providing access to the adjustment mechanisms without disassembling the iron yoke. The latter may result in adjustments of the tilt of the radial and lateral tie rods of the central cluster (shown in figure 21 of the assembly sequence).

- Design of the radiation shield may show that the space between the coil case and the cryostat is insufficient to prevent thermal shorts due to the displacements of the cold mass during cooldown and adjustments of the position of the cold mass after the field mapping. To increase that space it may be inevitable that the magnetic design will have to be modified to increase the gap between the coil winding and the iron yoke. Currently it was set to 10 cm.

- Another adjustment of the magnetic design is related to the fact that in the current design the winding is split into 4 quadro pancakes , each with cooling plates above and below, whereas the baseline magnetic design assumes a continuous winding with a uniform current density. Future magnetic models shall take account of this modification.

# 3. Winding Pack Design

This section provides a conductor and winding pack arrangement to match the present winding pack cross section of 31 cm tall by 16 cm wide, an overall current density of 34.68 A/mm$^2$ and a per sector stored energy of 53.3 MJ.

The conductor consists of 12-strands of 1.25 mm diameter NbTi superconductor with a copper to superconductor ratio of 1.3:1 soldered into a copper channel for stability and quench protection. With the stabilizer proposed, dump voltages during quench will be less than 600 V with a hot spot temperature less than 120 K in RRR50 copper. Achieving this performance will require quench detection using voltage taps and an active protection circuit utilizing a dump resistor of about 0.12 Ω to be switched across the terminals of the sector on detection of a normal zone. The current sharing temperature of the chosen conductor at the location of the 4.37 T peak field in the winding is 6.4 K when carrying the nominal 4887 A conceptual design current necessary to meet the ampere-turns requirement of the magnet. Thus, the superconductor temperature margin will be about 1.4 K when the conductor is operated at up to 5.0 K. This elevated operating temperature over the ~4.5 K temperature of the He coolant allows some design margin for ionizing-radiation-induced heat load into the winding pack. A cooling arrangement is also described, which assumes a 1.33 cm thickness for helium carrying cooling plates to remove heat generated in or absorbed by the winding pack. Optimization of the cooling plate design is a task for later.

## 3.1. Details of the arrangement

The 31 x 16 cm winding pack envelope is shown in Fig. 3-1 and consists of 4 sets of double pancake windings, interleaved with and contained at the top and bottom by a total of 5 cooling plates. Conceptual details of the double pancakes and cooling layer are further elaborated below.

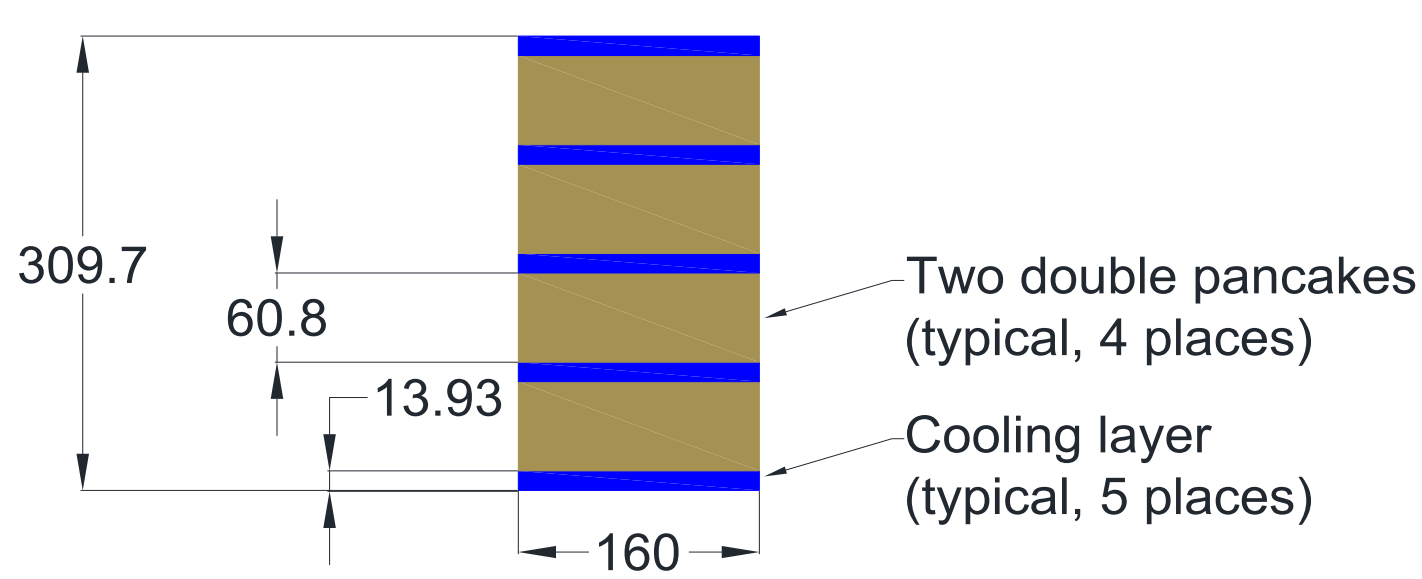

Fig. 3-1. Winding cross section sketch (dimensions in mm)

The conductor cross section is shown in Fig. 3-2 with a half-lap of 0.1 mm thick fiberglass tape overwrap, making a turn insulation thickness of 0.2 mm all around. The strand is 1.25 mm diameter NbTi with Cu:Sc=1.3:1, Supercon 54S43 or equivalent, formed into a Rutherford cable. The copper channel usually requires some cold work to support winding pack loads. At 20% cold work, for example, the minimum expected RRR in the copper is 55 in accordance with Fig. 3-3.

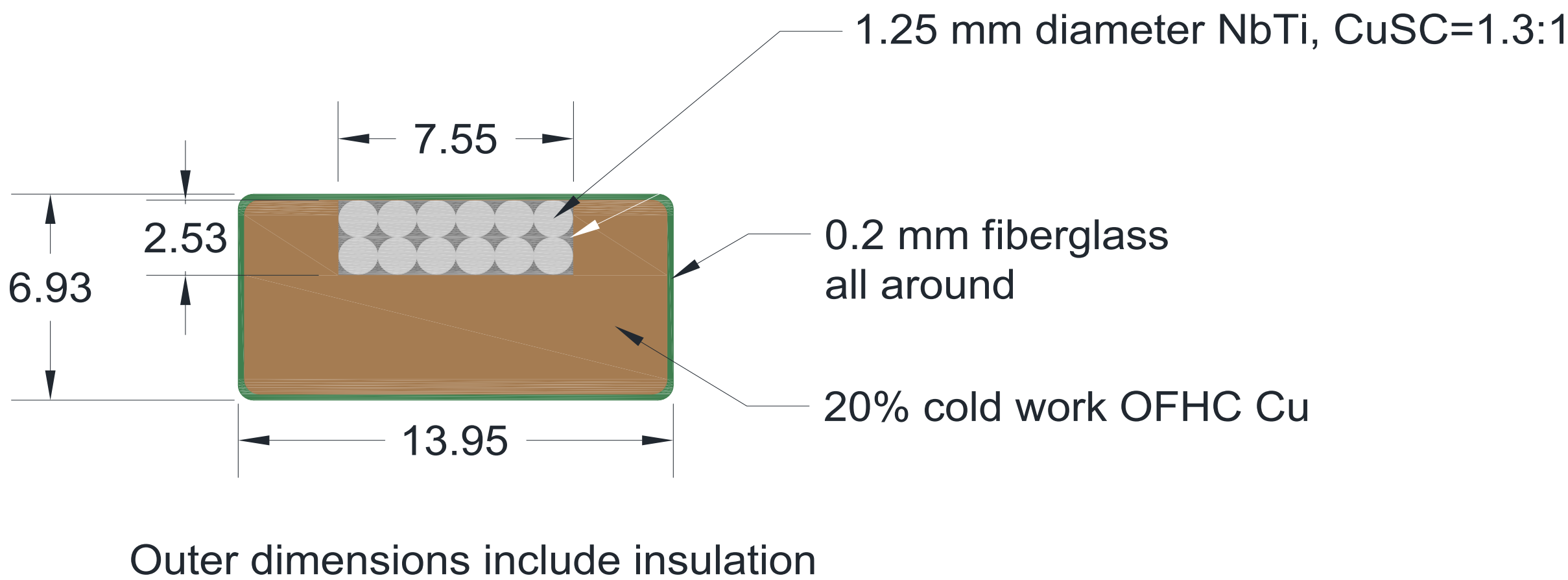

**Fig. 3-2. Cable in channel conductor (dimensions in mm)**

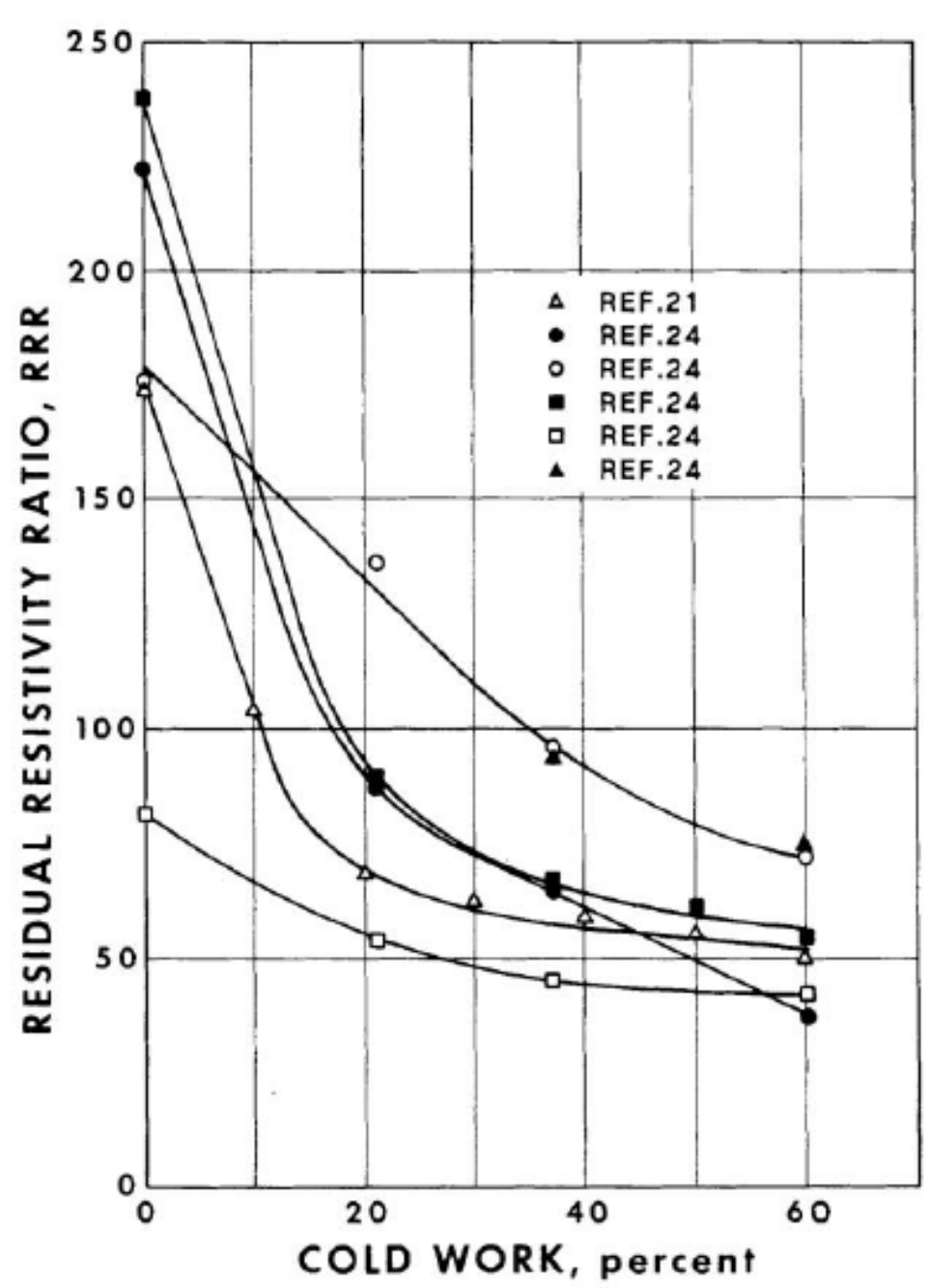

Figure 8.8. The residual resistance ratio, RRR, which is defined in the text, is shown as a function of the amount of cold work. Products were in plate (Reference 8.21) and wire form (Reference 8.24).

**Fig. 3-3. RRR vs. Cold Work %, from *Properties of Copper and Copper Alloys at Cryogenic Temperatures,* NIST Monograph 177**

The conductor will be wound as shown in Fig. 3-4 to form double pancakes, with the layer to layer transition occurring at the inside of the winding pack and jumpers between double pancakes at the outer dimension. A 2 mm space between the single pancake layers is allocated for the potential future addition of a quench heater covered on each side with a minimum of 0.5 mm of insulation. The need for such a heater will be established in a later phase of the project when a more detailed quench and quench stress analysis can be performed.

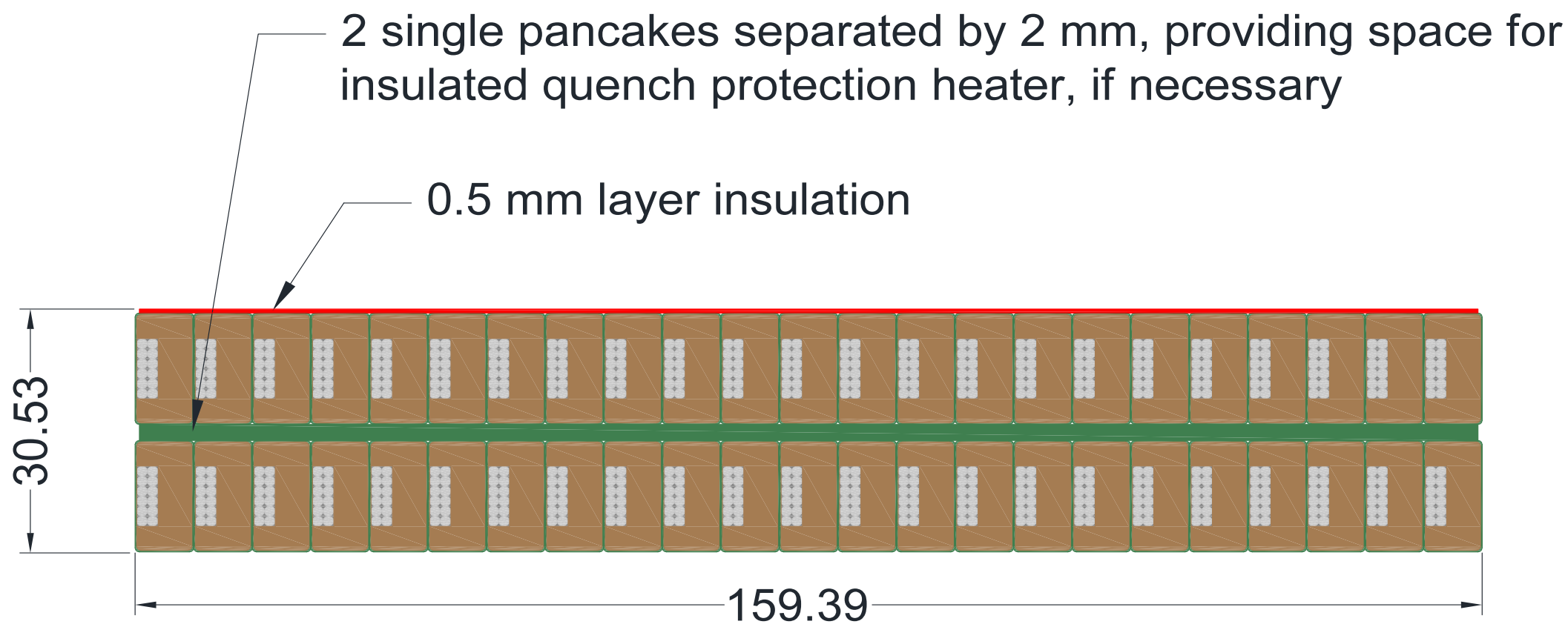

Fig. 3-4. Double pancake (~44 turns)

A stack of two pancakes, as shown in Fig. 3-5, comprise the two double pancakes shown in the Fig. 3-1 arrangement sketch.

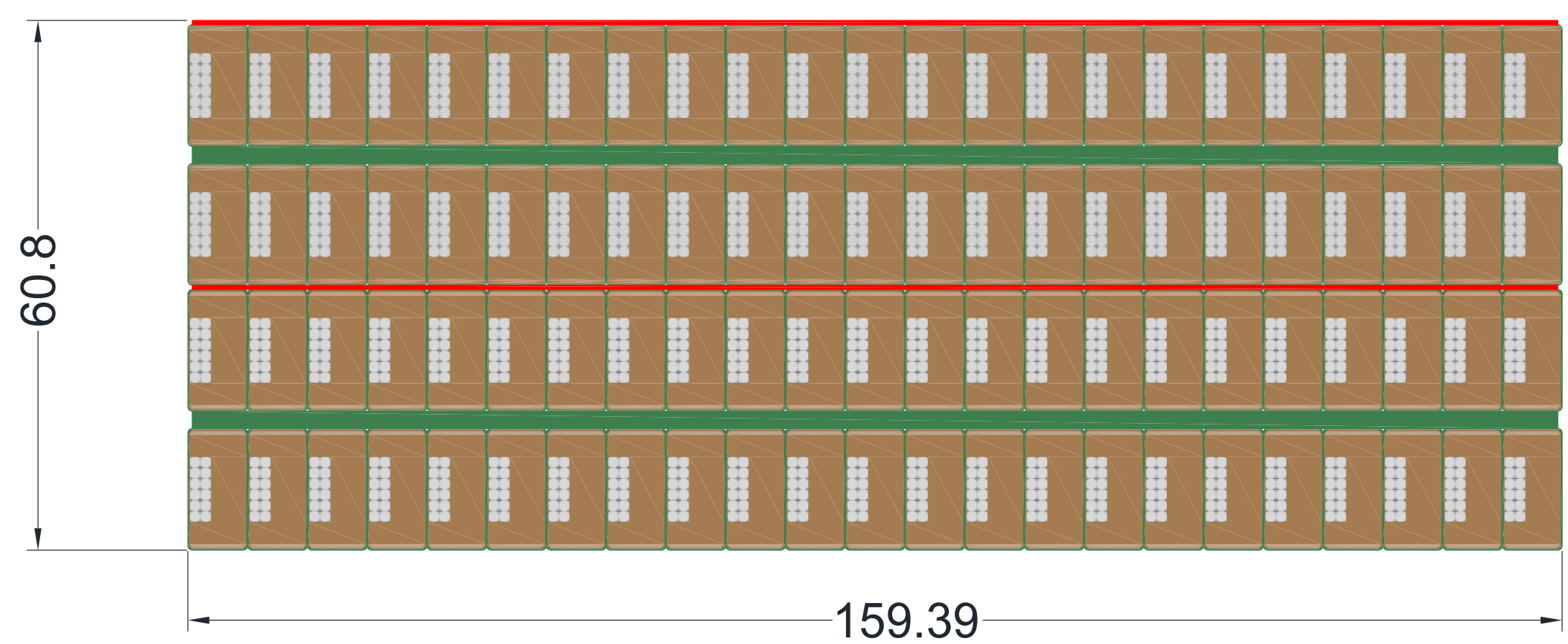

Fig. 3-5. Two double pancakes (~88 turns)

A concept for the stainless steel cooling plate is shown in Fig. 3-6. Stainless steel will provide the best thermal contraction compatibility with the copper-based winding. Additional details of cooling channel dimensions are contained in the cryogenic section of this report, but a possible approach is that the cooling grooves and blocking weld shown would spiral around the plate along the developed length of the winding. Each cooling plate accepts heat from the coil pancake on either side of the plate. Additional details of the heat loads including an estimate of the maximum allowable ionizing radiation are contained in the cryogenic section.

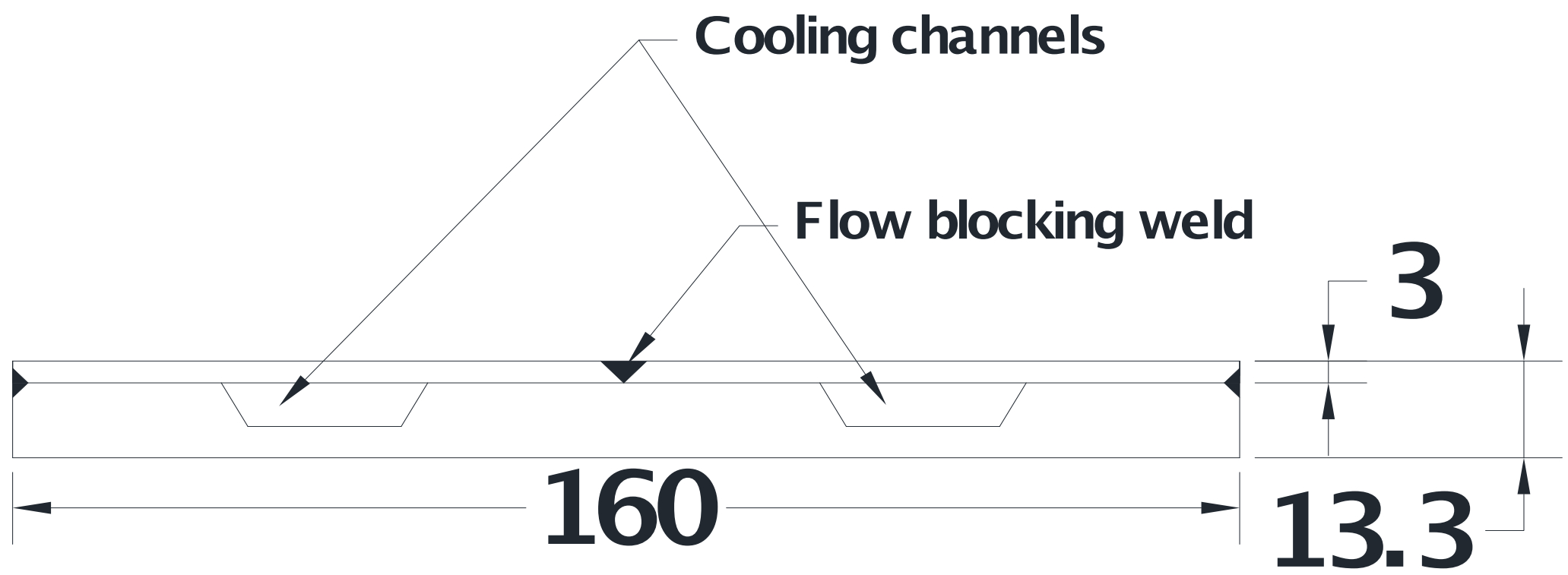

Fig. 3-6. Stainless steel cooling plate cross section

# 4. Structural Analysis

The conceptual structural design and analysis of the superconducting ring cyclotron (SRC) used for the Daeδalus experiments is presented. The structural design includes a manufacturable concept for: (1) the coil case and warm-to-cold supports that hold the superconducting coil in place, and (2) the cryostat surrounding the cold mass. The structural analysis demonstrates that the cryostat and the cold mass supports minimize displacements around the beam chamber and they are not overstressed during: (1) cool-down to 4K, (2) normal operation at 4K, and (3) fault condition at 4K. The geometry is assumed to be a 6 sector system where all concepts can be scaled to an 8 sector system. The electromagnetic (EM) simulations include the iron yoke and superconducting coils to generate the Lorentz body forces input to the coils in the mechanical analysis. The mechanical simulations of the cold mass couple the EM loads with thermal loads to calculate the displacements and stresses as well as the forces in the warm-to-cold struts. The forces in the warm-to-cold struts are used to size the cross section of the struts. The forces in the warm-to-cold struts and atmospheric pressure are applied to the cryostat to analyze its structural integrity.

## 4.1.Geometry

### 4.1.1. Coil Segments

The coil is divided into 9 segments for the EM and structural analyses as shown in Fig. 4-1. Either a Cartesian or cylindrical coordinate system orients each segment's winding direction. Tangency is required between coordinate systems on abutting segments in order to have continuity of the anisotropic mechanical material properties used to describe the coils in the structural analysis. The coil has a 16cm x 31cm rectangular cross section and has minor deviations between the coil shown in Fig. 4-1 and the coil design provided [1]. The minor deviations peak at 2.2cm at two locations, at 1.6cm at two different locations, and are <1.0cm for 70% of the coil length.

| SEGMENTS MADE FROM STRAIGHT LINES | | COORDINATES ON INNER PERIMETER(cm) | |
|---|---|---|---|
| | | X | Y |
| SEGMENT 1 | START | 149.134 | -4.87 |
| | END | 148.089 | 5.075 |
| SEGMENT 4 | START | 176.308 | 32.172 |
| | END | 484.959 | 145.957 |
| SEGMENT 8 | START | 373.909 | -101.241 |
| | END | 161.981 | -21.498 |

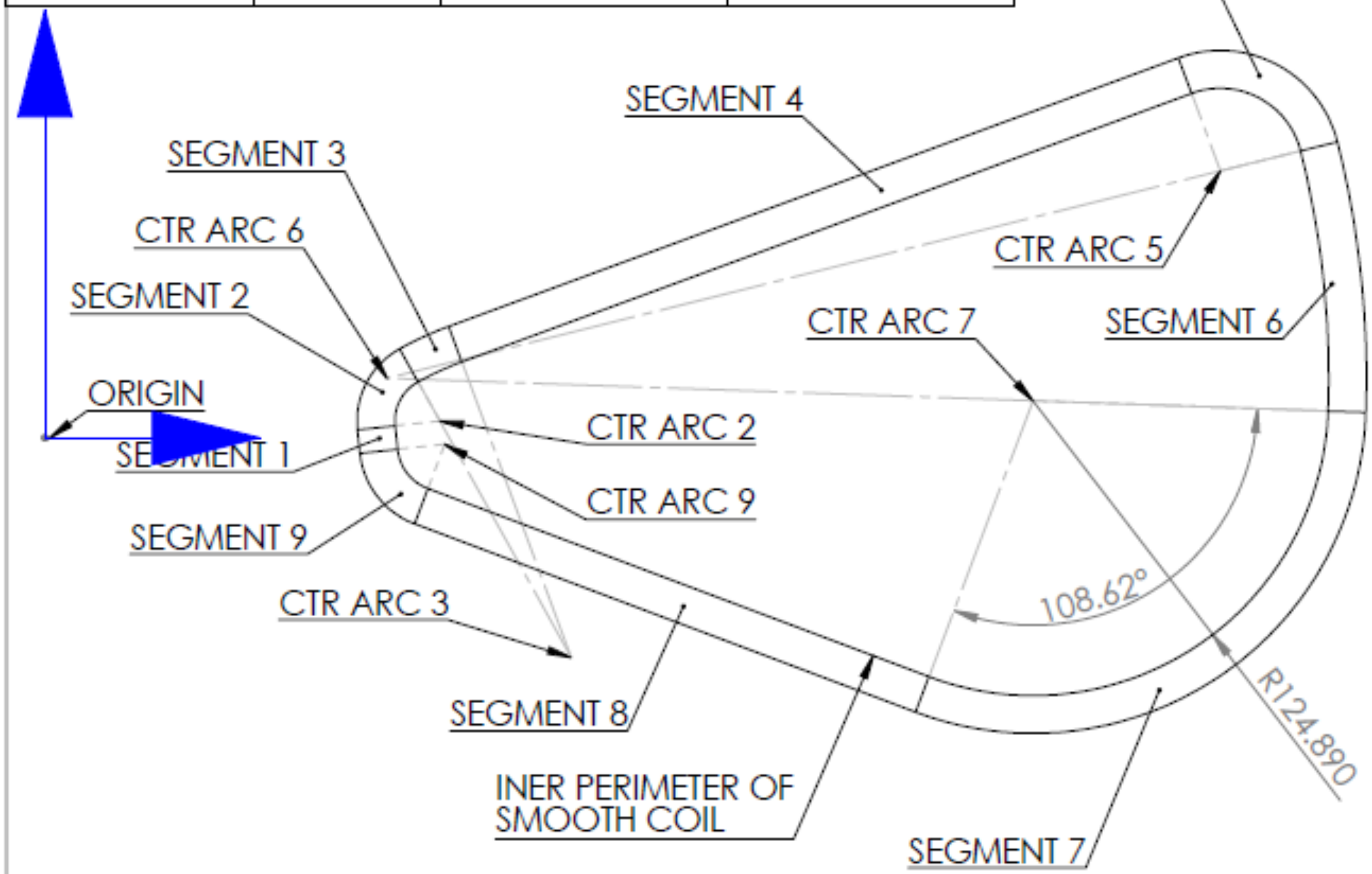

| SEGMENTS MADE FROM ARCS | ARC CENTER COORDINATES (cm) | | RADIUS OF ARC SEGMENT (cm) | ANGLE OF ARC SEGMENT (°) |
|---|---|---|---|---|
| | X | Y | | |
| SEGMENT 2 | 166.985 | 7.061 | 19 | 67 |
| SEGMENT 3 | 222.461 | -93.021 | 133.429 | 8.7635 |
| SEGMENT 5 | 497.09 | 113.052 | 35.07 | 96.2365 |
| SEGMENT 6 | 144.693 | 25.189 | 398.255 | 16 |
| SEGMENT 7 | 417.891 | 15.649 | 124.89 | 108.62 |
| SEGMENT 9 | 169.024 | -2.78 | 20 | 63.38 |

**Fig. 4-1. Description of the coil broken into 9 segments used in the EM and mechanical analyses.**

### 4.1.2. Coil Case

The coil case surrounds each 16cm x 31cm coil along its entire length with 3cm thick 316 Stainless Steel. It will be designed such that it can be manufactured by welding and/or bolting together piece parts around the coil; however, the design presented here is one continuous piece of material. A pair of coil cases within one sector of the SRC will connect to each other through the cold-to-cold posts highlighted in Fig. 4-2.

The coil case bears the tremendous loads generated by the charged coils while limiting their displacements. First, the 6cm tie plate supports the azimuthal forces that are bending the coil from its triangular shape into a circular shape. The 6cm additional coil case thickness in the lower right of Fig. 4-2 provides stiffness for the same bending action of the coil where there is no tie plate. Second, the axial loads created by the attraction between the pair of coils in each sector are supported by the cold-to-cold posts. These forces can lead to unacceptable axial deflections along the long lengths of the coil between the cold-to-cold posts which could interfere with the beam chamber passing through this location. The bending of the coil is minimized using the axial support structure highlighted in Fig. 4-2. This additional structure is a U-channel extending the long lengths of the coil support away from the median plane in order to increase its bending stiffness. Third, there is an additional plate used to stiffen the tie plate upon which the radial support boss is mounted. The radial load is supported using warm-to-cold struts connected to the radial support boss and an additional plate is necessary to stiffen the tie plate.

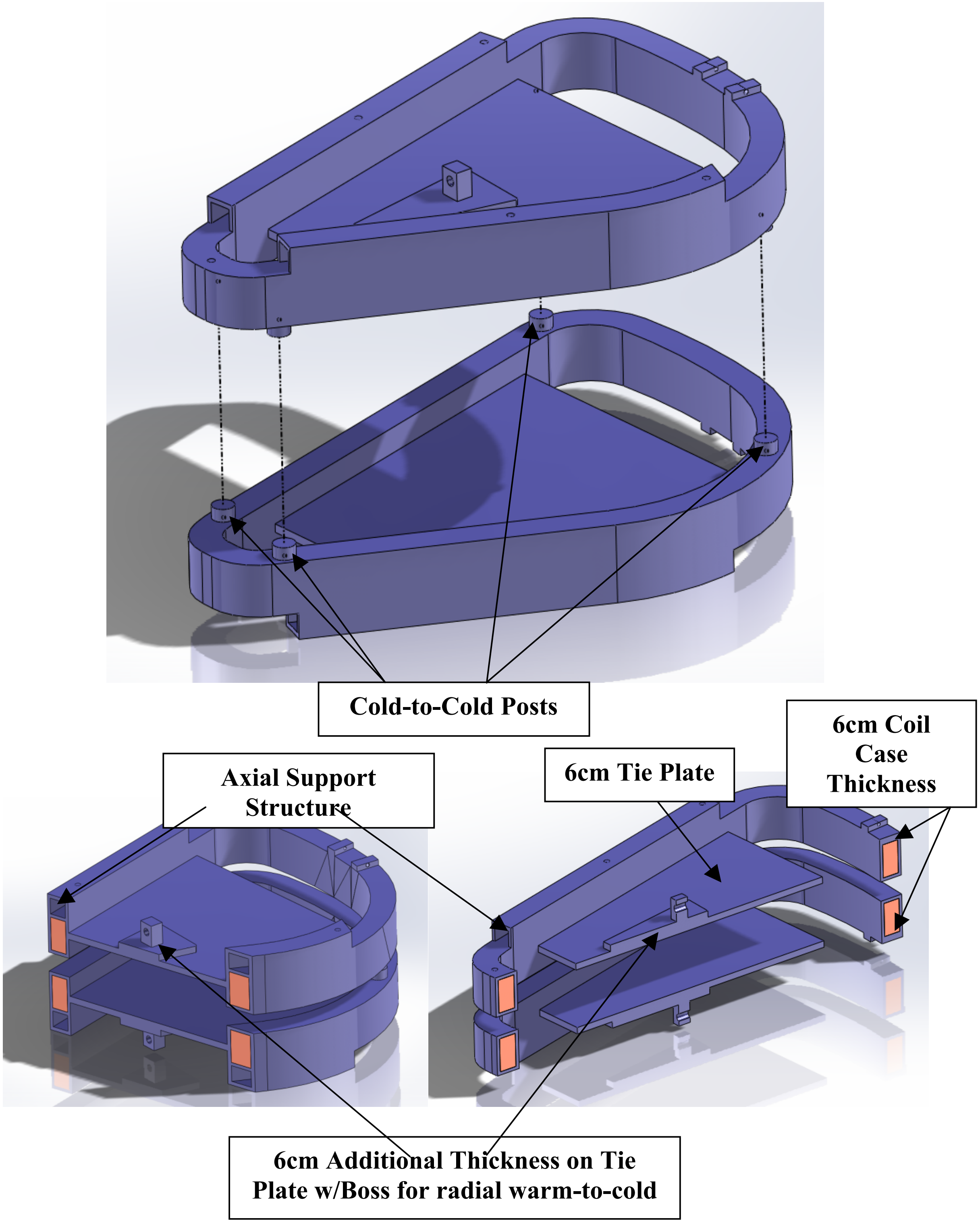

Fig. 4-2. Coil case for bottom superconducting coil in one sector of SRC. Dimensions are in cm.

### 4.1.3. Warm-to-Cold Support Struts

The cold mass is suspended from the room temperature (RT) cryostat using several support struts. These support struts are described as "warm-to-cold" because the warm end is attached to the RT cryostat and the cold end is attached to the cold mass. They are typically tension only members that require pre-tensioning during the manufacturing of the SRC. The advantage of tension-only struts is the buckling failure mode need not be considered.

The two coil cases and the warm-to-cold struts in one sector of the SRC are shown in Fig. 4-3, where the struts are numbered for reference. The suggested material for the struts is Nitronic 50. The four axial tension-only support struts (Struts #1, #2, #3, #4) are used to bear three loads: (1) the axial loads during fast dump of one coil while a simultaneous fault has produced a short in the other coil, (2) the weight of the cold mass, and (3) contraction of the cold mass away from the RT cryostat during cool-down. There is one radial warm-to-cold strut, Strut #9, which bears the large radial force of the coil when charged. There are two pairs of azimuthal support struts to bear the azimuthal load and the torque generated by the coils (Struts #5 & #6, #7 & #8).

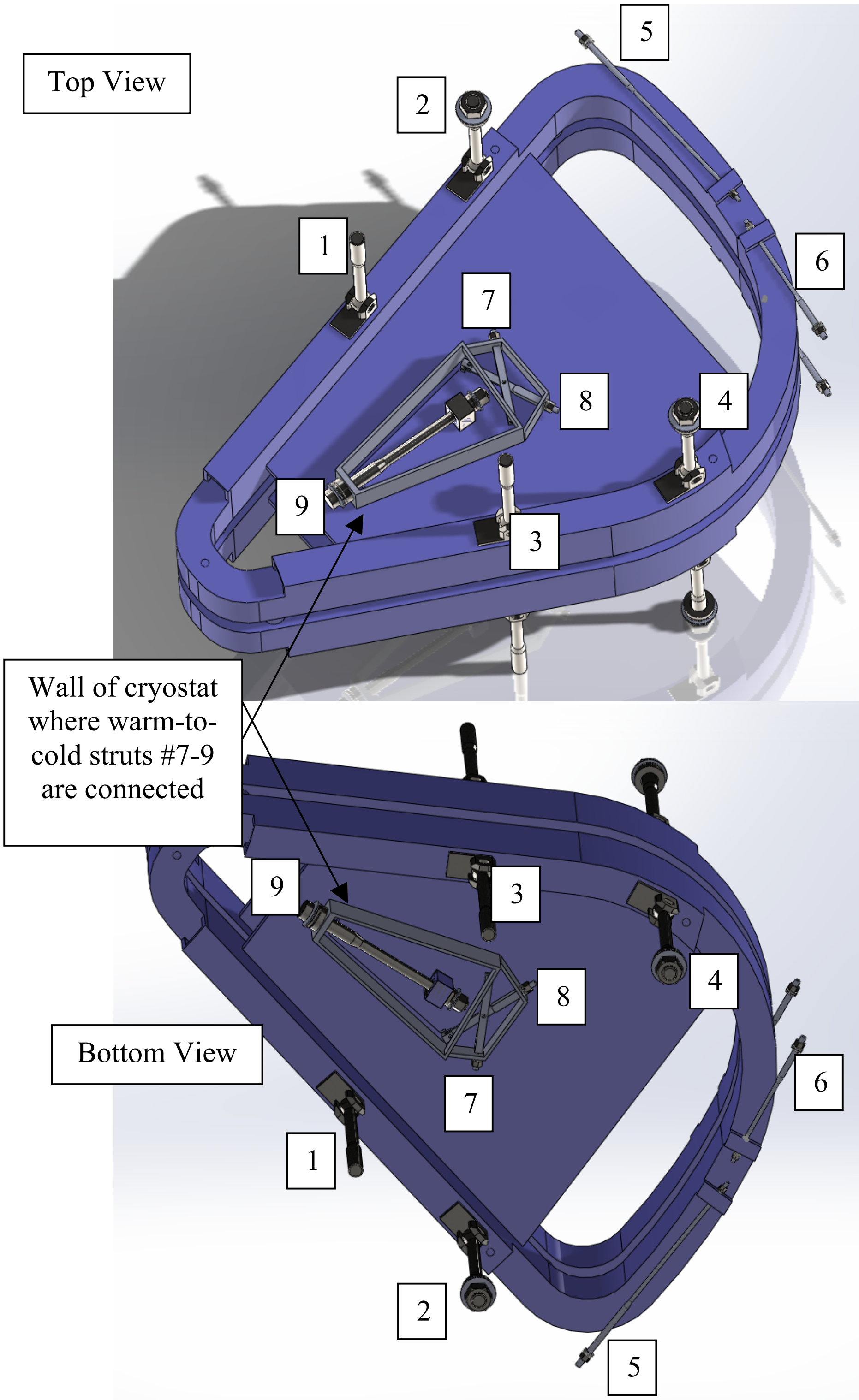

**Fig. 4-3. Top and bottom views of the coil cases in one sector showing numbered warm-to-cold support struts.**

#### 4.1.4. Cryostat

The cryostat seals the vacuum around the cold mass to allow the cooling system to reach is desired cryogenic temperature. It is made of stainless steel and is pictured in Fig. 4-4. It is 2 cm thick and is beefed up around the connections to the warm-to-cold supports. The locations the cryostat shell that are not 2cm thick are highlighted in Fig. 4-4. The large radial force experienced both during normal operation and during fault condition is reacted through a very small face of the cryostat. Ribs are added to distribute this load through more of the cryostat. The thickness of 10 cm on the front face of the cryostat around the tie plate may require removing some material from the tie plate. Fortunately, this is a region of low stress on the tie plate.

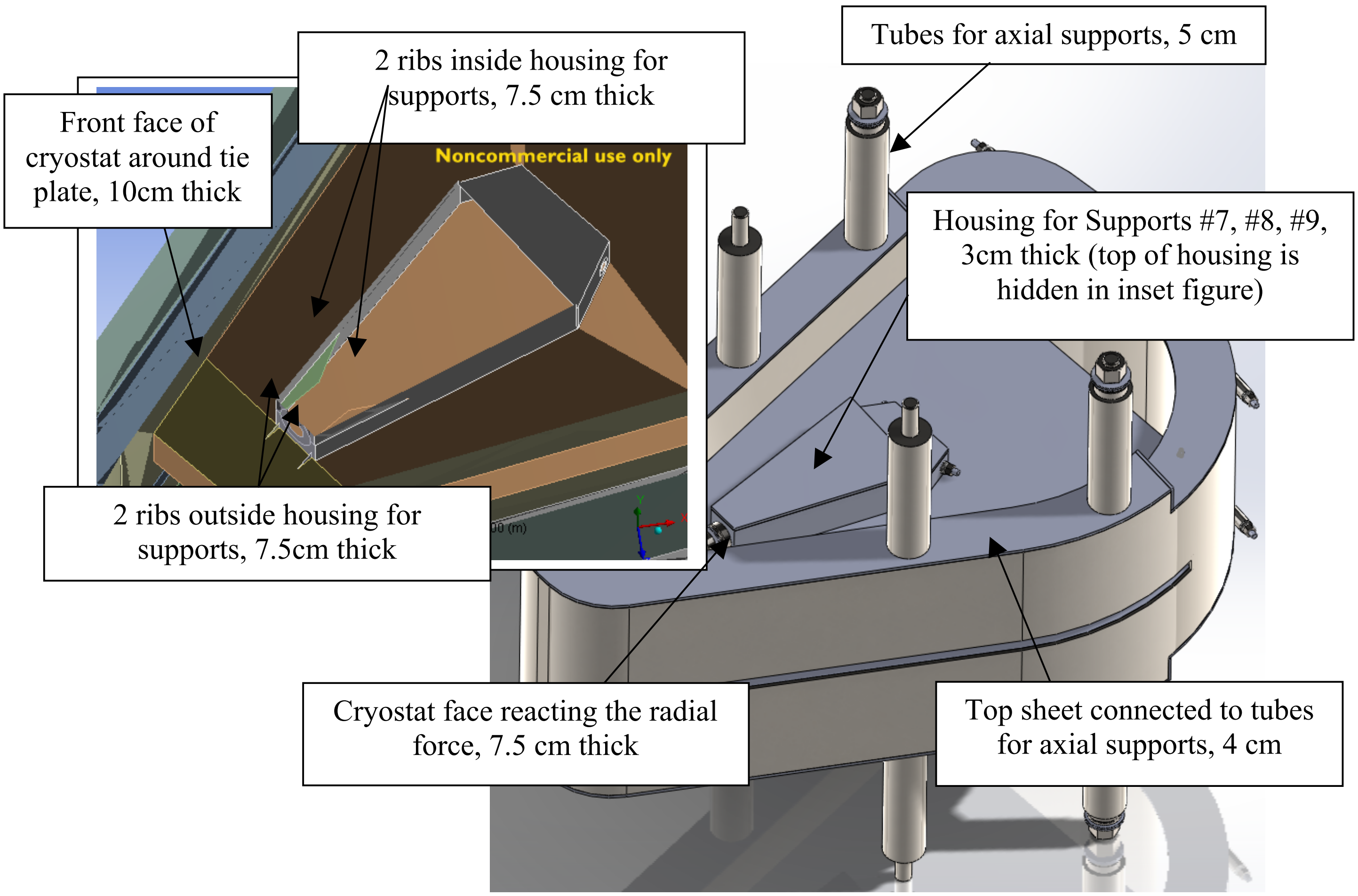

**Fig. 4-4. Top view of cryostat surrounding cold mass. Some of the warm-to-cold supports connections are shown. Inset picture is a sectioned view exposing the ribs used inside the housing to support the radial load on the cryostat.**

### 4.1.5. Entire Structure

The entire structure includes the iron yoke, coil case, coils, warm-to-cold supports and cryostat and is shown in Fig. 4-5. Fig. 4-5B hides the upper part of the iron yoke and the upper coil case is transparent, showing the copper colored coil. Figure 5C shows a sectioned view to capture how the warm-to-cold supports #7, #8, #9 fit into the cavity within the iron yoke. Because the iron is neglected in the structural analyses and the coil case is neglected in the EM analyses, it is important to see the clearances between the coil case, the cryostat and the iron. There is a minimum 10cm gap between the coil and the iron to allow space for the 3cm coil support and 7cm for the shield, insulation, cryostat and clearances between all parts. There is one location with less than 7cm gap between the iron and coil support that are highlighted in Fig. 4-5C. These are areas where some iron may need to be removed. This tradeoff between structural and magnetic design is most appropriate during the next design stage.

Fig. 4-5 labels the center of mass of the winding pack and coil case. It is located immediately behind Struts #7 & #8 and has a radial distance from the center of the SRC that is between Struts #1 and #2, or, using definitions described in Fig. 4-5: $r_1 < r_{cm} < r_2$. Note that Struts #1 and #3 are the same distance from the SRC centerline such that $r_1 = r_3$ and $r_2 = r_4$. The following is a list of the radii:

- $r_1$=325cm
- $r_{center\ of\ EM\ forces}$=392cm (described in further detail Section 4.1)
- $r_{center\ of\ mass}$=368cm
- $r_2$=443cm

The center of shrinkage of the cold mass is immediately in front of its connection to Strut #9. The mass of the various parts are summarized in Table 4-1, where a smeared winding pack density of 8000kg/m$^3$ was assumed. Note that the cold mass is 21 metric tons in each sector as this will be the additional loads the warm-to-cold supports must carry.

**Table 4-1. Mass of major components of SRC.**

| Part | Mass (metric ton) | | |
|---|---|---|---|
| | Half-Sector | Sector | 6 sector SRC |
| Iron Yoke | 415 | 830 | 4980 |
| Coil Case | 6.3 | 12.6 | 75.6 |
| Winding Pack | 4.32 | 8.64 | 51.84 |
| Cryostat | N/A | 13.5 | 81 |
| Total | 425.62 | 864.74 | 5188.44 |

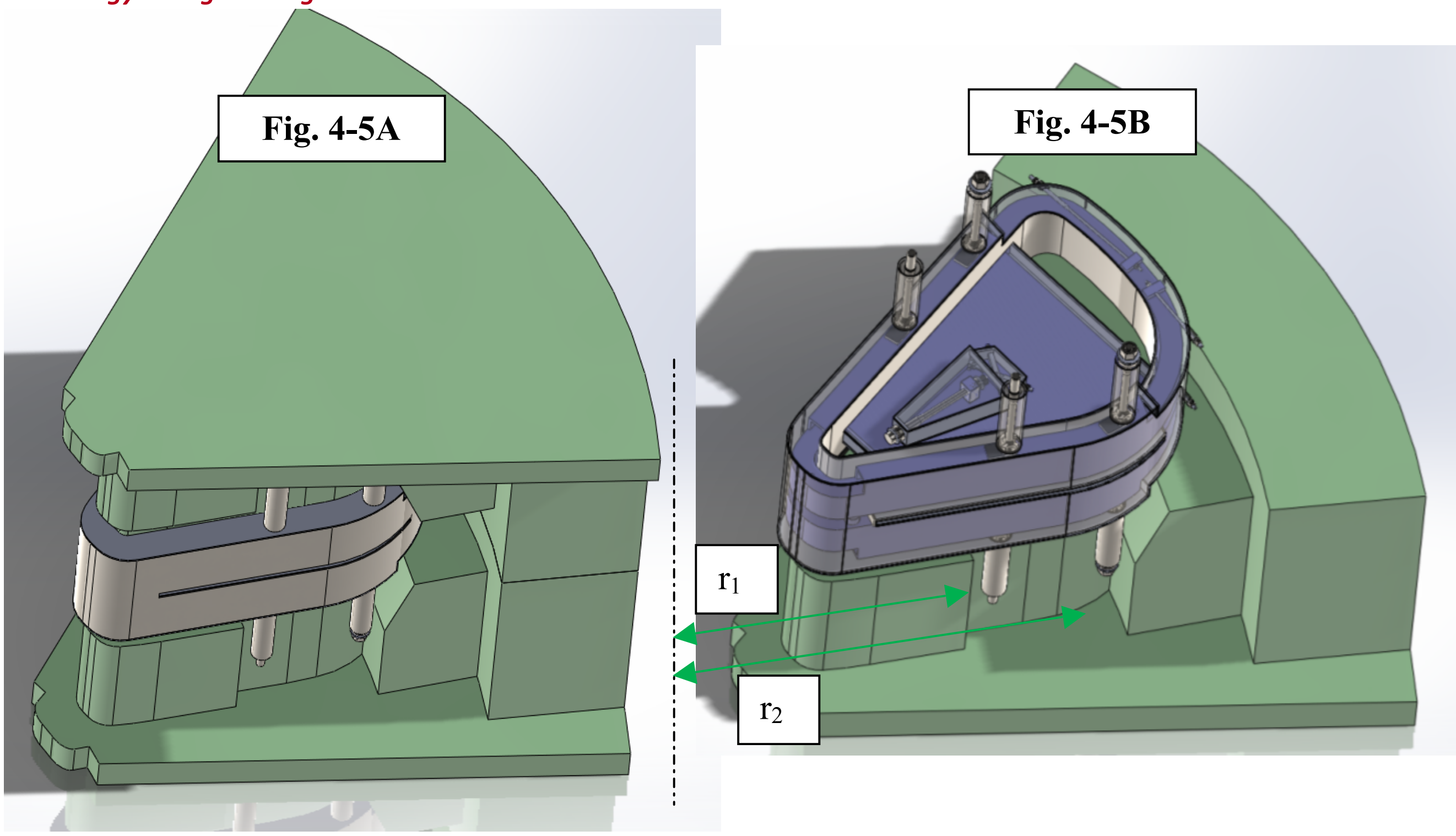

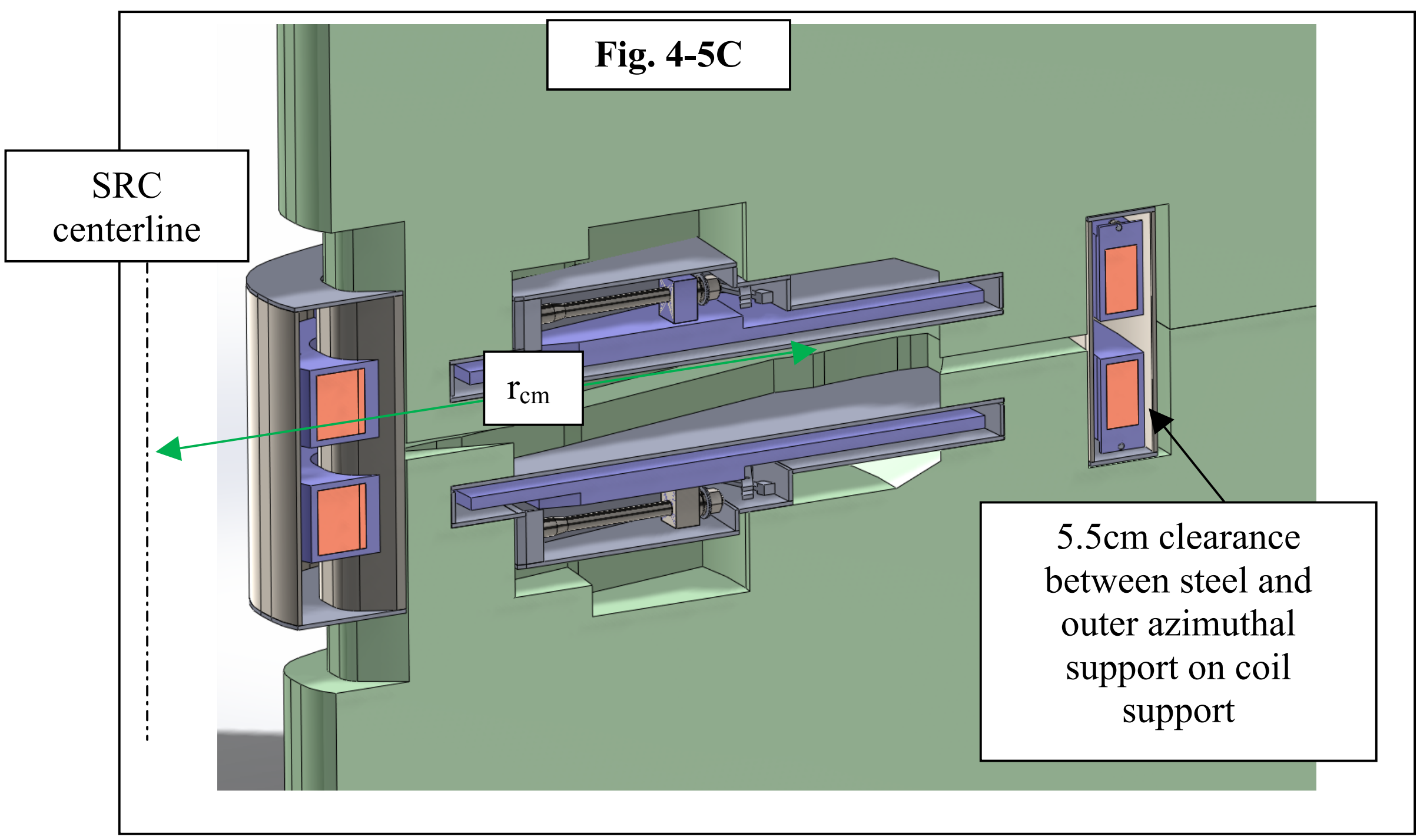

**Fig. 4-5. One sector of SRC shown with upper iron hidden and upper coil case transparent showing the copper coil. Inset is a section view showing clearances that are <7cm.**

## 4.2.Material Properties

The BH curve used for the iron yoke in the EM analyses is shown in Fig. 4-6. The coil in the electromagnetic analysis used default copper material properties and the surrounding air was modeled as vacuum.

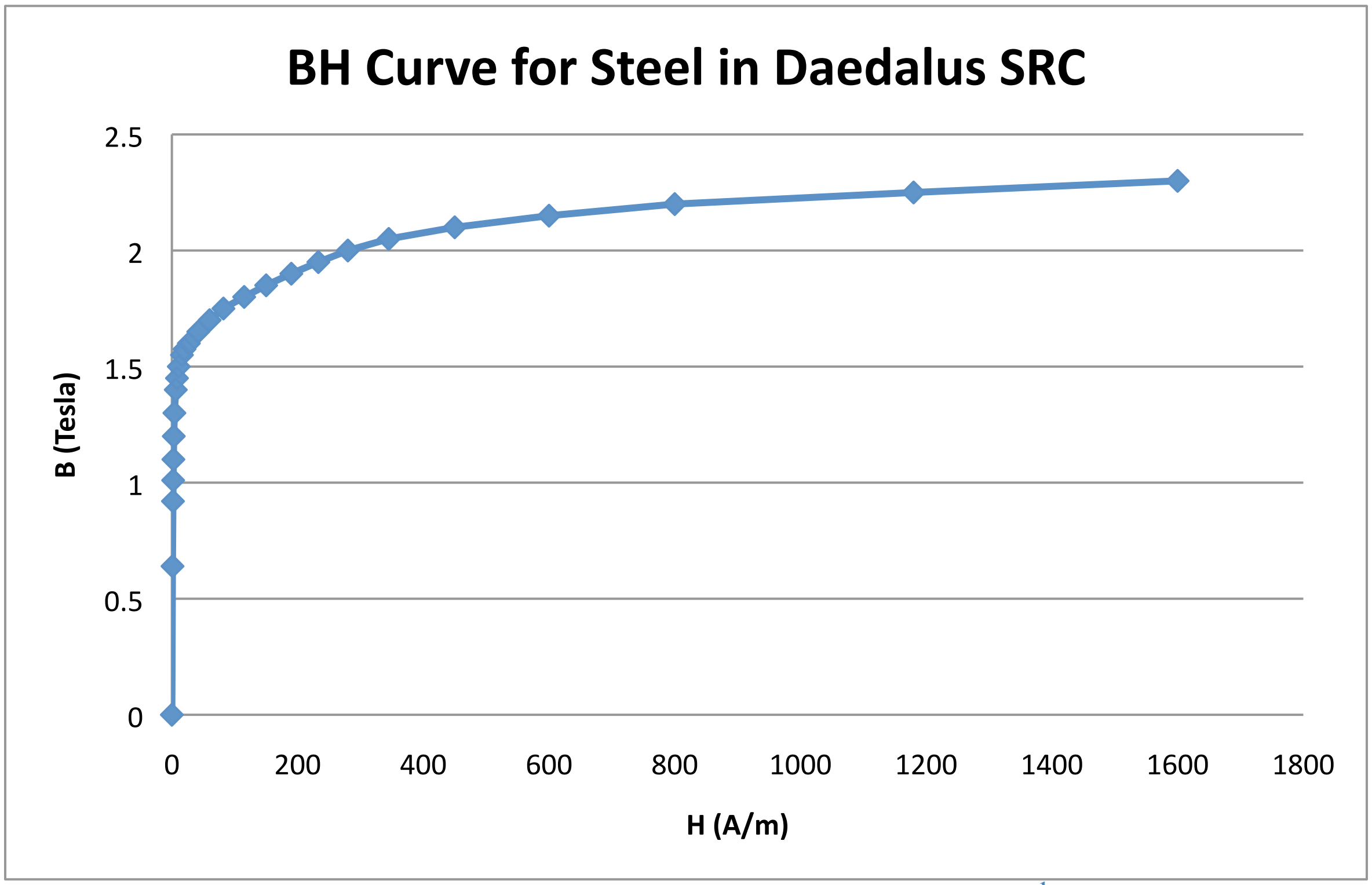

Fig. 4-6. BH curve used for SRC steel in EM analyses [1].

Two types of material behavior are used in the structural analyses: isotropic and orthotropic. The isotropic material is Stainless Steel 316 and its temperature dependent material properties are listed in Table 4-2.

The orthotropic material properties used here for the coil are taken from a project that employed a conductor in copper channel winding [2]. The winding pack material properties were determined using two methods. First, in the direction of the winding (labeled in Table 4-2 as the hoop direction), material properties are calculated from standard rule-of-mixtures formulas. These formulas determine smeared values based upon constituent values and weighs them based upon % of cross-sectional area each constituent represents. Second, the in-plane material properties were determined using a detailed 2D FEA model of many windings. The model loaded the winding in tension and compression to generate a load-deflection curve, from which the transverse elastic modulus was calculated. This procedure was repeated for the two transverse directions (labeled radial and axial in Table 4-2). The lower values of the transverse stiffness are due to the fiberglass turn wrap insulation in the winding pack. These material properties neglect the cooling tubes used to internally cool the winding pack [3]. The cooling tubes are stainless steel and comprise about 20% of the volume. This will increase the hoop stiffness and decrease the CTE in all directions (transverse stiffness is driven by the soft turn insulation will be unaffected).

The failure criteria used to compare against the resulting stress distributions in the support structure and within the winding pack considers the yield, tensile strength and safety factors specified by FIRE design document [4]. The stainless steel coil support has a primary membrane stress intensity failure criteria, Sm, that is the lower of $\sigma_{UTS}/3$ and $(2/3)\sigma_{yield}$., whereas the supported copper coil is the lower of $\sigma_{UTS}/2$ and $(2/3)\sigma_{yield}$. Table 4-3 lists the strengths and allowable stresses for stainless steel and Nitronic materials used for the coil supports and the struts.

**Table 4-2. Material properties used in mechanical analyses.**

| Material | Direction | Property | | | | |
|---|---|---|---|---|---|---|
| | | Elastic Modulus (GPa) | | | CTE (ppm/C) 22C to 4K | Nu |
| | | 4K | 22 C | 80 C | | |
| Stainless Steel 316[5] | N/A | 205 | 194 | 193 | 10.31 | 0.25 |
| Coil | Hoop | 103 | 93.8 | 90.2 | 11.5 | 0.33 |
| | Radial | 38.6 | 19.1 | 18.9 | 13.1 | |
| | Axial | 40 | 19.7 | 19.5 | 12.7 | |

**Table 4-3. Strength and allowable for various materials.**

| Material | Temperature (K) | $\sigma_{UTS}$ (MPa) | $\sigma_{yield}$ (MPa) | Sm (MPa) | 1.5*Sm (MPa) |
|---|---|---|---|---|---|
| Stainless Steel 316[6] | 296 | 613 | 276 | 184 | 275 |
| | 4 | 1379 | 992 | 460 | 689 |
| Stainless Steel 316 welded[6] | 296 | 482 | 324 | 161 | 241.5 |
| | 4 | 1110 | 724 | 366 | 550 |
| Nitronic 50[7] | 296 | 1113 | | 371 | 557 |
| | 77 | 1558 | 883 | 519 | 779 |
| OFHC, 20% cold work [5] | 4 | 450 | 330 | 220 | 330 |

## 4.3.Electromagnetic Analyses

The EM analyses presented here employ the Maxwell software to prepare the Lorentz body forces on the coils for inputs to the structural analyses. There are two EM analyses here, one to simulate normal operation with both coils fully charged, and a second to simulate a fault condition where one coil fast dumps and the other shorts. The fault condition assumes the normal operating current density in the shorted coil and zero current in the other coil. In reality the flux is conserved, so there will be additional current induced in the shorted coil – this is addressed in the structural analysis by adding a 1.3x multiplier to the imported Lorentz forces. The geometry used for the EM analyses is shown in Fig. 4-7 and it includes two coils, two pieces of steel and vacuum for one of the six sectors of the SRC. A matching boundary condition is applied to the surfaces forming the 60° wedge, shown in Fig. 4-7. This is typically used for periodic structures where the H-field at every point on one surface matches the H-field of every corresponding point on the opposite surface. The coils are assumed to be stranded conductors with an excitation of 1,720,000A·turns with current density of 34.68A/mm$^2$. [8]

The resulting flux density on the coils is shown in Fig. 4-8 as the peak flux is 4.45T (compared to 4.37T in the Vector Fields analysis [8]) on the inside surface of the coils when both are fully charged. If the upper coil has its current set to zero, the peak field on the lower [charged] coil is 3.96T and the flux on the coil surface for this case is shown in Fig. 4-9
.

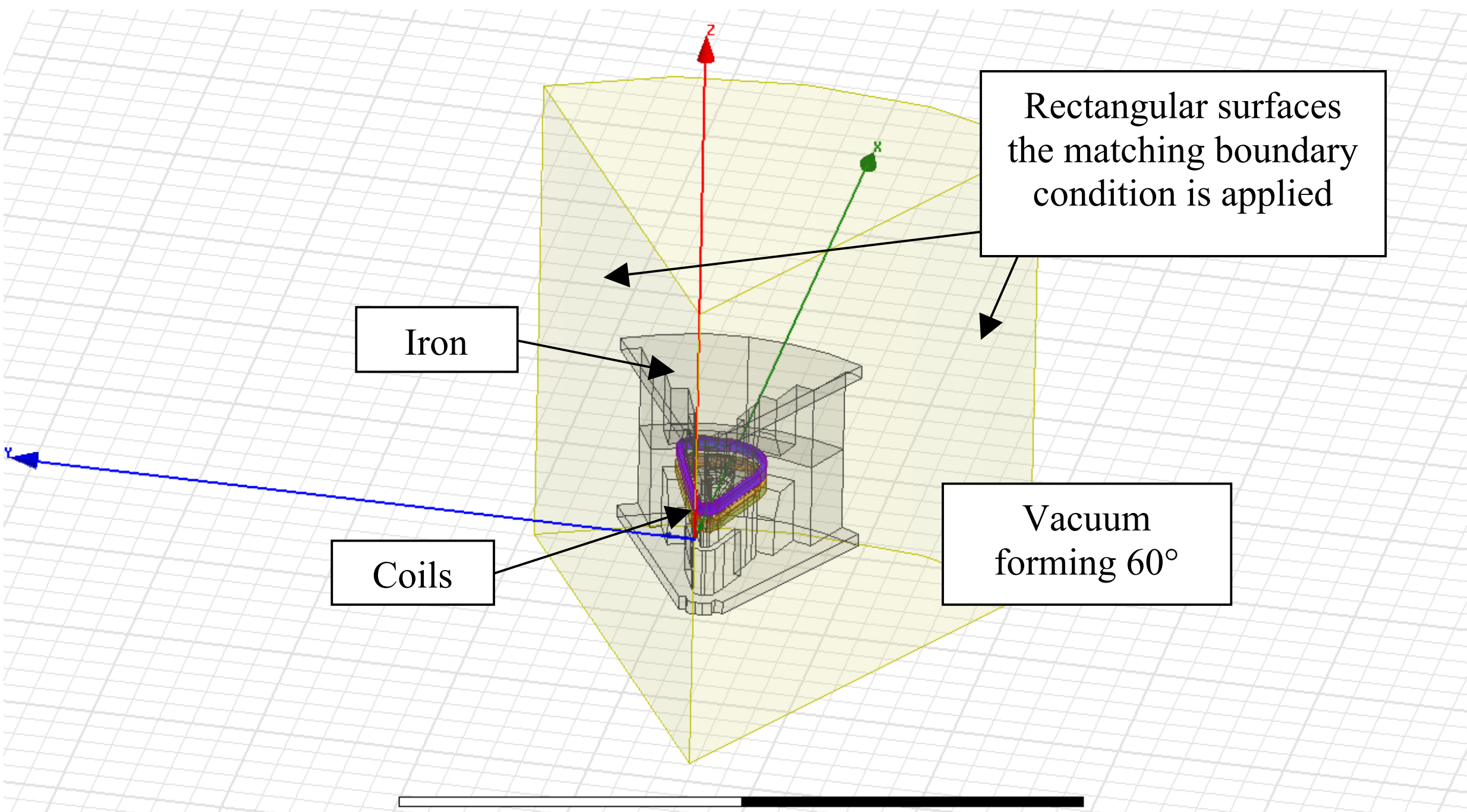

Fig. 4-7. Geometry of one sector used in EM analysis. Upper coil is highlighted in purple.

The EM analyses also calculate the resulting forces and torques on the coils and iron. It is the forces distributed through the coils that are imported to the structural analyses. The net forces on the coils are summarized in Table 4-4 using the coordinate system shown in Fig. 4-7. When both coils are charged, the center of the resulting forces is on the median plane (Z=0) at X=3.62m, Y=0.13m.

**Table 4-4. Net Lorentz body forces on coils imported to mechanical analyses.**

| Case | Coil | Lorentz Forces (MN) | | |
|---|---|---|---|---|
| | | Fx | Fy | Fz |
| Both Coils Charged | Upper | 2.09 | 0.092 | -7.1 |
| | Lower | 2.01 | 0.027 | 7.07 |
| Bottom Coil Charged | Lower | 1.61 | 0.11 | -4.63 |

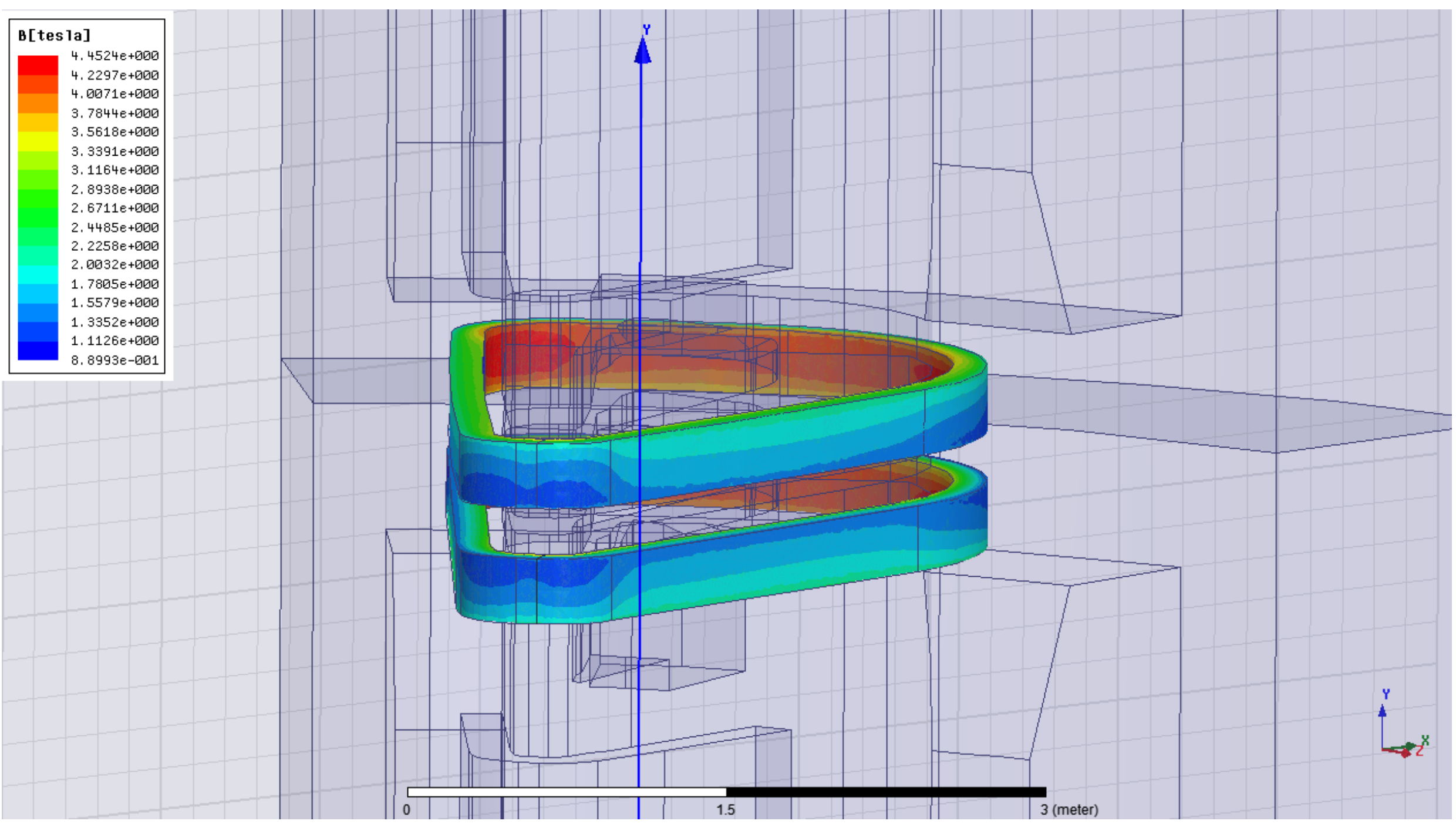

**Fig. 4-8. Flux density on surface of coils with both coils charged.**

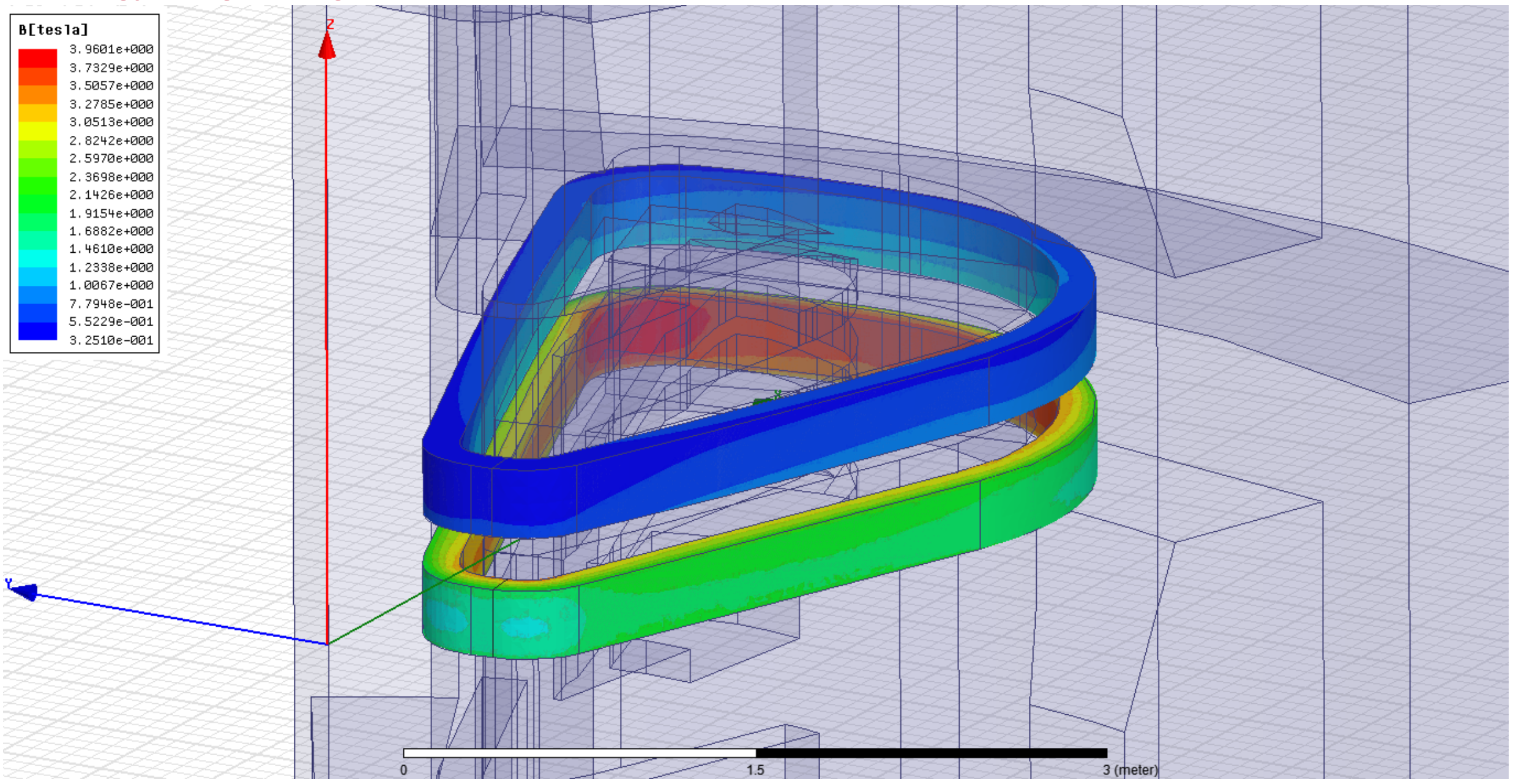

**Fig. 4-9. Flux density on surface of coils with upper coil current zero.**

## 4.4. Structural Analyses

### 4.4.1. Cold Mass

The structural analyses use ANSYS to calculate the system response of the cold mass resulting from pre-loading of the warm-to-cold support, gravity, the Lorentz body forces on the coils and from thermal loads. Fig. 4-10 shows the geometry input to ANSYS, which includes the coils (not seen in Figure), coil cases, and warm-to-cold support struts. The warm-to-cold support struts are modeled as linear springs with one end attached to the cryostat the other to the cold mass. Three analyses are performed: (1) cool-down to 4K, (2) cool-down + normal operation, and (3) cool-down + fault condition. Thermal loads apply a uniform 4K temperature to all volumes with a zero-stress temperature of 22C. All three analyses include gravity and pre-tensioning of Axial Struts #1-#4.

As previously mentioned, the support struts that suspend the cold mass from the cryostat are modeled as tension-only linear springs. One end of each spring attaches to the cold mass and is deformable, whereas the other end is attached to the RT cryostat and is assumed rigid. Thermal shrinkage of the struts is not included in the spring connection input data for the analyses. The eight (four top and four bottom) axial supports are pre-loaded with a 40,000N load to remove any initial compression in the four bottom supports due to the weight of the cold mass. There is no pre-load on any of the remaining struts. The spring stiffness, k=EA/L, assumes the struts have an elastic modulus E=200GPa (stainless steel and Nitronic 50 have same modulus). The length, diameter, and stiffness of each strut were estimated using initial analyses of the SRC and are listed in Table 4-5. The diameter of the rods is assumed to be constant along its length; however, there will be a smaller diameter at the cold end and a larger diameter at the warm end (cryostat) due to the temperature dependence of the material strength. It is important to note that the spring stiffness was increased and decreased by an order of magnitude with very little change in the resulting spring forces. Therefore, the ANSYS force data can be used to size struts that have reasonable changes from the initial estimation of Table 4-5 without requiring the model to be re-run with updated stiffness.

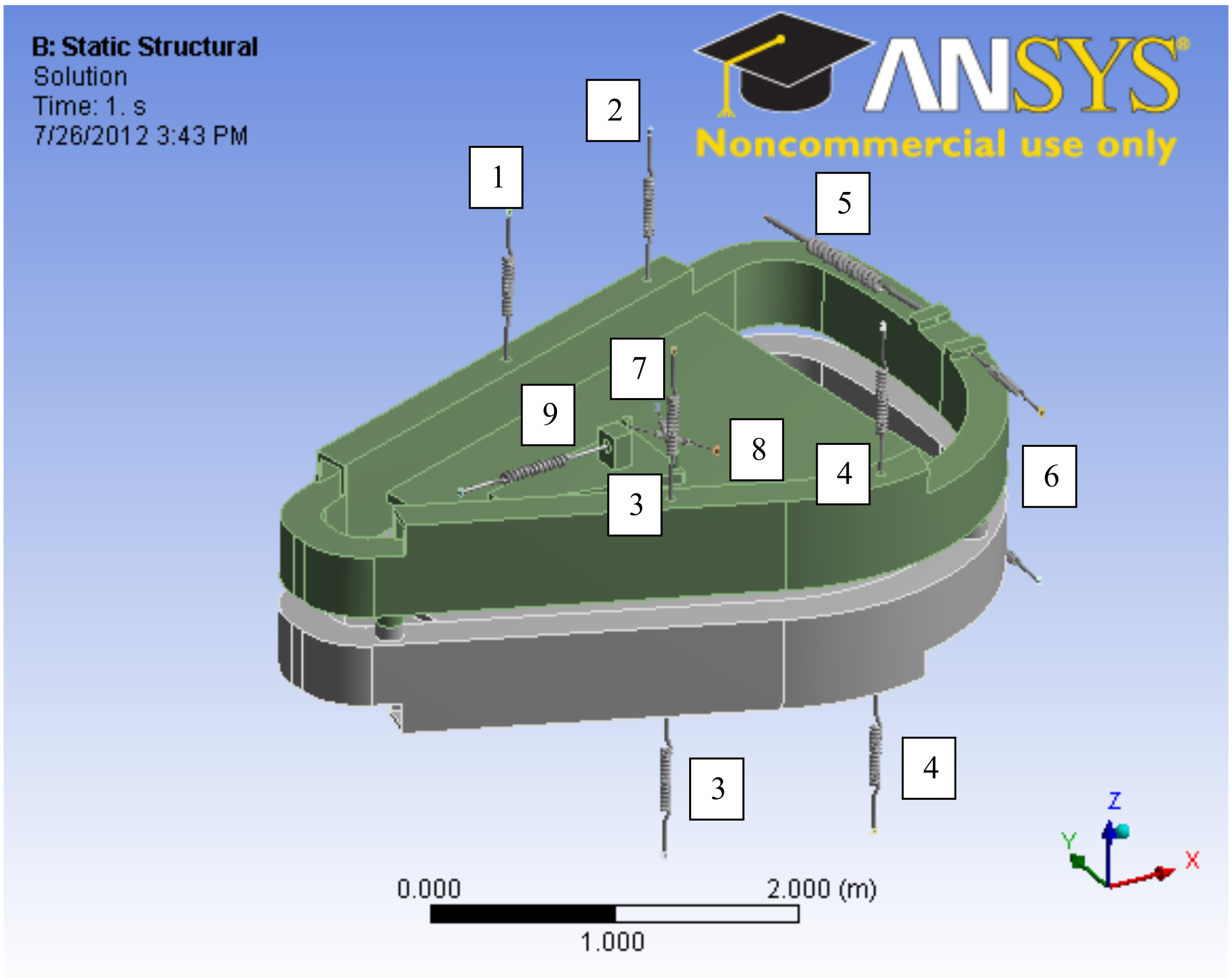

Fig. 4-10. Geometry used for mechanical analyses of cold mass..

Table 4-5. Estimated spring stiffness for tension only struts used in simulations.

| Strut # | L (35cm) | Avg ∅ (cm) | Kavg (N/m) |
|---|---|---|---|
| 1 | 90 | 5.01 | 4.39E+08 |
| 2 | 90 | 9.13 | 1.45E+09 |
| 3 | 90 | 4.90 | 4.18E+08 |
| 4 | 90 | 8.29 | 1.20E+09 |
| 5 | 134 | 1.71 | 3.43E+07 |
| 6 | 77 | 1.00* | 2.04E+07 |
| 7 | 54 | 1.00* | 2.91E+07 |
| 8 | 54 | 2.09 | 1.28E+08 |
| 9 | 95 | 7.25 | 8.70E+08 |

*:Minimum diameter of 1cm

The winding pack is initially in intimate contact with the coil case and their surfaces are simulated with frictional sliding contact along all four sides. Frictional contact allows for compressive load transfer (N), shear load transfer up to $F_{MAX}=\mu N$, and separation (no tensile loads are transmitted). A coefficient of friction of $\mu=0.2$ was assumed for all analyses. The magnet design from a thermal perspective does not require bonding of the winding pack to the coil case because cooling will be done via cooling tubes routed within the winding pack.

The displacements and stresses are used to evaluate the cold mass. The displacements are of interest for two reasons. First, the coil position creates the desired magnetic flux distribution and displacements should be minimized. Second, there will be tight clearances between the iron, cryostat, shield and coil support and we want to prevent the displacements during cool-down and coil charging from closing any initial gaps between the parts. This especially true in the region around the beam chamber.

#### 4.4.1.1. COLD MASS DISPLACEMENTS DURING NORMAL OPERATION

The resulting axial displacement field for the cold mass is shown in Figure 11 for (A) coold down to 4K + gravity + pretention in axial warm-to-cold struts, (B) EM loads from normal operation + (A), and (C) EM loads from fault condition + (A). the axial displacements are the z-direction of the coordinate system highlighted in Fig. 4-11. Notice the gap between the two coil cases abutting the beam chamber opens by 1.53+1.61=3.14mm upon cool down. During normal operation this gap closes, but remains 0.88+0.96=1.84mm wider than at room temperature. During the fault condition, both coils are pulled downward. The outer edge of the upper coil moves 16mm (dark blue) and the gap around the beam chamber closes by 10.63-6.33=4.3mm. While the nominal spacing between the coil cases is 18cm, there will be tight clearances among the radiation shield and cryostat around the beam chamber that could be impacted by these large displacements.

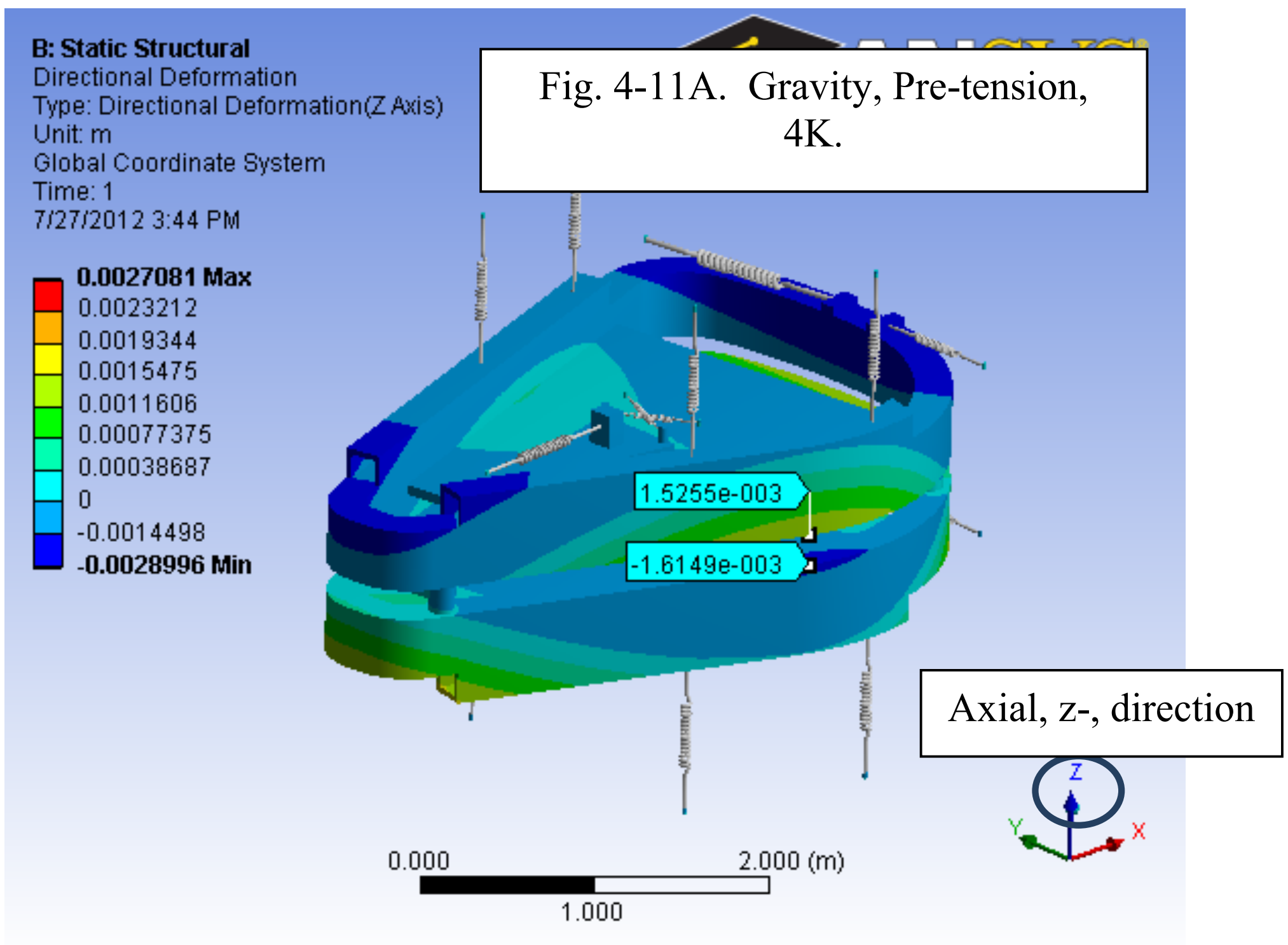

Fig. 4-11A. Gravity, Pre-tension, 4K.

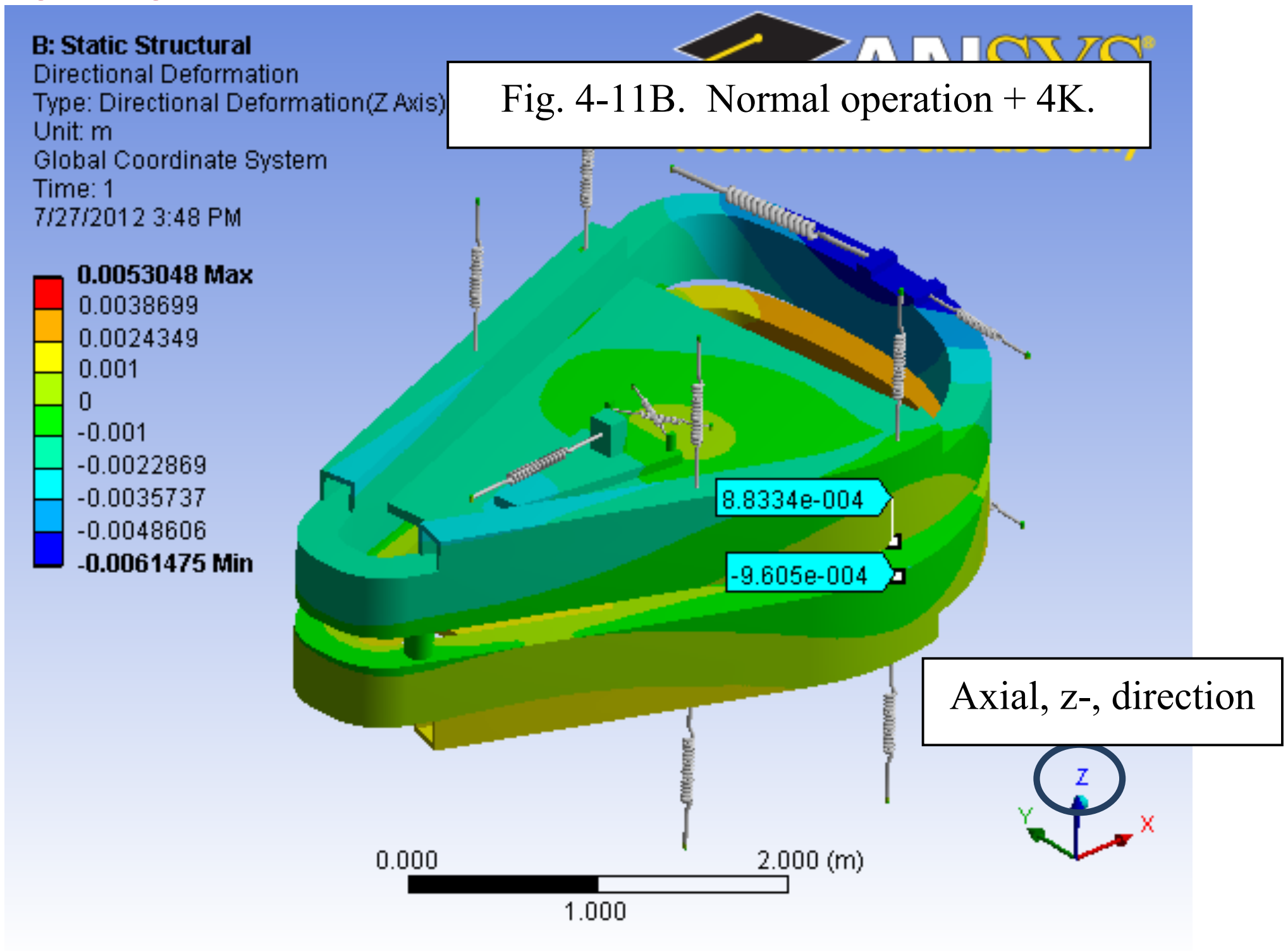

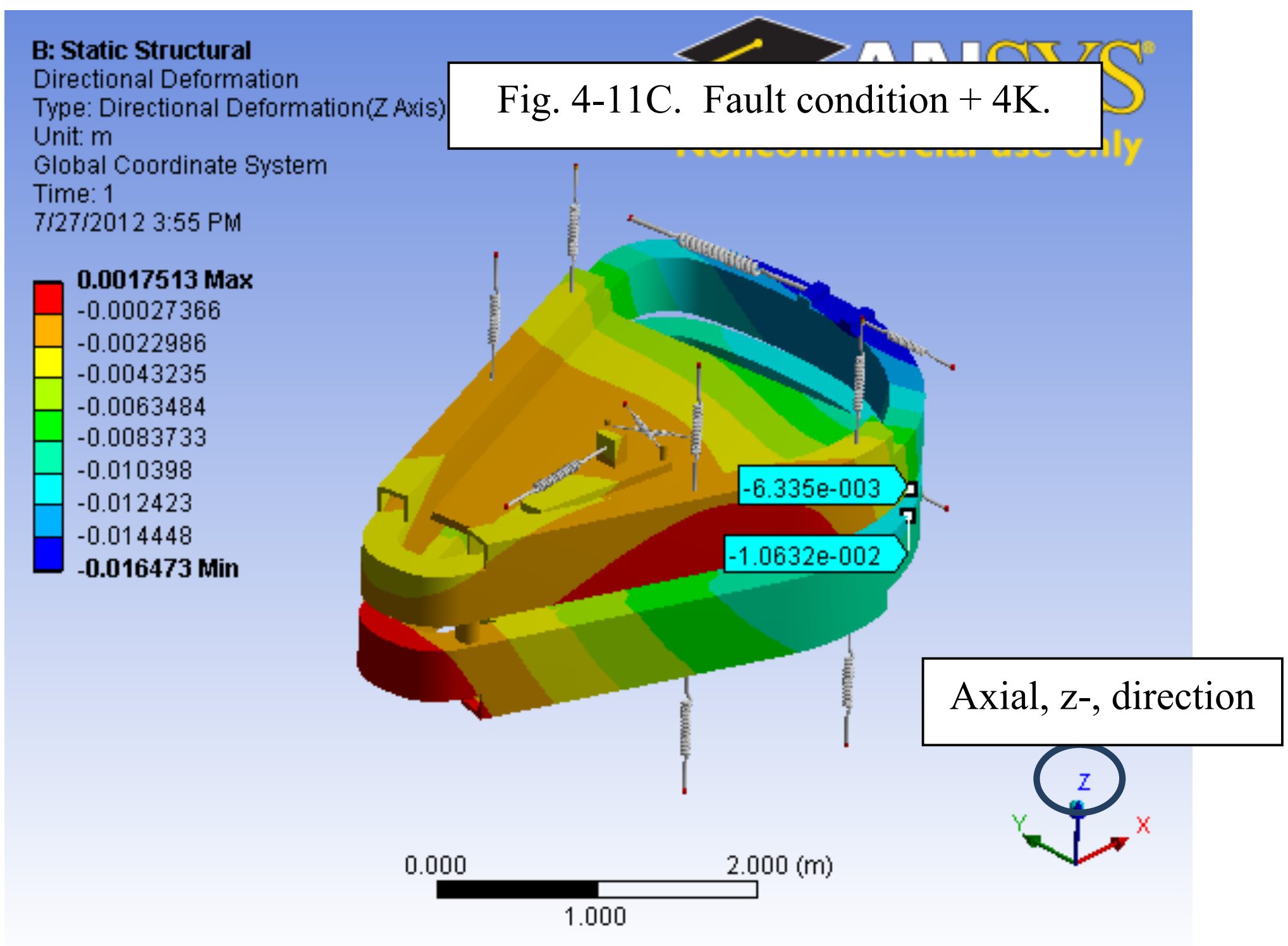

**Fig. 4-11. Axial displacement fields on coil cases for (A) cool-down, (B) 4K + normal operation and (C) 4K + fault condition.**

Fig. 4-12 shows a magnified and sectioned picture to describe the relative deformations of the coil within the coil case. The color contours are the azimuthal displacements (y-direction on Fig. 4-13) and they indicate 3.86-3.13=0.73mm and 1.03-0.86=0.67mm separation between the inside of the coil and the coil case at the locations highlighted. There is approximately 0.25mm axial separation between the coil and coil case along the sides opposite the median plane. These separations approximate the amount of sliding between 2 sides of the coil and coil case.
Fig. 4-13 is similar to Fig. 4-12, except the section plane is rotated 90°. Fig. 4-13 indicates the separation between the coil and coil case is minimal radially – 1.08-1.07=0.01mm. Therefore, there is no significant sliding between the coil and coil case at these locations. Fig. 4-14 shows the relative displacements at the locations shown in Fig. 4-12. Fig. 4-14 indicates the coil is in contact with the coil case on only one side along Segment #7 (refer to Fig. 4-1 for segment numbering).

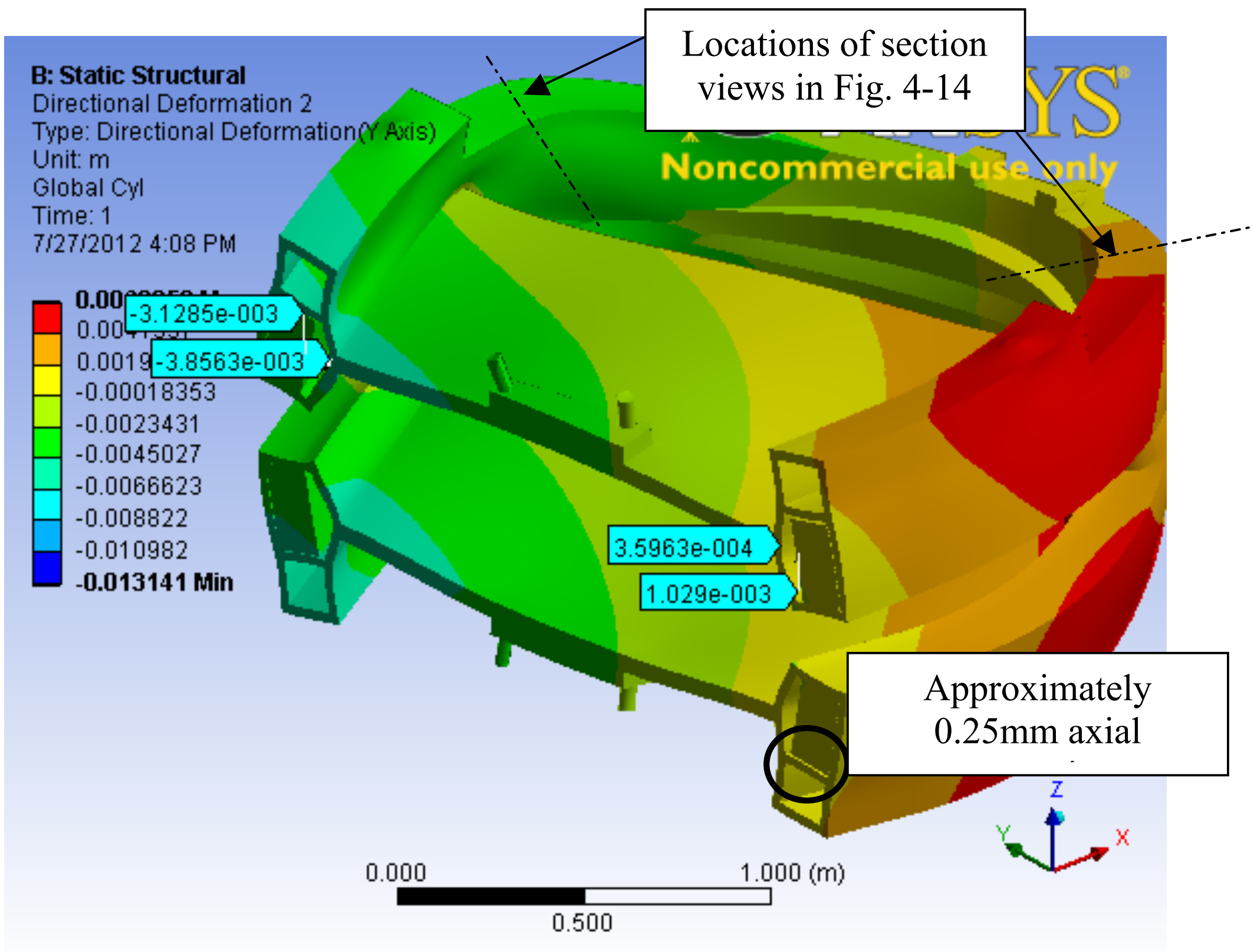

Fig. 4-12. Cold mass sectioned to show displacement of coil within coil case for 4K + normal operation. Displaced structure is at 53x magnification of actual displacements. Color contours depict azimuthal displacements.

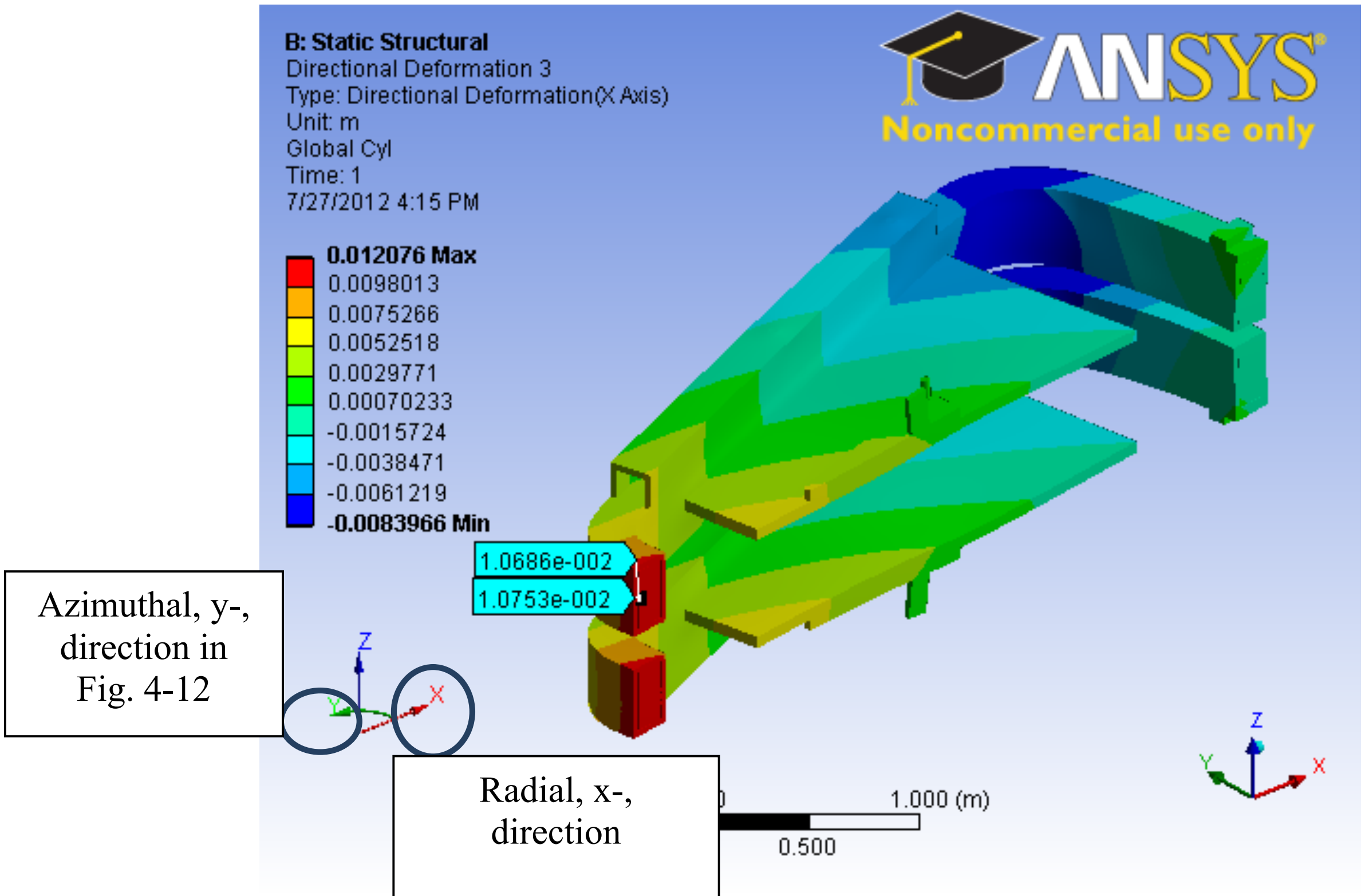

**Fig. 4-13. Cold mass sectioned to show displacement of coil within coil case for 4K + normal operation. Displaced structure is at 53x magnification of actual displacements. Color contour depicts radial displacements.**

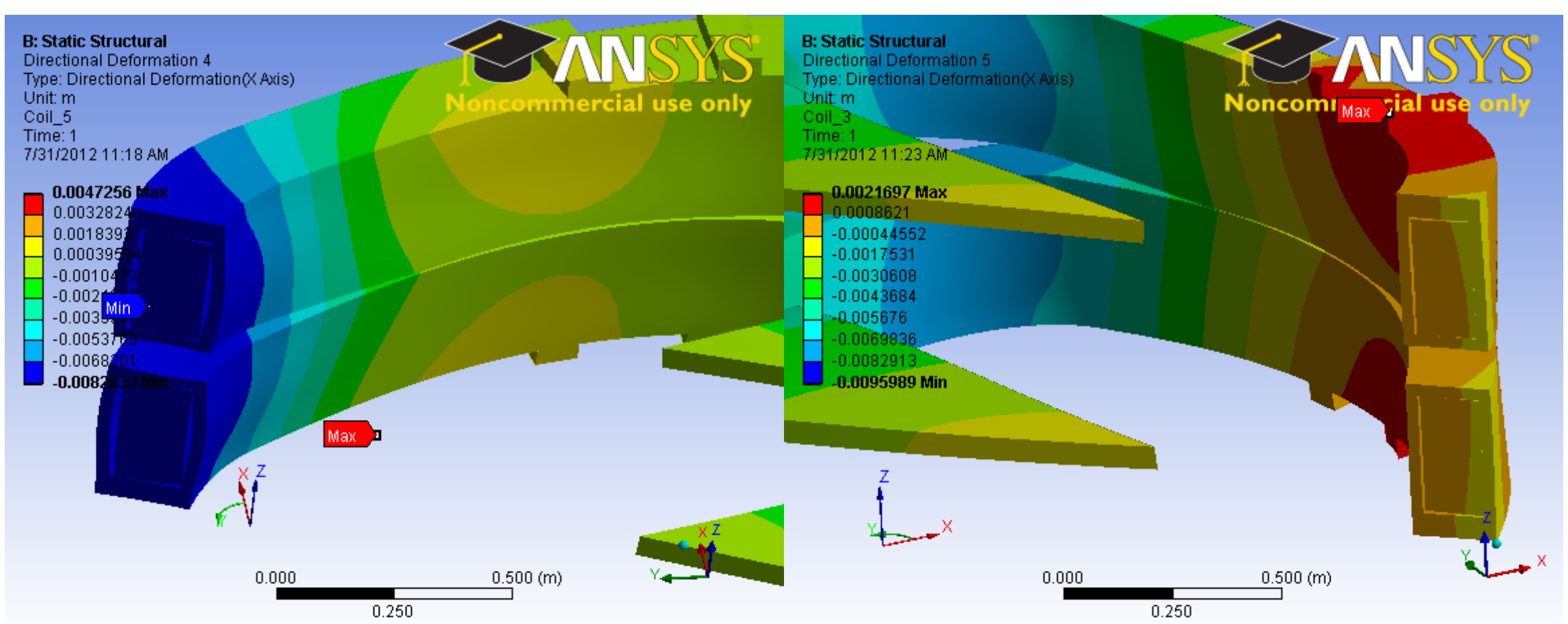

**Fig. 4-14. Cross sections showing the motion of the coil within the coil case at locations shown in Fig. 4-12. Section of the left indicates coil is in contact with coil case on only one side.**

### 4.4.1.2. WINDING PACK RESPONSE

The coil is allowed to slide within the coil case and the combination of sliding with large shear can cause localized heating and/or delamination of fiberglass ground wrap leading to coil quench. Therefore, it is important to know the amount of shear between the bodies while they slide against each other. Fig. 4-15is a contour plot of the xz-shear stresses using the coordinate system attached to Segment #4 (Fig. 4-15A) and Segment #8 (Fig. 4-15B). These are segments that were shown to slide the most. Fig. 4-15A & B show values of the surface shears on the sides that sliding occurs stress and they indicate all surface shears are <5MPa and that they are typically <2.5MPa.

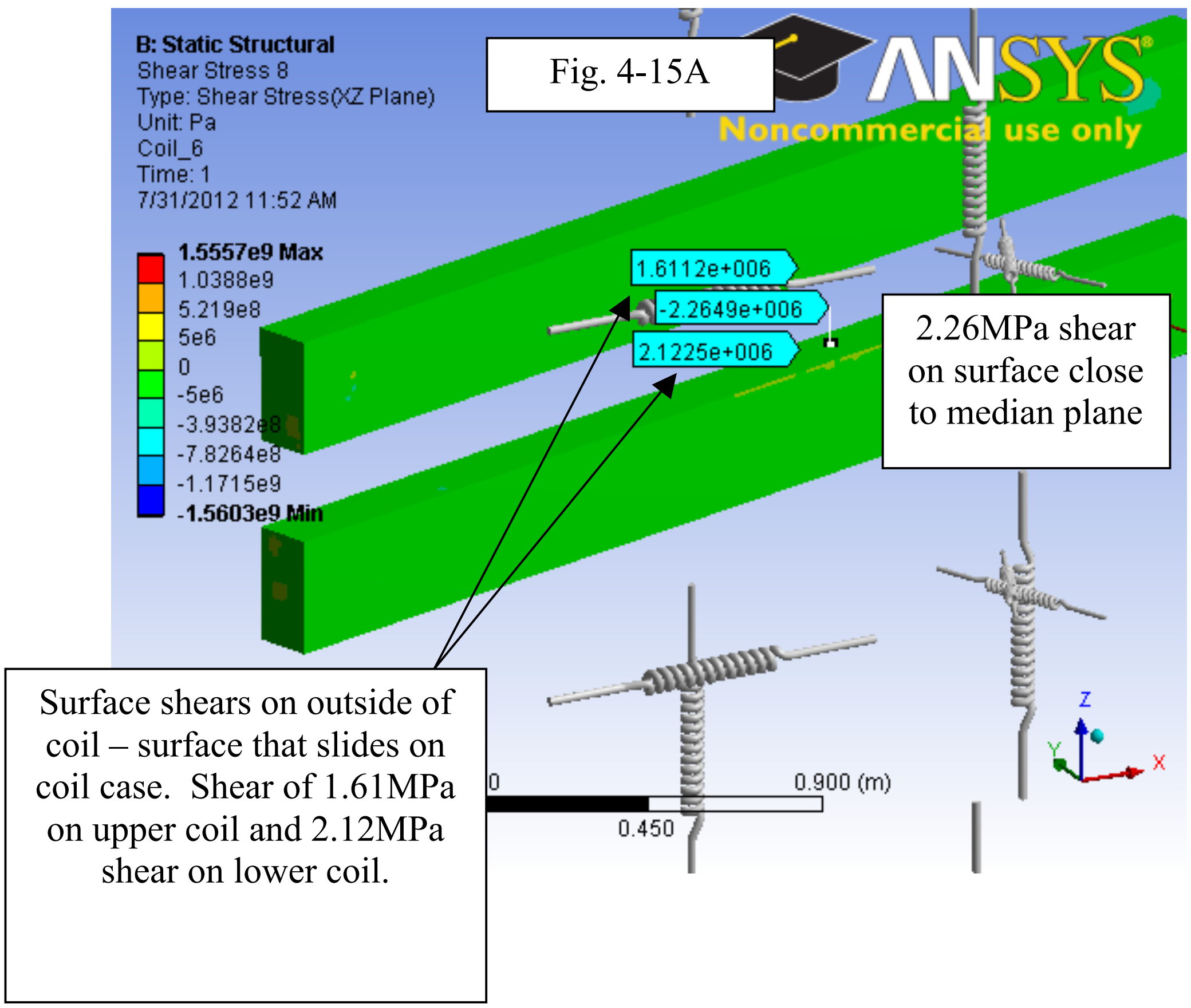

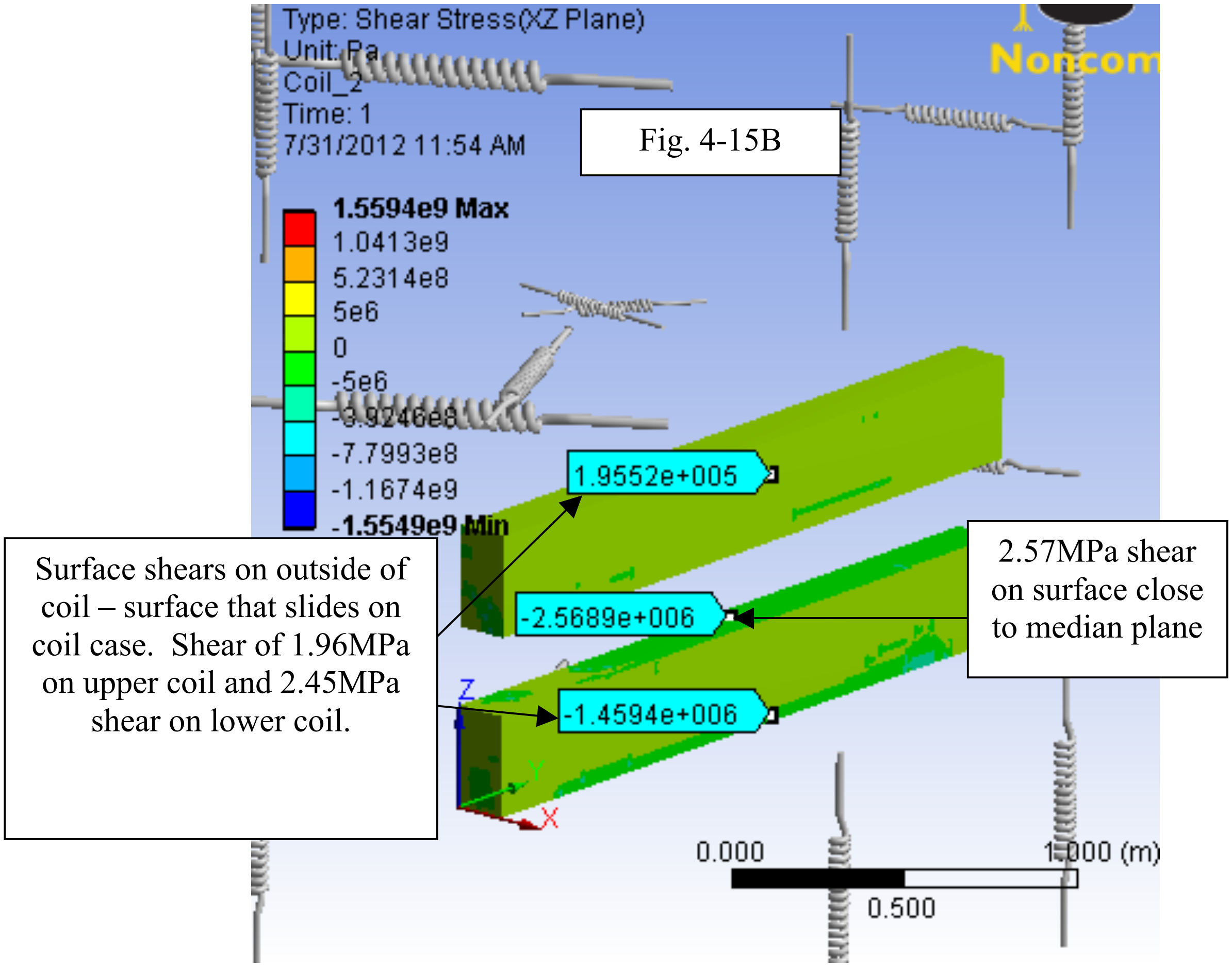

Fig. 4-15. Surface shear on (A) Segment #8, and (B) Segment #4 during normal operation.

The winding pack design was introduced in Section 3. It consists of alternating several double pancake windings with stainless steel cooling plates. The layers are assembled and fastened together using vacuum pressure impregnation (VPI) to prevent sliding between the layers during cool-down. It is critical the integrity of the fiberglass is analyzed within the winding pack using a detailed substructure analysis. As seen in Table 4-2, stainless steel and the coil in the hoop direction have a small difference in the thermal expansion, which is especially important in the winding direction where the contact length is large. The substructure analysis can optimize the detailed structure of the winding pack in order to maintain the integrity of the fiberglass.

At this level of design, the integrity of the fiberglass ground wrap around the outside surface of the winding pack can be analyzed. The failure criteria used to evaluate the fiberglass are the stresses leading to delamination of the ground insulation. The failure criteria used compares: (1) the shear force to a combination of the bond strength and compressive forces, and (2) any tensile forces normal to the surface. The primary response on the surface of the coils using the smeared properties is converted into stresses in a thin layer of ground wrap using the following process: (1) stresses normal to the surface of the smeared coil are equal to normal stresses in ground wrap because the normal stresses are continuous across boundaries, and (2) shear strains on the surface of the smeared coil are scaled using the shear modulus, G, of fiberglass to estimate the shear stresses in the ground wrap because shear strains are continuous across boundaries. Both the normal and shear failure criteria are applied to the estimated ground wrap stresses. The result is greater than 99% of the ground wrap on the outer surface of the coil pass the criteria – a very encouraging result for a preliminary design.

The von Mises strains on the winding pack are shown in Fig. 4-16. Fig. 4-16A shows the contour of the von Mises strains for 4K + normal operation and the peak strain is 0.23%. This von Mises strain is mostly due to tensile strains along the direction of the winding (sometimes referred to as hoop strains). This is equivalent to a 326MPa stress in the winding and a *very* conservative assumption can be made that the copper bears all loads in the winding pack. The 326MPa can then be compared to the 1.5*Sm=330MPa allowable for copper in bending, so the stresses in the winding are acceptable during normal operation.

Fig. 4-16B shows the von Mises strains on the winding pack for 4K + fault condition and the strains slightly decrease. Therefore, the winding pack can withstand the fault condition.

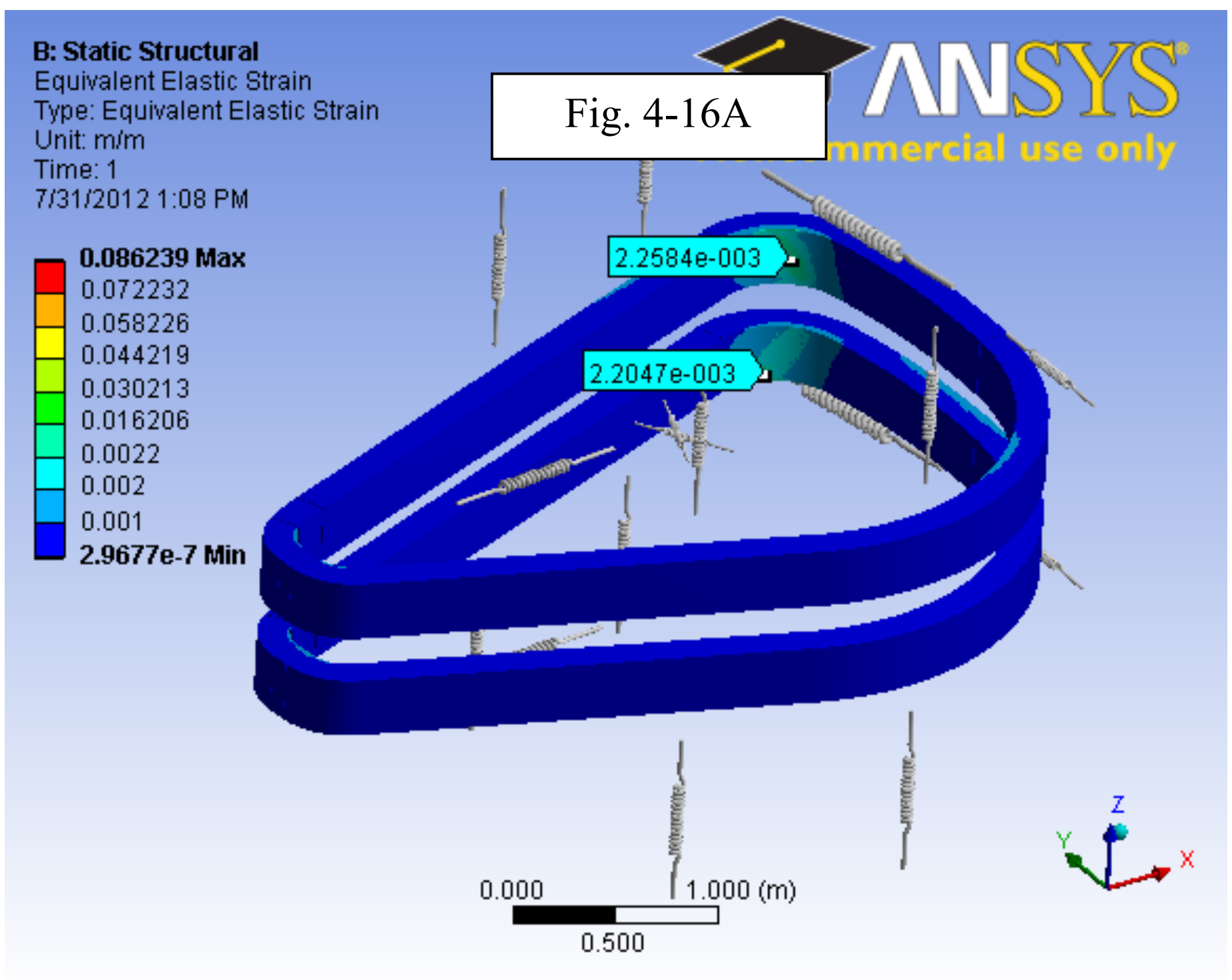

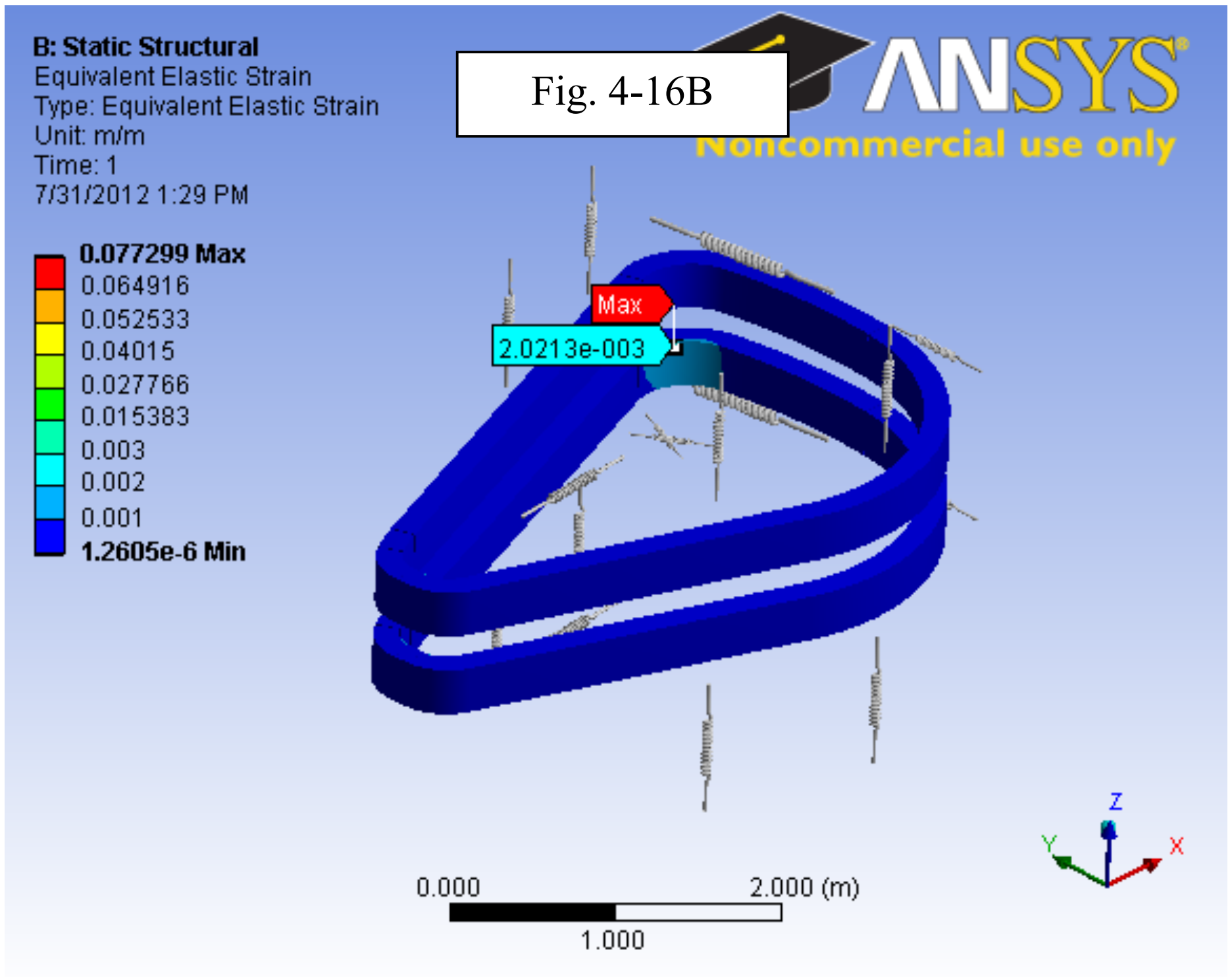

Fig. 4-16. Von Mises strains on winding pack for (A) 4K + normal operation and (B) 4K + fault condition.

### 4.4.1.3. COIL CASE STRESSES

Fig. 4-17 & Fig. 4-18 show the von Mises stress contours on the coil case for 4K + normal operation and 4K + fault condition, respectively. The red color indicates stresses greater than the allowable for stainless steel in bending (1.5*Sm=689MPa), the orange indicates stresses greater than the allowable for welded stainless steel in bending (1.5*Sm=550MPa), the yellow color indicates stresses greater than allowable for stainless steel membrane stresses (Sm=460MPa, and light green indicated stresses greater than allowable for welded stainless steel membrane stresses (Sm=366MPa). The stresses are the highest in the area immediately surrounding the locations the warm-to-cold struts are connected to the coil supports. These are singularities and typically are not reliable data.

In normal operation, Fig. 4-17A indicates that the highest stress away from the connections to the supports is 530MPa. Fig. 4-17B shows that this stress does not go through the 3cm thickness of the coil case; therefore, it is not classified as a membrane stress and is a bending, or secondary stress. The stress of peak 523MPa in the coil case is at a safe level during normal operation for both welded and unwelded stainless steel; however, it may be prudent to keep welded joints clear of the ID of Segment #5.

During the fault condition, Fig. 4-18A indicates that the highest stress away from the connections to the supports is 522MPa. Fig. 4-18B indicates that this stress does not go through the thickness of the coil case, thereby classifying it as a secondary stress. This stress is slightly less than during normal operation, so the system will continues to safely bear the loads during fault.

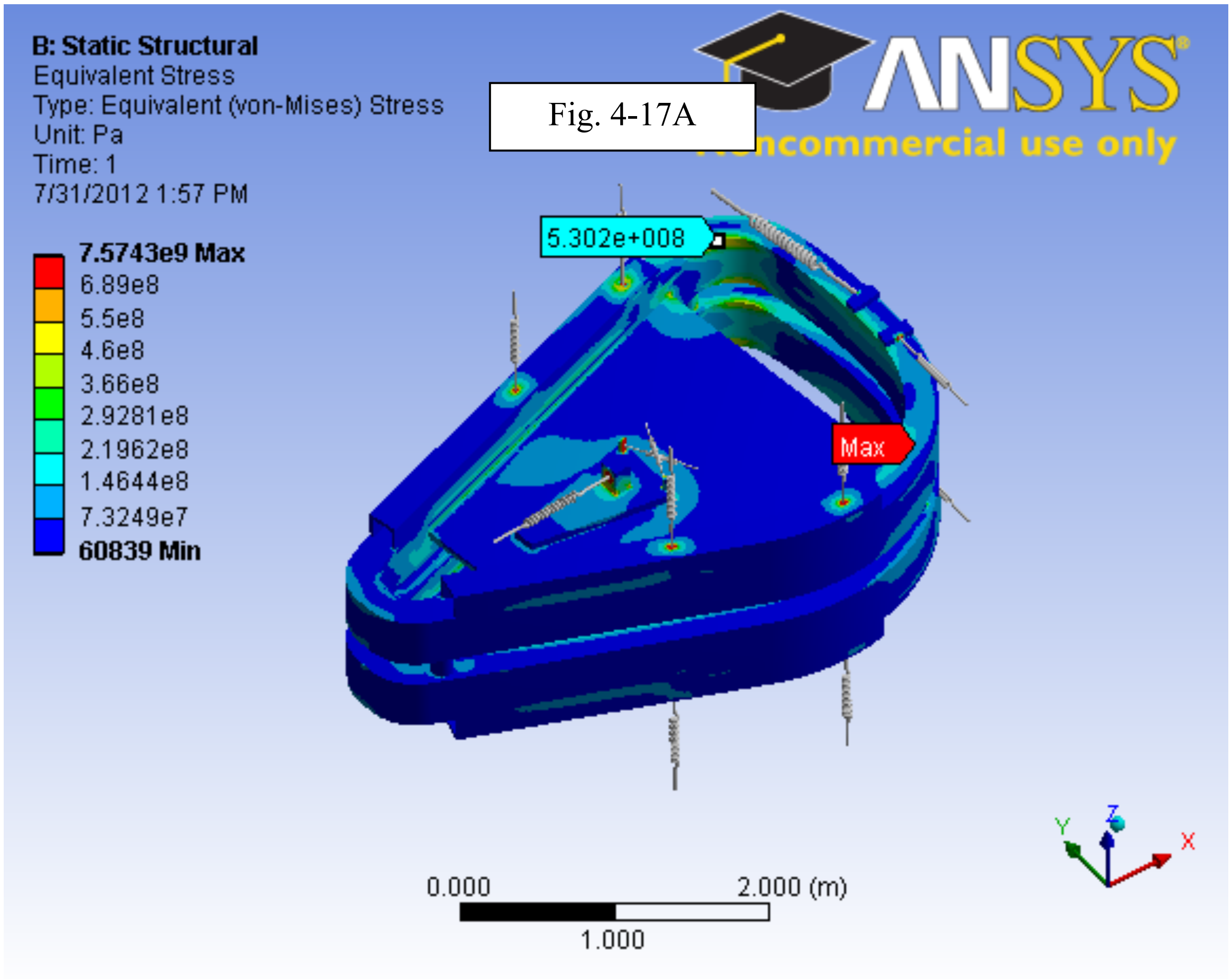

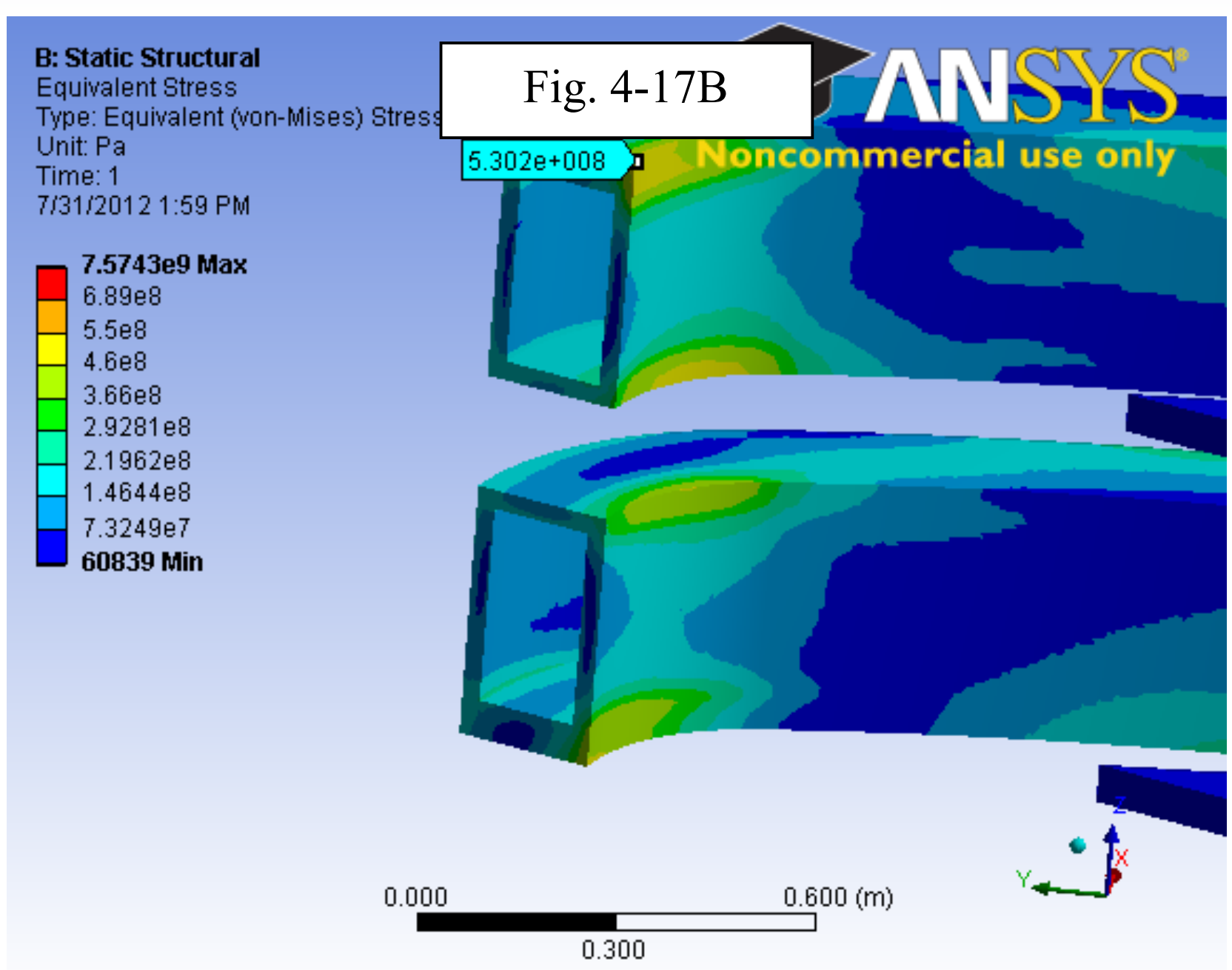

**Fig. 4-17. von Mises stress contour on the coil case with 4K + normal operating EM loads.**

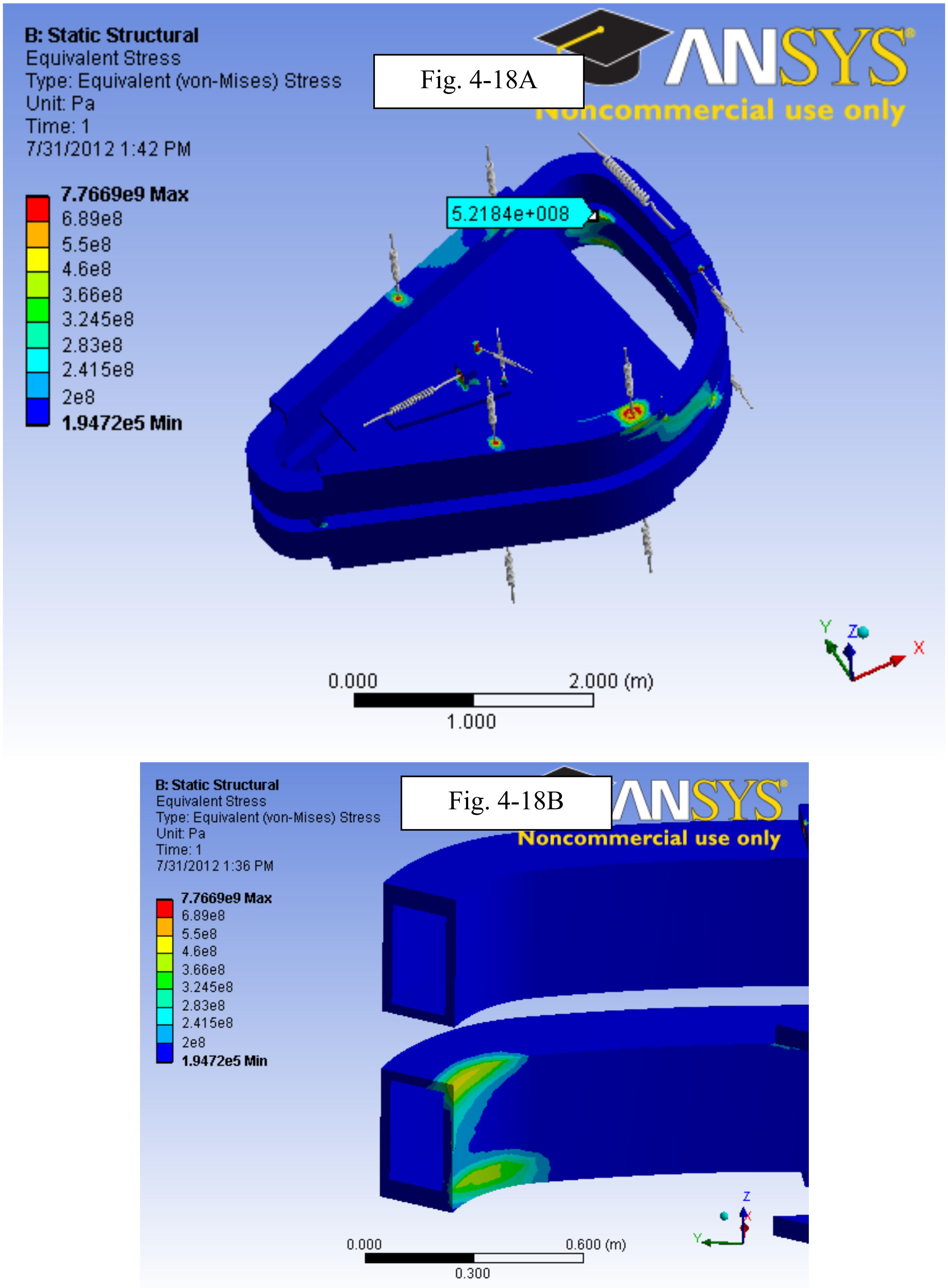

Fig. 4-18. von Mises stress contour on the coil case with 4K + fault condition EM loads.

### 4.4.2. Warm-to-Cold Strut Sizing

The warm-to-cold support struts will carry the weight while maintaining the position of the cold mass. The peak load they carry during all conditions is critical to properly size each strut in order to preserve their structural integrity, while minimizing the heat load to the liquid helium. Warm-to-cold supports are commonly sized by mechanical strength and then sized by EM stiffness separately, and the design uses the worst-case of the two methods. The warm-to-cold struts are sized only by mechanical strength here because supports in three orthogonal directions are being employed.

The sizing of the warm-to-cold struts uses the largest force in each strut among: (1) cool-down to 4K, (2) normal operation, and (3) fault condition. Table 4-6 summarizes the loads in each strut from each of the three analyses. The fault condition simulates a short in the lower coil during fast dump in the upper coil. As previously mentioned, the flux is conserved during this event, so the current in the lower coil increases from the 1,720,000A·turns being applied to the lower coil. To account for this increased current, hence increased body load, the EM loads are scaled by 1.3x when imported from Maxwell into ANSYS. Table 4-6 reports the load in each warm-to-cold support strut for each of the three conditions and then the maximum load from these three conditions. This maximum load is the design load of each support strut.

**Table 4-6. Summary of the strut forces.**

| | | Resulting Loads (MN) | | | | | | Maximum Load (MN) |
|---|---|---|---|---|---|---|---|---|
| Operating Condition | | N/A | | Normal | | Fault | | |
| Applied Loads | | Thermal | | EM & Thermal | | EM & Thermal | | |
| Strut Location | | Upper | Lower | Upper | Lower | Upper | Lower | |
| Strut # | 1 | .21 | .15 | .59 | .57 | .62 | 0 | 0.62 |
| | 2 | .59 | .54 | 1.02 | .93 | 2.85 | 0 | 2.85 |
| | 3 | .17 | .12 | .40 | .61 | .42 | 0 | 0.61 |
| | 4 | .32 | .28 | .51 | .58 | 2.35 | 0 | 2.35 |
| | 5 | .01 | .01 | .09 | .08 | .1 | .1 | 0.10 |
| | 6 | .01 | .01 | 0 | 0 | .01 | 0 | 0.01 |
| | 7 | 0 | 0 | 0 | 0 | 0 | 0 | 0 |
| | 8 | 0 | 0 | .13 | .10 | .03 | .15 | 0.15 |
| | 9 | .04 | .04 | 1.88 | 1.75 | 1.8 | .52 | 1.88 |

Table 4-7 summarizes the minimum diameter warm-to-cold strut required to bear the predicted peak load for each warm-to-cold support strut. For example, Nitronic 50 radial Strut #9 requires a minimum ∅6.79cm at the cold end and ∅8.03cm at the warm end. It is recommended that a minimum ∅1.0cm rod be used for Struts #6 & # 7 as they typically do not carry any tensile loads, but should be included in the design for any unforeseen circumstances. Nitronic 50 rods

are available up to ∅10in, so a monolithic rod can be machined to all of the required diameters listed in Table 4-7.

It should be noted that the support struts do not necessarily need to be circular. The design uses non-circular cross-sections for Struts #7 & #8. The diameter, Dmin, in Table 4-7 is a characteristic dimension as if each strut had a circular cross-section. For example, the cross sectional areas of Struts #7 & #8 are equal to $\pi(Dmin/2)^2$.

The shrinkage of the struts during cool-down and the flexing of their connection to the cryostat will affect the resulting load in each strut. While the effect of each may cancel each other out, these loads should be considered in the next design phase. Additional stresses in the struts due to bending are also neglected; however, spherical washers are used to connect the struts to both the cryostat and the cold mass.

**Table 4-7. Minimum required diameters of warm-to-cold struts to bear maximum loads in Tables 6.**

| Strut # | Maximum Load (MN) | Temperature (K) | Dmin | | Strut Orientation |
|---|---|---|---|---|---|
| | | | cm | in | |
| 1 | 0.62 | 4 | 3.90 | 1.54 | Axial |
| | | 296 | 4.61 | 1.82 | |
| 2 | 2.85 | 4 | 8.36 | 3.29 | |
| | | 296 | 9.89 | 3.89 | |
| 3 | 0.61 | 4 | 3.87 | 1.52 | |
| | | 296 | 4.58 | 1.80 | |
| 4 | 2.35 | 4 | 7.59 | 2.99 | |
| | | 296 | 8.98 | 3.54 | |
| 5 | 0.1 | 4 | 1.57 | 0.62 | Azimuthal |
| | | 296 | 1.85 | 0.73 | |
| 6* | 0.01 | 4 | 1.00 | 0.39 | |
| | | 296 | 1.00 | 0.39 | |
| 7* | 0.01 | 4 | 1.00 | 0.39 | |
| | | 296 | 1.00 | 0.39 | |
| 8 | 0.15 | 4 | 1.92 | 0.76 | |
| | | 296 | 2.27 | 0.89 | |
| 9 | 1.88 | 4 | 6.79 | 2.67 | Radial |
| | | 296 | 8.03 | 3.16 | |

*:It is recommended that a minimum diameter of 1cm be used for the warm-to-cold struts.

### 4.4.3. Cryostat

The geometry of the cryostat was described in Section 2.4. Fig. 4-19 depicts the loads applied to the cryostat from the warm-to-cold struts during normal operation. Atmospheric pressure is applied to all outside faces and gravity is considered. In reality, the cryostat will rest on the iron yoke and its loads be reacted and distributed through the iron. In the simulation, the faces of the cryostat that would come in contact with the iron yoke are constrained with frictionless supports. Frictionless supports constrain displacement normal to the face and allow in-plane sliding. These faces of contact between the cryostat and iron yoke should be designed to be nominally abutting one another – otherwise the cryostat will shift under the tremendous loads from the warm-to-cold struts until contact occurs.

The geometry is meshed with 3D shell elements, so that the thickness of the cryostat is not explicitly modeled. The advantage of this approach is to reduce the elements comprising the structure to a reasonable number. The disadvantage of this approach is shell elements do not carry normal stress (plane stress). This prevents accurate results when high forces are applied normal to the shell plane, such as at the transitions from vertical to horizontal sheets – several of such transitions exist in the cryostat.

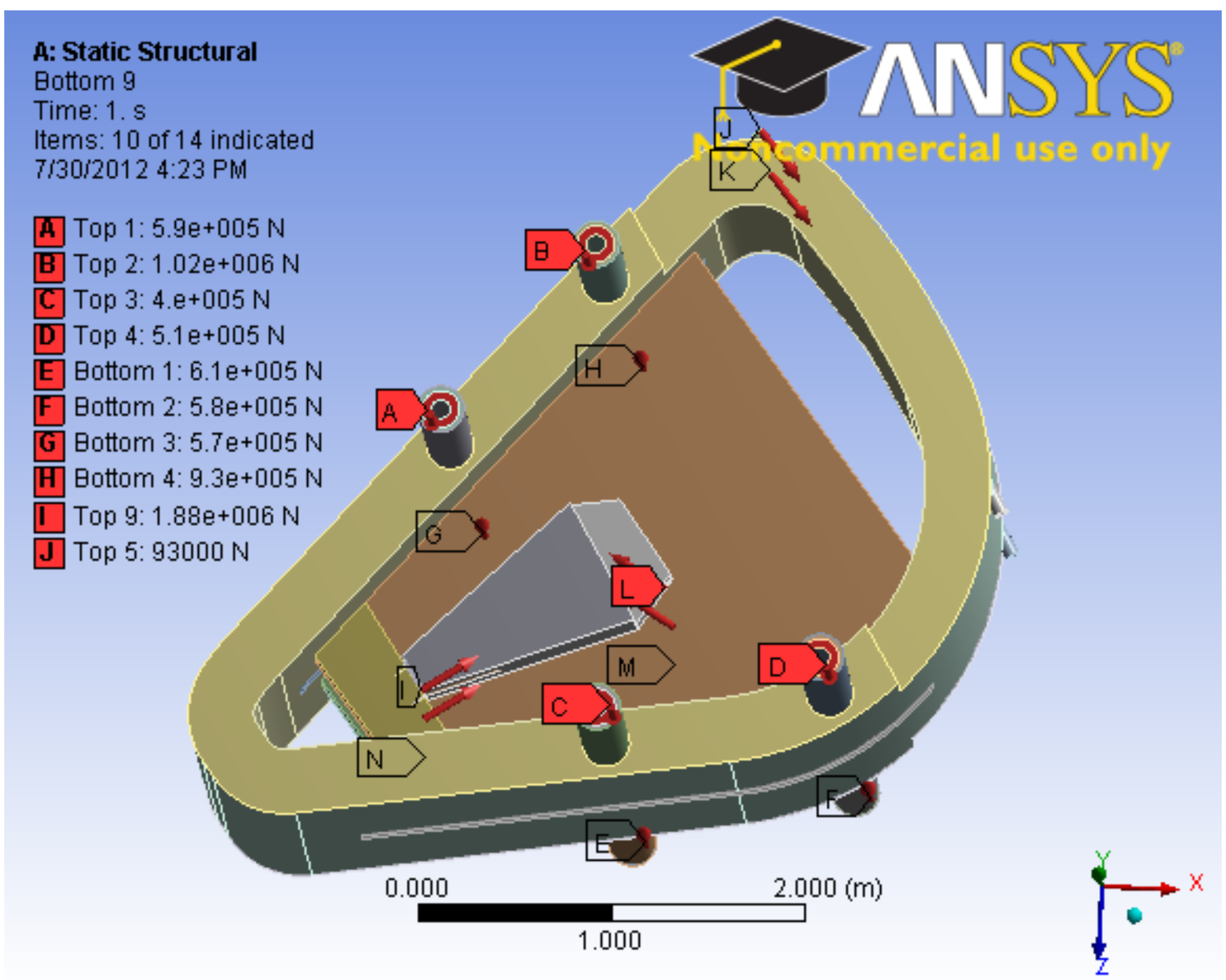

Fig. 4-19. Forces applied to the cryostat from warm-to-cold support struts during normal operation.

Fig. 4-20 and Fig. 4-21 show the von Mises stress in the cryostat during normal operation and fault condition, respectively. The red color indicates stresses greater than the allowable for stainless steel in bending (1.5*Sm=275MPa), the orange indicates stresses greater than the allowable for welded stainless steel in bending (1.5*Sm=241MPa), the yellow color indicates stresses greater than allowable for stainless steel membrane stresses (Sm=184MPa, and light green indicated stresses greater than allowable for welded stainless steel membrane stresses (Sm=161MPa). The allowable stresses for the cryostat are so much lower than for the coil case because the coil case is at 4K and the cryostat is at RT.

The stresses in the cryostat do not vary significantly through its thickness, so the stresses in Fig. 4-20 & Fig. 4-21 are membrane stresses. The cryostat shell that is in the yellow, orange, and red color exceeds the allowable stress. Notice that most of these areas are at the 90° transition where vertical shells connect to horizontal shells. As previously mentioned, the data in these regions are not physical, but these regions can be strengthened very easily using fillets and/or ribs. The concerning location in both Fig. 4-20 & Fig. 4-21 have peak stresses highlighted as 350MPa and 422MPa, respectively. The other concerning location is at the 413MPa stress point highlighted in Fig. 4-21. This is a location the will need to be strengthened in the next stage of the design.

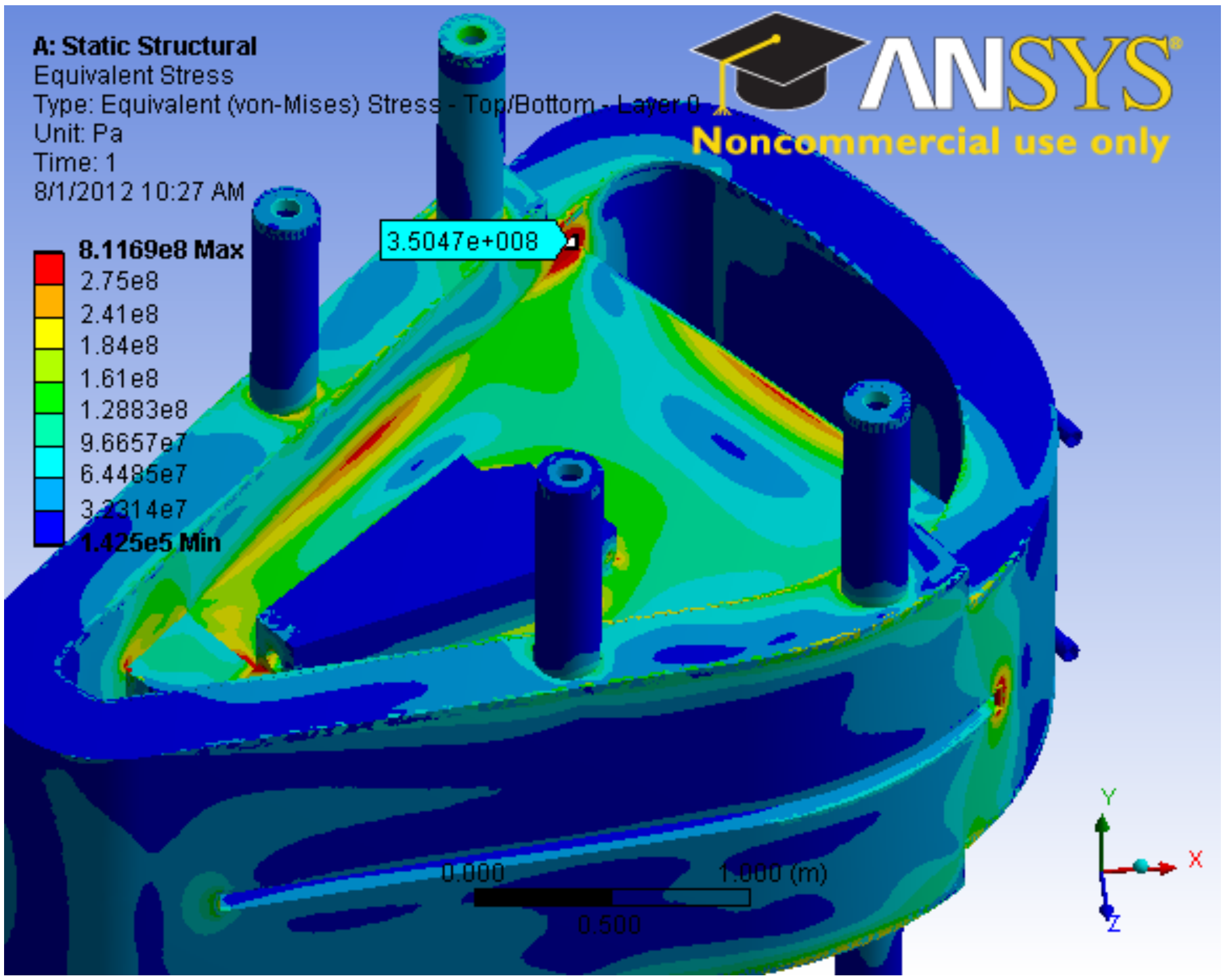

Fig. 4-20. Von Mises stress contours on cryostat, normal operation.

The tubes surrounding the axial warm-to-cold struts bear compressive loads that can lead to buckling of the tube. This can easily be prevented using external ribs and they must be designed such that there is no interference with the iron. The buckling analysis of the tubes for both normal operation and fault condition should be performed with a more detailed design of the cryostat. This detailed design should conform to the iron at the points of contact and include additional support to bear the higher stresses located in Fig. 4-20 & Fig. 4-21.

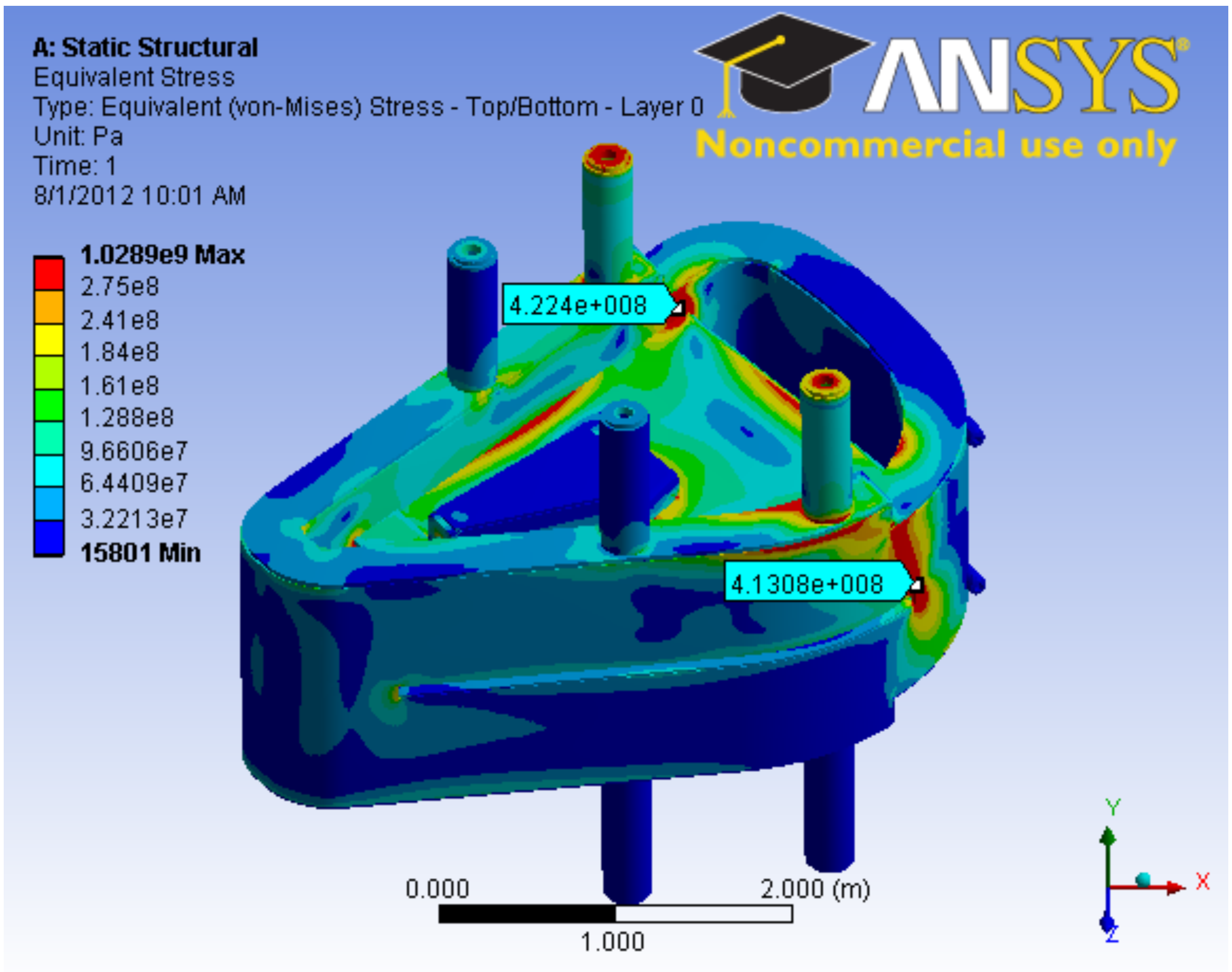

Fig. 4-21. von Mises stresses on cryostat during fault condition.

## 4.5.Summary

A conceptual design for the coil case, warm-to-cold struts and cryostat for the SRC is presented and analyzed. The structural analysis indicates the designs will safely bear cool-down loads as well as EM loads during normal operation and fault condition. The mechanical simulations yield much data that indicate:

- The coil supports carry the Lorentz forces on the coils without being over-stressed,
- The deflections of the coil supports for EM and thermal loads are reasonable,
- Slip planes between the winding pack and the coil support will be necessary,
- May need to sacrifice some of the iron around the azimuthal warm-to-cold struts to allow for necessary clearances,
- Warm-to-cold struts require solid rods with large diameter and length – the long length may require additional iron material to be sacrificed,
- A buckling analysis should be performed on the cryostat, especially on the tubes surrounding the axial warm-to-cold struts to assist in designing ribs.

## 5. Cryogenic Refrigeration System

The superconducting coil and associated cold mass are cooled to 4.5K via forced flow of supercritical helium (SCHe) within a closed loop at 3.0atm. A schematic of the flow path is shown in Fig. 5-1. SCHe enters each of 12 magnet coils at 4.5K, 3.0atm, and a flow rate of 15g/s. Upon entering the coil, the flow splits into 5 parallel paths at nominally 3 g/s each and accepts a thermal load of approximately 3.2W. The flow is recombined at the end of the coil and exits at 4.6K without any ionizing heat load into the coil. The SCHe then passes through a series of heat exchangers at the base of 7 structural supports where it accepts a load of 16.2W, exiting slightly above 4.97K. Flow exiting the support heat exchangers then passes through a pump where it accepts a load of 1.6W, exiting slightly below 5.0K. Calculation of the pump load assumes a pump efficiency of 75% and a total pressure drop per coil of 10kPa. Finally, the SCHe enters the refrigerator heat exchanger where it exchanges heat with a 4.2K liquid Helium bath. The SCHe enters slightly below 5.0K and exits at 4.3K. The heat exchanger is a simple coiled 12.7mm inner diameter tube 1.4m in length.

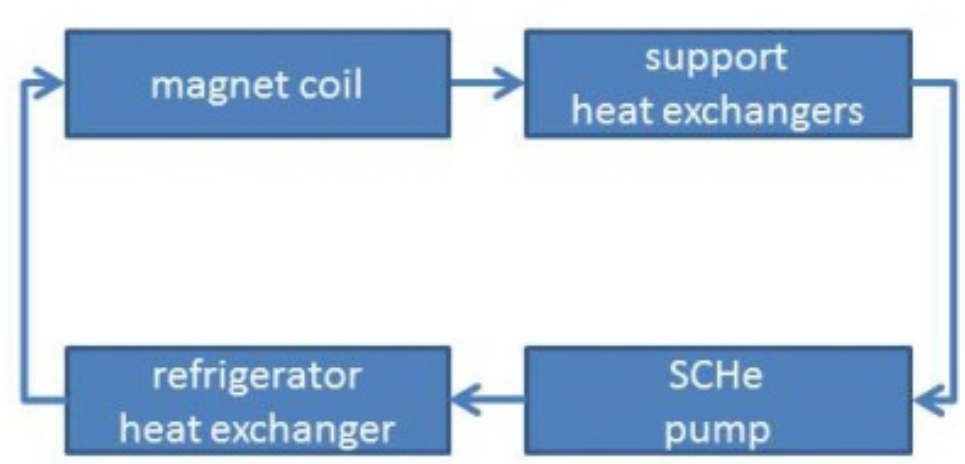

Fig. 5-1. Flow path schematic for the SCHe loop.

The breakdown of the thermal load in the magnet coil consists of contributions from 5kA current leads, MLI insulation, and heat from 9 mechanical supports. The current leads are binary current leads, with a warm section vapor-cooled by atmospheric liquid nitrogen (LN2), and a cold section composed of high temperature superconductor (HTS). The warm section heat load is 125W per lead, corresponding to 5.6L/h of LN2. The conduction heat leak through HTS is 1.2W per lead pair, with only 2 leads per sector (i.e. 1 lead per coil). The MLI around the cold mass consists of two stages: an LN2 intercepting layer at 77K and a 4.5K layer at the cold mass. The heat leak from 300K that is intercepted at 77K is 30W. The heat leak from 77K to 4.5K that is deposited into the SCHe inside the coil is 0.5W. The remaining heat leak to the SCHe in the coil comes from 9 supports: 7 supports with 77K intercepts located at 1/3 of their length as well as 4.6K heat exchangers, and 2 supports with77K intercepts only (i.e. no 4.6K heat exchangers). The 2 supports without heat exchangers have a smaller diameter and only contribute 0.5W to the SCHe in the coil. The remaining 7 supports combine for a total heat leak of 1.6W. This assumes the conductance of the heat exchanger is 10W/K and the thermal resistance from the bottom of the support to the SCHe flow is on the order of 1K/W. The LN2 load for all 9 supports is 201W. This amounts to a 3.2W (0.6W + 0.5W + 2.1W) heat load to the SCHe flowing in the magnet coil.

The overall refrigeration need per coil is 371W at 77K (125W for a single current lead, 30W for the cryostat intercept, 201W for the supports, and 15W for transfer line leakage) and 22W at 4.2K (3.2W for the magnet coil, 16.2W for the supports, 1.6W for the pump, and 1.0W for transfer line leakage). For 6 sectors, each with 2 coils, this results in a total of 4.45kW at 77K and 264W at 4.2K. The total flow rate of SCHe is 180g/s. Table 5-1 summarizes the heat loads.

In addition, the tolerable heat load due to ionizing radiation has been estimated based on maintaining a superconductor temperature martin 1.5 K in the presence of a steady, uniform heat generation throughout the coil. Based on the 1.5K temperature margin limit, the allowable uniform heat load is approximately 46μW/g.

**Table 5-1. Summary of heat loads*.**

| Component | | Cold Stage Load (W) | Warm Stage Load (W) |
|---|---|---|---|
| Coil | cryostat | 0.5 | 30 |
| | supports (after intercepts) | 2.1 | - |
| | leads | 0.6 | 125 |
| Supports | heat exchangers | 16.2 | 201 |
| Pump | | 1.6 | - |
| Transfer line | | 1.0 | 15 |
| | Total | 22.0 | 371 |

*: Excludes ionizing radiation

## 5.1.Supercritical Helium Pump and System Pressure Drop

Cooling for the coil is provided by the forced flow of supercritical helium (SCHe) in a closed loop. SCHe was chosen as the heat exchange fluid over two-phase helium in order to avoid flow maldistribution and quench problems associated with potential vapor collection locations within the magnet coil as described by Morpurgo [1]. System pressure drops in the SCHe stream are due to wall friction in the coil and transfer lines, pressure drop in bends and valves, as well as flow resistance in the heat exchangers.

The flow path within a single coil is shown in Fig. 5-2. SCHe passing through the cryostat outer wall enters the coil via a plenum (not shown), where it is divided into 5 parallel channels. The dimensions of the channel are provided in Table 5-2 along with flow characteristics and fluid properties within the channel. Helium fluid properties are calculated at 4.5K and 3.0atm using EES [2] property routines.

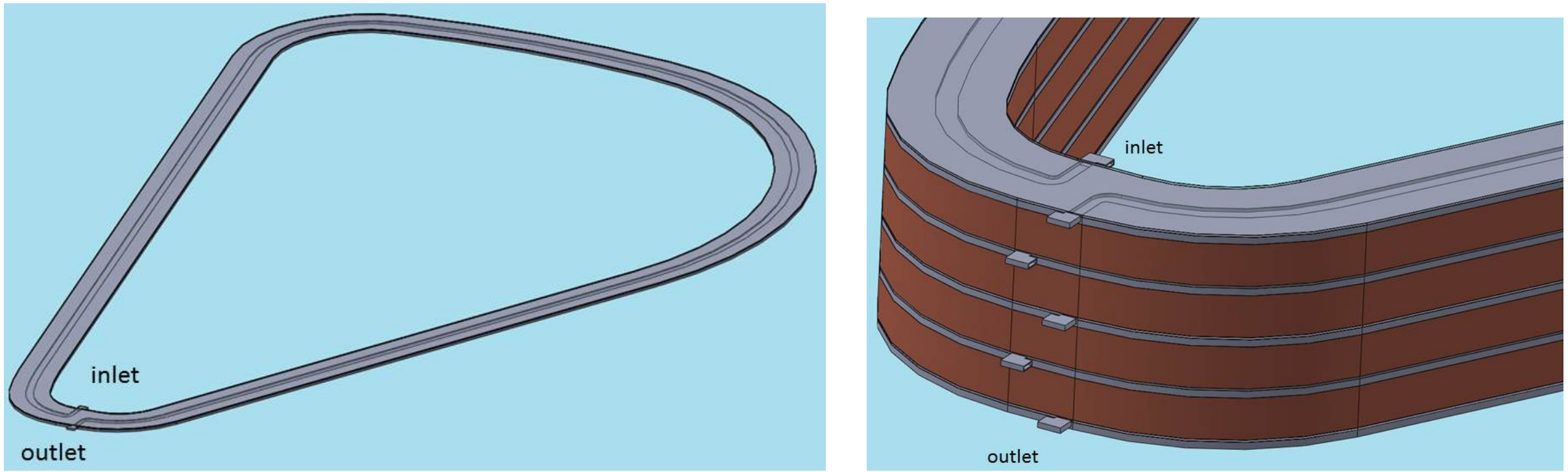

Fig. 5-2. Flow path and distribution in the magnet coil.

It can be seen from Table 5-2 that the flow within the coil is turbulent, Re = 34900. The pressure drop in a single channel of the magnet coil is calculated using the correlation by Zigrang and Sylvester [3] and results in a total pressure drop of 40.0Pa across the coil.

Flynn [4] describes a supercritical helium pump designed by Barber-Nichols, Arvada Colorado. The pump has a head rise of 61m at a flow rate of 250g/s with an inlet pressure of 3.9atm and a temperature of 4.5K. Assuming the usage of a similar pump, the pressure drop in the coil, as well as any transfer lines, is negligible. This means that the pressure drop across heat exchangers and valves will drive the size of the pump.

Table 5-2. SCHe flow parameters at 4.5K and 3.0atm within a single coil channel.

| | | | |
|---|---|---|---|
| Channel height | $H_{ch}$ | 5.0 | mm |
| Channel width | $W_{ch}$ | 30.0 | mm |
| Channel length | $L_{ch}$ | 11.0 | m |
| Hydraulic diameter | $D_h$ | 8.57 | mm |
| Density | $\rho$ | 152 | $kg/m^3$ |
| Viscosity | $\mu$ | 4.92E-6 | kg/m-s |
| Mass flow rate | $m$ | 3.0 | g/s |
| Velocity | $v$ | 0.132 | m/s |
| Reynolds number | Re | 34900 | - |
| Pressure drop | $\Delta P$ | 40.0 | Pa |
| Heat transfer coefficient | $h$ | 221 | $W/m^2$-K |

## 5.2.MLI Heat Load

A significant heat load to the magnet coil arises from the transfer of heat via radiation from the 300K outer wall of the cryostat to the 4.5K inner cold mass. Vacuum pressure impregnation of the coil ensures that the heat load to the cold mass will be uniformly distributed and shared by each of the five SCHe streams providing cooling, thereby preventing local hotspots and alleviating the risk of quench. The heat leak to the 4.5K cold mass is minimized with MLI shielding and a 77K heat intercept. The intercept is situated approximately halfway between inner and outer walls of the cryostat. The intercept is cooled by the forced flow of low vapor quality two-phase nitrogen near ambient pressure. The two-phase stream also provides cooling to the support heat exchangers discussed below.

The heat leak to the 77K intercept is estimated from data found in Iwasa [5]. For 60 layers of MLI insulation in a 2.5cm vacuum gap (i.e. 24 layers/cm), the heat leak is 2.5W/m$^2$. The surface area of the intercept is estimated assuming a rectangular cross-section of 34.5cm by 20.0cm and a total length of 11m. This results in a heat leak of 30W to the 77K intercept.

The heat leak to the 4.5K cold mass from the 77K intercept is estimated assuming a leak rate of 42mW/m$^2$ [6]. This leak rate can be accomplished with 3-5 layers of MLI between layers of EM #425 adhesive aluminum tape. The surface area of the cold mass is estimated assuming a rectangular cross-section of 31.0cm by 16.0cm and a total length of 11m. This results in a heat leak of less than 0.5W to the 4.5K cold mass.

## 5.3. Structural Supports Heat Load

The primary heat leak to the magnet coil is due to heat conduction through the cold-mass structural supports. The nine structural supports for a single magnet coil are shown in Fig. 5-3. For each support, a portion of the heat load from 300K is intercepted at 77K via a two-phase liquid nitrogen stream. This occurs via a heat exchanger located at approximately 1/3 of the way from the room temperature end of the support. An additional heat exchanger at the cold end of the support deposits heat into the SCHe stream after it has picked up a load from the magnet coil.

In order to calculate the heat transferred to the heat exchanging fluids, and the heat leakage to the magnet coil, it is assumed that the heat exchangers have a conductance of 10W/K and that the resistance to heat transfer after the 4.6K heat exchanger to the magnet coil SCHe stream is on the order of 1.0K/W. The resistance to heat transfer for the top third and bottom two-thirds of each support is provided in Table 5-3. In addition, the heat intercepted by the heat exchangers, the heat leak into the magnet coil, and the average temperature drop from the support to the heat exchange fluid are shown in Table 5-3. The total heat flow intercepted by forced flow of two-phase nitrogen in the heat exchangers is 201W, with a maximum temperature difference between the fluid and the support of 6.1K at support 2. The total heat flow intercepted by the SCHe stream in the heat exchangers after it has left the magnet coil is 16.2W, with a maximum temperature rise of 0.52K, again at support 2. The remaining heat transferred to the magnet coil SCHe stream is 2.05W.

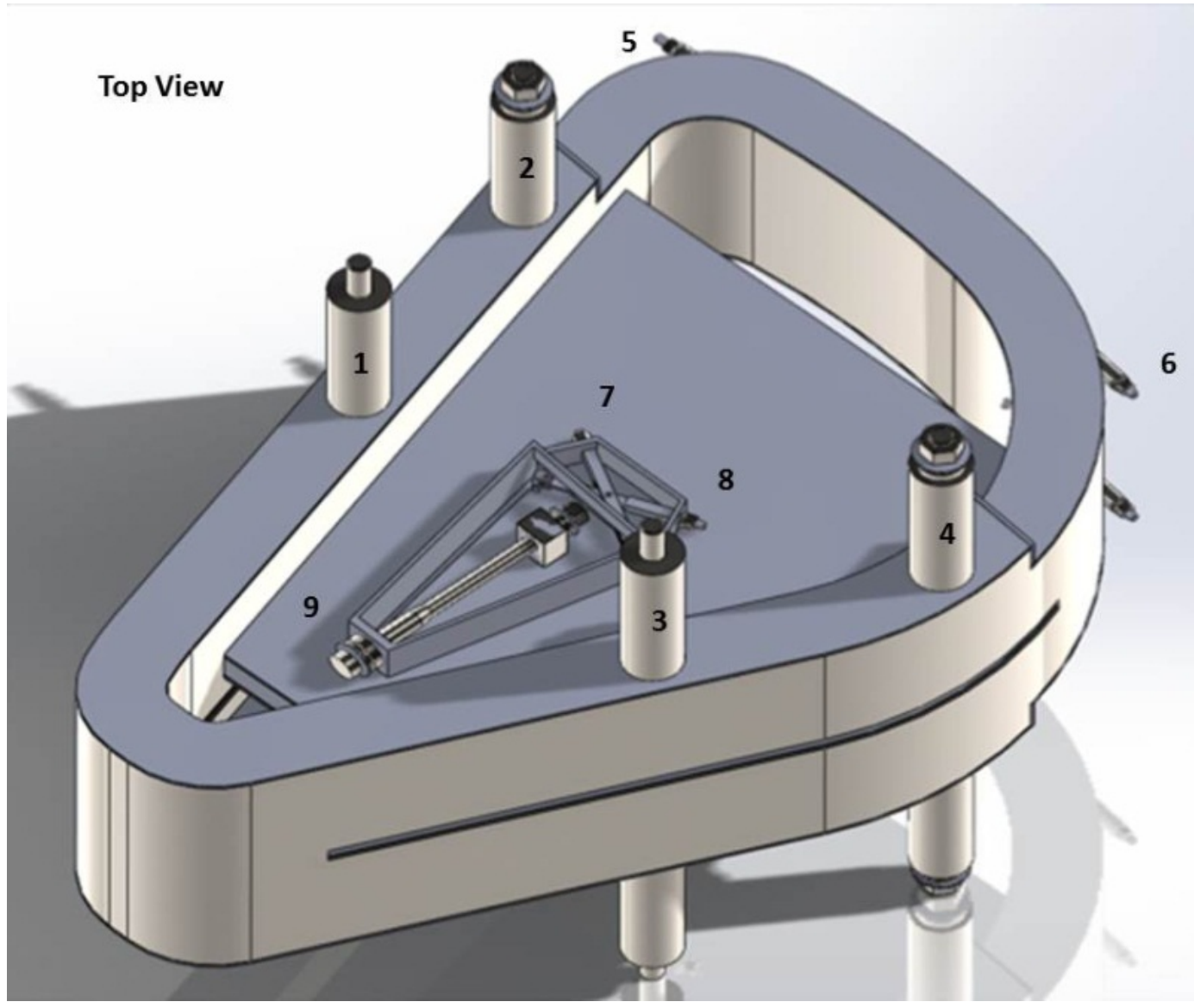

Fig. 5-3. Top view of the cryostat and the nine structural supports for the top coil.

Table 5-3. Heat leakage to magnet coil via conduction through supports. Note, supports 5 and 6 do not have a heat exchanger at 4.6K

| Support | $R_{top1/3}$ (K/W) | $R_{bot2/3}$ (K/W) | $q_{77K}$ (W) | $q_{4.6K}$ (W) | $q_{leak}$ (W) | $\Delta T_{77K}$ (K) | $\Delta T_{4.6K}$ (K) |
|---|---|---|---|---|---|---|---|
| 1 | 10.8 | 45.3 | 18.9 | 1.5 | 0.15 | 1.89 | 0.15 |
| 2 | 3.3 | 13.7 | 61.0 | 5.2 | 0.52 | 6.10 | 0.52 |
| 3 | 11.3 | 47.5 | 18.0 | 1.4 | 0.14 | 1.80 | 0.14 |
| 4 | 3.9 | 16.6 | 50.5 | 4.2 | 0.42 | 5.05 | 0.42 |
| 5 | 138.1 | 579.6 | 1.5 | 0.0 | 0.13 | 0.15 | 0.00 |
| 6 | 56.7 | 237.9 | 3.6 | 0.0 | 0.31 | 0.36 | 0.00 |
| 7 | 39.8 | 166.8 | 5.2 | 0.4 | 0.04 | 0.52 | 0.04 |
| 8 | 37.1 | 155.7 | 5.5 | 0.4 | 0.04 | 0.55 | 0.04 |
| 9 | 5.4 | 22.8 | 37.0 | 3.0 | 0.30 | 3.70 | 0.30 |
| TOTAL | - | - | 201 | 16.2 | 2.05 | - | |

Due to the configuration of the supports and the flow of SCHe within the magnet coil, it is possible, in a worst case scenario, that the SCHe flowing in the top channel of five (see Fig. 5-2), would intercept all of the heat load flowing into the coil from supports 1 through 4, without transferring any of the heat to the other four streams. This scenario does not seem entirely unlikely due to the amount of insulation between flow channels in the coil. If this were to occur, the temperature rise of the SCHe with a flow rate of 3g/s, and a heat capacity of 3.17kJ/kg-K from a 1.23W load (the sum of loads from supports 1 through 4) would be approximately 0.13K.

## 5.4. Current Leads Heat Load

Each sector comprised of two magnet coils has a single set of current leads. The current leads for this application are binary leads, with a normal copper lead between 300K and 77K, and a high temperature superconducting (HTS) lead between 77K and 4.5K. Binary leads are chosen for two reasons: they reduce the amount of refrigeration necessary at 4.5K, displacing the load to 77K requiring less electrical input, and they simplify the SCHe stream by allowing it to be a closed loop (i.e. no Helium boil off). In the binary lead configuration, atmospheric liquid nitrogen at 77K intercepts most of the heat traveling down the lead via conduction and generated within the lead during operation. This requires refrigeration of 25W/kA per lead, per Chang [7]. For a pair of 5kA leads, this results in a 77K heat load of 125W per lead. The remaining heat travels through the HTS to the SCHe stream at 4.5K. HTS-110 [8] currently manufactures HTS leads. Though they do not manufacture 5kA leads off-the-shelf, extrapolating from their data for 2kA leads and below, the heat leak to the cold stage from 5kA leads will be approximately 1.2W per pair of HTS leads. This means that the heat leak per coil managed by SCHe within the coil is 0.6W.

## 5.5.Ionizing Radiation Load

During operation of the magnet it is possible for ionizing radiation to penetrate through the magnet coil and deposit a thermal load that will need to be managed by the flow of SCHe through the stainless steel cooling channel. Potential local hotspots that could lead to quench are avoided via vacuum pressure impregnation of the magnet coil, which ensures a continuous and uniform conduction path within the coil and between the coil and the cooling channel. In order to estimate this load, a steady state, 2D finite element thermal model of a unit cell of the magnet coil was created and solved under the constraint that the warmest conductor temperature is allowed to rise to 5K while the coil is immersed in a uniform body heat load (μW/g). The 5K operating temperature in the conductor will still enable approximately 1.5K of temperature margin [9] in the superconductor. The unit cell geometry is provided in Fig. 5-4. The model consists of a single turn of a double pancake layer and assumes that each double pancake layer experiences the same thermal conditions. On top of the double pancake layer is a layer of steel, modeling the SCHe channel. The boundary conditions on the left, right and bottom sides of the model are adiabatic. The top boundary, which corresponds to the flow of SCHe through the stainless steel is a convection boundary, with a specified heat transfer coefficient and fluid temperature of 4.5K. To model the ionizing radiation, a uniform generation load at each node in the model is applied. The materials shown in Fig. 5-4 correspond to the copper channel, A2, the insulation between conductor turns, A3, the stainless steel cooling channel, A4, and an effective thermal region, A1, composed of the composite superconductor and copper wire, and solder. The effective thermal conductivity of the region is calculated by summing the area-weighted average of conductivity for each material. The conductivity for each material at 4K was taken from Iwasa [5] and is displayed in Table 5-4 along with the corresponding area. The result is an effective conductivity of 187W/m-K.

Table 5-4. Conductivity and area for materials in the unit cell model evaluated at 4K.

| Material | Area ($cm^2$) | Conductivity (W/m-K) |
|---|---|---|
| Copper | 8.32 | 420 |
| NbTi | 6.40 | 0.17 |
| Solder | 4.38 | 15.0 |
| SS304 | - | 0.2 |
| Insulation | - | 0.1 |

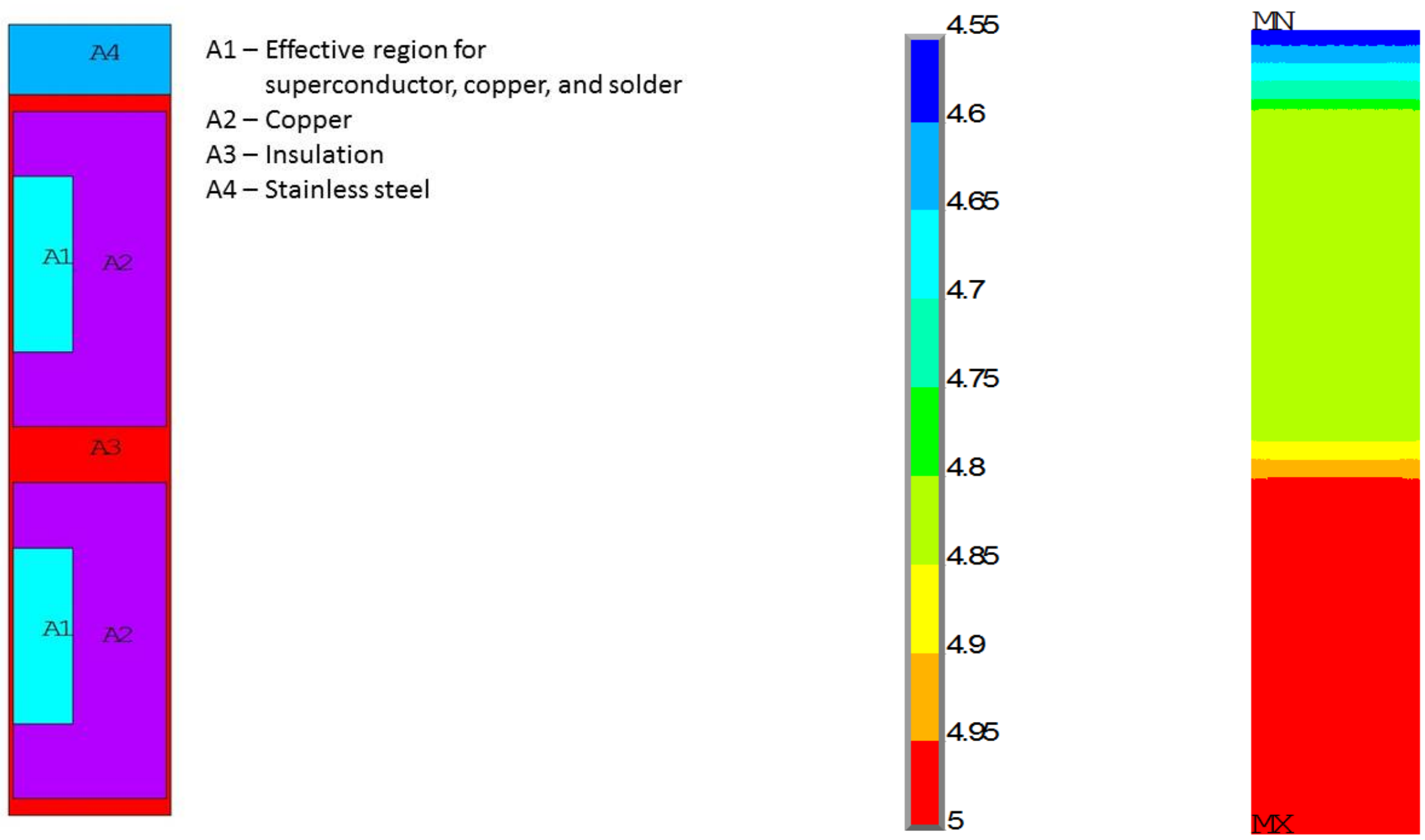

Fig. 5-4. Unit cell geometry and temperature contours for heat generation of 410W/$m^3$ and heat transfer coefficient of 250W/$m^2$-K.

Fig. 5-5 plots the maximum temperature rise in the unit cell as a function of the uniformly applied heat generation rate within the model for various values of the heat transfer coefficient along the top stainless steel boundary. The results show that for a heat transfer coefficient of 250W/$m^2$-K, a heat generation rate of approximately 410W/$m^3$ results in a maximum coil temperature of 5K. With the density of copper being approximately 8900kg/$m^3$, this can also be expressed as 46μW/g. Fig. 5-4 plots the temperature contours for this calculation.

The heat transfer coefficient in the channel is calculated from the correlation by Gnielinski [10]. For the flow described in Table 5-4, the heat transfer coefficient is approximately 220W/$m^2$-K.

**Fig. 5-5. Maximum coil temperature as a function of heat generation for values of heat transfer coefficient within the cooling channel.**

## 5.6.References

# 6. Recommended R&D Leading to Sector Fabrication

The conceptual design for the Daedalus magnet sector, as reflected in this report, has already benefitted from an early, positive interaction with the Daedalus team. This interaction focused on a directed effort to define coil fabrication rules that would render the coil to be more easily manufactured, resulting in lower risk. This interaction and the evolved rules are discussed in more detail in the Coil Design Rules subsection of the Magnetic Design section of this report. We believe that the conceptual design developed here, which largely abides by these rules, will pay dividends in reducing both manufacturing and magnet performance risk going forward. Nevertheless, despite the coils largely adhering to the fabrication rules, there are some additional R&D tasks, related to the specific conceptual design developed here, that will further reduce risk. These tasks are discussed below, in no particular order.

As a general approach to R&D, we note that it is often useful to engage industry in small industrial subcontracts to study specific issues. In such cases, industry can simply supply the materials needed or, depending on their capabilities, might also be involved in testing. If testing is not within the industrial capability, additional small subcontracts can be placed with testing laboratories, possibly including MIT. This approach could be explored in some of the tasks discussed below.

The ideal time to perform this suggested R&D is prior to, or at latest in parallel with the Final Design activity for the sector magnet. Note that the schedule given in the Cost and Schedule volume assumes the R&D precedes the Final Design.

## 6.1.Support Rod Test and Intercept Design

The large diameter axial support rods are the largest source of heat leak to 4.5 K. These rods will probably be fabricated from Nitronic 50, a material with good strength and which has been used previously in magnet warm to cold supports. The rationale for the choice of the diameter of these rods is given in the structural section of this report. The diameter, however, is outside the "off the shelf" range of manufacturing for these rods. In addition, it is suggested that the rods be designed with heat intercepts at 77 K and at 4.5 K to reduce the heat leak into the coil. It is suggested that the combination of these specific issues be explored further in an R&D study.

## 6.2.Coil Case Fabrication and Fit-up

One of the risks that is always present in superconducting magnets, especially those that are conduction cooled, is that small relative motions can be sufficient to induce magnet quench from frictional heating. This risk can be amplified in magnet systems where the dimensions are large, as in the case of the Daedalus sectors, which have turn lengths on the order of 11 m. To combat this issue, we have chosen stainless steel coil cases that are ideally tightly fitted to the winding pack/cooling plate assemblies. Stainless steel has a thermal contraction coefficient from room temperature to 4.5 K that is close to that of copper. This choice of the case material, together with proper clamping of the winding by the case, will largely mitigate the risk of relative motion. However, the detailed implementation of the clamping vis-à-vis both the case welding design/sequence and the coil epoxy impregnation approach is a good candidate for further R&D study.

There are two separate but related issues to be explored here: 1) Getting the tight fit between the coil and the case, and 2) Developing a case welding procedure, which does not jeopardize the coil assembly, with consideration that the coil will already be installed inside the case when the welds are made. The case welds will likely be full penetration welds, so qualifying the welder and the procedure, probably with the aid of instrumented mockups including temperature sensors, will be necessary prior to performing the actual welds on the deliverable components.

## 6.3. Helium flow arrangement

Our conceptual design has chosen to cool the 5 winding pack cooling plates in parallel with supercritical helium at 3.0 atm. A question that invariably arises is whether series or parallel flow is the better arrangement. With parallel flow, heating in the plates nearer the outlet does not accumulate heat from other plates nearer the inlet as it does in a series arrangement. However, parallel flow often raises issues of flow stability and the ability to properly control the flow. In our design, the pressure drop through the plates is low and the sensitivity of the coil temperature to the heat transfer coefficient, which is a function of flow velocity, is shown in the cryogenic section to be relatively small. A further evaluation of these two approaches may still be warranted.

## 6.4. Tie-plate/cryostat assembly sequence

Our design uses a structural tie plate between the two long inner walls of the coil case to restrain the electromagnetic loads which tend to force the coil into an overall circular shape. These tie plates are at 4.5 K, and must be bridged by the inner wall of the nitrogen shield and cryostat, while maintaining minimal heat leak to 4.5 K and simultaneously maintaining the structural integrity of the cryostat. While we have presented a high-level conceptual approach to this assembly, further definition of the details would greatly benefit from a mockup of this interface region.

## 6.5. Magnet quench performance

The quench protection approach for the conceptual design employs a dump resistor that is switched into the coil circuit while, at the same time, the power supply is disconnected. This action is triggered by an active quench detection system based on voltage taps whose signals are processed through electronics that are designed to distinguish between inductive and resistive voltages. The quench analysis done to date is a simply based on $\int_0^\infty J_{Cu}^2 dt$ ($J_{Cu}$ = current density on the conductor copper stabilizer, dt = differential time) under the conservative assumption that the time constant for the dump is given by $L/R_{dump}$ (L = coil inductance, $R_{dump}$ = resistance of the dump resistor). The winding pack design section of this report gives the resulting hot spot temperature from this analysis as 120 K, and this is believed to be reasonably conservative. Because of the large size (~11 m equivalent circumference) of the Daedalus coils however, there is some concern that some sections of the coil can remain cool relative to the hot spot at or near the end of the transient. This condition leads to thermal stresses that should really be explored further by running a 3-d quench analysis on the winding and inputting that temperature distribution into a thermal FEA analysis. The results of this combined analysis can reveal expected tensile and shear stress magnitudes in the winding insulation system during a magnet quench. If these stresses are too high, mitigation techniques should be employed. One mitigation approach that could be successful is to use heaters in the winding pack that would be energized upon quench detection and spread the heating more uniformly into the winding. This would reduce differential thermal stress. As briefly explained in the winding pack section of this report, we have left space within the winding pack design to place these heaters if necessary. An early understanding of the quench performance would benefit the later detailed design of the winding pack.